\documentclass[table,xcdraw]{article}
\usepackage{amsmath}
\usepackage{amssymb}
\usepackage{amsthm}
\usepackage{txfonts}
\usepackage{authblk}
\usepackage{breakcites}
\usepackage{hyperref}
\usepackage{hyphenat}

\newtheorem{theorem}{Theorem}[section]
\newtheorem{lemma}[theorem]{Lemma}
\newtheorem{example}[theorem]{Example}
\newtheorem{definition}[theorem]{Definition}
\newtheorem{corollary}[theorem]{Corollary}
\newtheorem{proposition}[theorem]{Proposition}
\newtheorem{remark}[theorem]{Remark}
\newtheorem{notation}[theorem]{Notation}
\newtheorem{conjecture}[theorem]{Conjecture}

\usepackage{graphicx}
\usepackage[all]{xy}
\usepackage{lscape}
\usepackage{pdflscape}

\usepackage{stmaryrd}

\usepackage{cmll}

\usepackage{comment}

\usepackage{amscd}
\usepackage{amssymb,amsmath,amsthm}
\usepackage{mathptmx}
\usepackage{mathrsfs}
\usepackage{color}
\usepackage{xspace}
\usepackage{bussproofs}
\EnableBpAbbreviations

\usepackage{tikz}

  \usepackage{txfonts}

\usepackage{bpextra}

\usepackage{enumerate}

\usepackage{centernot}

\usepackage{mathtools}

\usepackage{hyperref}

\usepackage{cleveref}

\usepackage{relsize}

\usepackage{caption}

\usepackage{multirow}

\usepackage{array}
\newcolumntype{C}[1]{>{\centering\arraybackslash}p{#1}}
\newcolumntype{L}[1]{>{\arraybackslash}p{#1}}

\renewcommand{\emph}{\textbf}

\newcommand{\Prop}{\mathsf{Prop}}

\newcommand{\nomi}{\mathbf{i}}
\newcommand{\nomj}{\mathbf{j}}

\newcommand{\cnomm}{\mathbf{m}}
\newcommand{\cnomn}{\mathbf{n}}

\newcommand{\marginnote}[1]{\marginpar{\raggedright\tiny{#1}}}

\newcommand{\val}[1]{[\![{#1}]\!]}
\newcommand{\descr}[1]{(\![{#1}]\!)}
\renewcommand{\phi}{\varphi}

\newcommand{\bba}{\mathbb{A}}
\newcommand{\blhd}{{\blacktriangleleft}}
\newcommand{\brhd}{{\blacktriangleright}}
\newcommand{\oz}{\overline{z}}

\def\aol{\rule[0.5865ex]{1.38ex}{0.1ex}}

\def\pdra{\mbox{$\,>\mkern-8mu\raisebox{-0.065ex}{\aol}\,$}}

\newcommand{\starfor}{{/\!\!}_{\star}}
\newcommand{\circfor}{{/\!}_{\circ}}

\newcommand{\starback}{\backslash_{\star}}
\newcommand{\circback}{\backslash_{\circ}}

\newcommand{\upsE}{\raisebox{1ex}{\rotatebox{180}{$e$}}}

\usetikzlibrary{arrows}
\usepackage[shortlabels]{enumitem}
\setlist[enumerate]{topsep=.5mm}

\begin{document}
\title{Modal reduction principles across relational semantics}
\author[4,5]{Willem Conradie}
\author[1]{Andrea De Domenico}
\author[1]{Krishna Manoorkar\footnote{Krishna Manoorkar is supported by the NWO grant KIVI.2019.001 awarded to Alessandra Palmigiano.}}
\author[1,2]{Alessandra Palmigiano}
\author[1]{Mattia Panettiere}
\author[3]{Daira Pinto Prieto} 
\author[1]{Apostolos Tzimoulis}
\affil[1]{School of Business and Economics, Vrije Universiteit, Amsterdam, The Netherlands}
\affil[2]{Department of Mathematics and Applied Mathematics, University of Johannesburg, South Africa}
\affil[3]{ILLC, University of Amsterdam, the Netherlands}
\affil[4]{School of Mathematics, University the Witwatersrand, Johannesburg, South Africa}
\affil[5]{National Institute for Theoretical and Computational Sciences (NITheCS), South Africa}



\maketitle

\begin{abstract}
The present paper establishes systematic connections among the first-order correspondents of Sahlqvist modal reduction principles in various relational semantic settings, including crisp and many-valued Kripke frames, and crisp and many-valued polarity-based frames (aka enriched formal contexts).  Building on  unified correspondence theory, we aim at introducing a theoretical environment which makes it possible to: (a) compare and inter-relate the various frame correspondents (in different relational settings) of any given Sahlqvist modal reduction principle; (b) recognize when first-order sentences in the frame-correspondence languages of different types of relational structures encode the same ``modal content"; (c)  meaningfully transfer and represent well known relational properties such as reflexivity, transitivity, symmetry, seriality, confluence, density, across different semantic contexts. These results can be understood as a first step in a research program aimed at making correspondence theory not just (methodologically) unified, but also (effectively) {\em parametric}.   
\end{abstract}
\tableofcontents

\section{Introduction}

Sahlqvist theory \cite{sahlqvist1975completeness,vB83,vanBenthem2001} identifies a large, syntactically specified class of modal formulas that are both canonical and elementary -- two highly desirable but generally undecidable properties. Over the last four decades,  many different authors have have been extending this theory to an ever widening range of non-classical logics by. Although starting as a model-theoretic investigation, algebraic and topological connections and methods have long been noted and employed in Sahlqvist theory (see e.g.\ \cite{Sambin:Vacarro:89,Jo94,DeRijke:Venema:95,GNV}). A recent line of investigation \cite{conradie2012algorithmic,CoGhPa14,ConPalSou,CoPa:non-dist}, named `unified correspondence', has  pursued this further and, by isolating the key algebraic and order-theoretic properties on which Sahlqvist results turn, made it possible to uniformly obtain Sahqvist correspondence and canonicity results for a wide range of non-classical logics including all those algebraically captured by varieties of normal lattice expansions (LE-logics) \cite{CoPa:non-dist} but also some non-normal logics \cite{impossible:palmigiano:2017, pseudo:conradie:2015}. These results were made possible by a general and uniform definition of Sahlqvist formulas and inequalities (and their generalization, the inductive formulas/inequalities \cite{Goranko:Vakarelov:2006}) which depends only on the order-theoretic properties of the algebraic operations interpreting connectives. This makes the definition applicable not just to individual logics, but to classes of logics in various signatures and on various propositional bases, e.g.,~classical, intuitionistic, distributive, non-distributive \cite{CoPa:non-dist,palmigiano2020constructive}. It also extends to e.g.\ many-valued modal logics \cite{fitting1992many,BritzMScThesis}, hybrid logics \cite{ConRob}, and the modal mu-calculus \cite{CoGhPa14}. 

The identification of the syntactic shape of (generalized) Sahlqvist axioms across signatures provided a springboard from which to establish systematic connections between various (independently established) methods for proving canonicity of classes of formulas/(in)equalities. This has lead to  the {\em methodological unification} of these methods, which include Sambin-Vaccaro's canonicity-via-correspondence \cite{Sambin:Vacarro:89}, Ghilardi-Meloni's constructive canonicity \cite{GhMe97}, J\'onsson's algebraic canonicity based on canonical extensions \cite{Jo94}, and, more recently, Venema's canonicity via pseudocorrespondence \cite{VenemaPseudo}, and De Rudder-Hansoul's canonicity in the context of subordination algebras \cite{Rudder2020subordination}. The algorithmic method for canonicity  adopted by the unified correspondence approach   directly generalizes the algorithmic method of \cite{Conradie:et:al:SQEMAI}, and is therefore based on Sambin-Vaccaro's  canonicity-via-correspondence argument.  However, like J\'onsson's canonicity, the unified correspondence argument is algebraic, and proves the preservation of generalized Sahlqvist axioms under canonical extensions by exploiting their order-theoretic properties as term functions to reduce them equivalently to expressions in a pure (i.e.~atomic proposition-free)  hybrid modal language where propositional variables have been eliminated in favour of  nominal and co-nominal variables. 
Moreover, as argued in \cite{palmigiano2020constructive}, because nominal (resp.~conominal) variables can be interpreted both non-constructively, \`a la J\'onsson, as completely join-irreducible (resp.~meet-irreducible) elements of the canonical extensions, and also constructively, as closed (resp.~open elements) of the canonical extensions, and because filters and ideals of the original algebra bijectively correspond to closed and open elements of their canonical extensions, the two-step process in Ghilardi-Meloni's argument for  constructive canonicity  can be reproduced within the unified correspondence argument for canonicity. In a similar vein, the algebraic and algorithmic canonicity arguments  in \cite{pseudo:conradie:2015, slanted}  successfully subsume and generalize Venema's canonicity via pseudocorrespondence and De Rudder-Hansoul's canonicity in the context of subordination algebras.

Another significant feature of the  unified correspondence approach is the {\em modularization} of correspondence theory. In particular, rather than using first-order correspondents interpreted on (general) frames, the hybrid modal language in which  correspondents are calculated 
is naturally interpreted in perfect algebras or in the canonical extensions of the class of algebras  associated with the given logic, independently of any choice of relational semantics, but the pure equivalents thus obtained can be transformed into first-order correspondents by applying the standard translation for any particular relational semantics appropriately linked by duality to the algebraic semantics.
This means that, in the case of those non-classical logics (such as positive modal logic \cite{Dunn:Pos:ML}, intuitionistic  and bi-intuitionistic modal logic \cite{fischerservi1977modal, rauszer1974formalization} and substructural logics \cite{GaJiKoOn07}) for which more than one type of relational semantics has been defined,  the algebraic approach makes it possible to develop correspondence theory simultaneously for each type of semantics. 

The methodological unification and modularization offered by unified correspondence theory allows one to determine the correspondent of a given (generalized) Sahlqvist axiom within different logical settings and across different relational semantics. For example, we may read the well-known reflexivity axiom as a formula of classical, intuitionistic, distributive, non-distributive, positive or many-value modal logic  and apply the unified correspondence machinery to determine its correspondent with respect to each relational semantics available in each of these  settings. It then becomes natural to ask whether and how these various first-order correspondents can be systematically related to each other. This question becomes all the more interesting in light of recent progress with generalizing Kracht's inverse correspondence theory for classical modal logic \cite{kracht1993completeness} to an analogously unified inverse correspondence theory for (D)LE-logics \cite{UnifiedInverse2022}.

One of the first answers to this question can be found in \cite{BritzMScThesis}, where it is shown that the many-valued first-order frame correspondents on Heyting algebra-valued Kripke frames \cite{fitting1992many} of all Sahlqvist formulas are syntactically (but not semantically) identical their first-order frame correspondents on ordinary crisp Kripke frames. 

In the context of non-distributive modal logics, a preliminary step toward answering this question was made in \cite{conradie2021rough}, where a framework unifying rough set theory and formal concept analysis is proposed, based on 
unified correspondence theory. 
The main tool exploited in \cite{conradie2021rough} is an embedding, defined as in  \cite{moshier2016relational}, from the class of sets to the class of formal contexts (cf.~\cite[Definition 3.3]{conradie2021rough}) which {\em preserves} the Boolean (powerset) algebra associated with each set, in the sense that the concept lattice of any formal context in the image of the embedding  is order-isomorphic to the powerset algebra of its pre-image (cf.~\cite[Proposition 3.4]{conradie2021rough}). This embedding extends (cf.~\cite[Proposition 3.7]{conradie2021rough}) to a {\em complex-algebra-preserving} embedding from the class of Kripke frames to the class of {\em enriched formal contexts}, or {\em polarity-based frames} (defined in \cite{conradie2016categories,TarkPaper2017}, cf.~Definition \ref{def:polarity:based:frm}), relational structures consisting of a formal context enriched with relations, each defined on the basis of duality-theoretic considerations (discussed in \cite{conradie2020non}), and supporting the interpretation of a modal operator. In \cite{conradie2021rough}, the study of the preservation properties of this embedding leads to the understanding that a connection can be established (cf.~\cite[Sections 3.4 and  4]{conradie2021rough}) between the first-order conditions on Kripke frames and those on enriched formal contexts defined by the best-known modal axioms which have also been  considered in the rough set theory literature. 
Besides providing a positive, if circumscribed, answer to the question posed above, this result also provided the principled motivation for the introduction of the notion of {\em conceptual approximation space} (cf.~\cite[Section 5]{conradie2021rough}) as an environment unifying rough set theory and formal concept analysis. 


Other concrete examples pointing to the existence of a systematic connection  between the first-order correspondents of well-known modal axioms (such as $\Box p\to p$ and $\Box p\to\Box\Box p$, especially in their epistemic readings) cropped up very recently in relation with a different relational semantic setting for non-distributive modal logic (namely, 
the {\em graph-based frames}) and have been discussed in \cite{conradie2016categories,TarkPaper2017, conradie2021rough,graph-based-wollic,vanBenthem2001}.



In the present paper, 
the above-mentioned connection established in \cite{conradie2021rough} is generalized syntactically from a finite set of modal axioms  to the class of {\em Sahlqvist modal reduction principles} \cite{van1976mrp} (cf.~Section \ref{sec:mrp}), and is also extended to the following four semantic settings: crisp and many-valued Kripke frames (cf.~Definition \ref{def:kripkelframe}), and  crisp and many-valued polarity-based frames (cf.~Definitions \ref{def:polarity:based:frm} and \ref{def:polarity:based:frm:mv}). 
Specifically, we show preliminarily that, in each of these semantic settings, the first-order correspondent of any Sahlqvist modal reduction principle can be represented as a pure inclusion of binary relations. With this formal common ground established, we can naturally and systematically relate a Sahlqvist MRP's correspondent in any of these four settings to its correspondent in any of the others, following the natural relationships/embeddings that that link them. 
%
%
In particular, 
\begin{enumerate}
\setlength{\itemsep}{0.2pt}
        \setlength{\parskip}{0pt}
        \setlength{\parsep}{0pt}
\item it is known from \cite{BritzMScThesis} that the first-order correspondents of all Sahlqvist formulas on many-valued Kripke frames  are verbatim ``the same'' as  their  first-order correspondents on classical Kripke frames;
\item we partially generalize  \cite{BritzMScThesis} by showing that the first-order correspondents of all Sahlqvist modal reduction principles on many-valued {\em polarity-based} frames  are verbatim ``the same'' as  their first-order correspondents on crisp {\em polarity-based} frames (cf.~Theorem \ref{thm:verbatim crisp mv});
\item we also generalize \cite[Proposition 4.3]{conradie2021rough}  by showing that the first-order correspondents of all Sahlqvist modal reduction principles on crisp {\em polarity-based} frames  are the liftings of  their first-order correspondents on crisp {\em Kripke} frames (cf.~Theorem \ref{prop: sahlqvist lifting});
\item combining  the items above, it also follows that the first-order correspondents of all Sahlqvist modal reduction principles on many-valued {\em polarity-based} frames  are the liftings of  their  first-order correspondents on many-valued {\em Kripke} frames.
\end{enumerate}

We conjecture that, with similar techniques, the present results can be further extended from Sahlqvist modal reduction principles to a suitable class of (generalized) Sahlqvist formulas and inequalities (more on this in the conclusions). 
These results systematically link different logics with overlapping languages via mathematical connections between their interpreting semantic environments. Results like these allow one to transfer theorems and insights from one logic to another, something that can be especially useful when the theory of one of the logics is much better developed and understood than that of the other(s). This is indeed markedly the case in the present paper, where we transfer results from the correspondence theory of classical modal logic, which has been intensively studied for more that half a century, to the comparatively very recently introduced LE-logics of polarity-based frames. But beyond giving us a means to transfer from the old to the new, such systematic connections between logics can be helpful to develop theory for whole families of logics simultaneously, provided the appropriate mathematical connections between the semantics are identified and captured into appropriate parameters. In this line, the authors conjecture that it would be possible, for example, to prove a parameterized version of the well-known Goldblatt-Thomason theorem \cite{goldblatt1975axiomatic} and of the frame-theoretic constructions underpinning it, thereby characterizing the modally definable classes of enriched polarities \cite{conradie2018goldblatt}, graph-based frames \cite{conradie2020non}, intuitionistic Kripke frames \cite{fischerservi1977modal} and classical Kripke frames, shifting between these characterizations simply by selecting the appropriate value of the parameter.         

\paragraph{Structure of the paper.} In Section \ref{sec:preliminaries}, after introducing some notation which we will use throughout the paper (cf.~Section \ref{ssec: notation}), we collect preliminaries on:  normal $\mathrm{LE}$-logics, their associated polarity-based frames and complex algebras, as well as the expanded language of the algorithm ALBA which is used to compute the first-order correspondents of (generalized) Sahlqvist  $\mathrm{LE}$-axioms (cf.~Section \ref{ssec:LE-logics}); how the latter language is used to compute and describe the ALBA outputs of Sahlqvist modal reduction principles (cf.~Section \ref{sec:mrp}); the connection, discussed in \cite{conradie2021rough}, between Kripke frames and polarity-based frames for any given normal modal signature (cf.~Section \ref{ssec: lifting}); the many-valued polarity-based semantics for normal $\mathrm{LE}$-logics, already discussed in \cite{conradie2021rough,conradie2019logic} (cf.~Section \ref{ssec:MV-polarities}).

In Section \ref{sec: composing crisp}, we introduce and study the properties of  several notions of compositions of relations in the settings of Kripke frames and of polarity-based frames. Thanks to these definitions,   suitable propositional languages $\mathsf{PRel}_{\mathcal{L}}$ and $\mathsf{KRel}_{\mathcal{L}}$ can be introduced  (cf.~Section \ref{ssec:prel and krel}) in which first-order conditions on Kripke frames and on polarity-based frames can be represented.

In Section \ref{sec:crisp}, we show that  the ALBA outputs of all Sahlqvist modal reduction principles on Kripke frames and on polarity-based frames can be represented as term-inequalities of the languages $\mathsf{PRel}_{\mathcal{L}}$ and $\mathsf{KRel}_{\mathcal{L}}$. Represented in this way, a systematic connection can be established between them (cf.~Section \ref{ssec: lifting}).

In Section \ref{sec:MV}, we show that Sahlqvist modal reduction principles have ``verbatim the same'' first-order correspondents on {\em crisp} and on  {\em many valued}  polarity-based frames. This is done  by showing that  the first-order correspondents of any Sahlqvist modal reduction principles on {\em crisp} and on {\em many-valued} polarity-based frames can be equivalently represented as {\em the same} $\mathsf{PRel}_{\mathcal{L}}$-inequality.
Finally, we conclude in Section \ref{sec:conclusions}.

\setlength{\abovedisplayskip}{1.7mm}
\setlength{\belowdisplayskip}{1.7mm}
\setlength{\abovedisplayshortskip}{1.7mm}
\setlength{\belowdisplayshortskip}{1.7mm}

\section{Preliminaries}
\label{sec:preliminaries}

In the present section, after introducing notation we will use throughout the paper, we provide an overview of the background theory, context, and motivations  of LE-logics and their polarity-based semantics.  We then recall the notion of inductive modal reduction principles and compute and represent their first-order frame correspondents in an expanded modal language. We illustrate how the class of Kripke frames can be embedded into the class of polarity-based frames. Lastly, we recall the environment of many-valued polarity-based frames.


\subsection{Notation}
\label{ssec: notation}
For all sets $A, B$ and any relation $S \subseteq A \times B$, we let, for any $A' \subseteq A$ and $B' \subseteq B$,
$$S^{(1)}[A'] := \{b \in B\mid  \forall a(a \in A' \Rightarrow a Sb ) \} \quad \mathrm{and}\quad S^{(0)}[B'] := \{a \in A \mid \forall b(b \in B' \Rightarrow a S b)  \}.$$
For all sets $A, B_1,\ldots B_n,$ and any relation $S \subseteq A \times B_1\times \cdots\times B_n$, for any $\overline{C}: = (C_1,\ldots, C_n)$ where $C_j\subseteq B_j$ for all $1 \leq j \leq n$, we let, for any $A'\subseteq A$ and $1\leq i\leq n$,
\begin{equation*}\label{eq:notation bari}
\overline{C}^{\,i}:  = (C_1,\ldots,C_{i-1}, C_{i+1},\ldots, C_n)
\quad \mbox{ and } \quad
\overline{C}^{\,i}_{A'}: = (C_1\ldots,C_{i-1}, A', C_{i+1},\ldots, C_n).
\end{equation*}
When $B_i\supseteq C_i: = \{c_i\}$ and $A\supseteq A': =\{a'\} $, we write $\overline{c}$ for $\overline{\{c\}}$,  and $\overline{c}^{\,i}$ for $\overline{\{c\}}^{\,i}$, and $\overline{c}^{\,i}_{a'}$ for $\overline{\{c\}}^{\,i}_{\{a'\}}$.
We also let:
\begin{enumerate}
\setlength{\itemsep}{0.2pt}
        \setlength{\parskip}{0pt}
        \setlength{\parsep}{0pt}
    \item $S_i \subseteq B_i \times B_1 \times \cdots \times B_{i-1} \times A \times B_{i + 1} \times \cdots\times B_n$ be defined by
	$(b_i, \overline{c}_{a}^{\, i})\in S_i \ \mbox{ iff }\ (a,\overline{c})\in S$.
	\item $S^{(0)}[\overline{C}] := \{a \in A\mid  \forall \overline{b}(\overline{b}\in \overline{C} \Rightarrow aS \overline{b} ) \}$, and $S^{(i)}[A', \overline{C}^{\,i}] := S_i^{(0)}[\overline{C}^{\, i}_{A'}]$.
\end{enumerate}

\subsection{LE-logics and their polarity-based semantics}
\label{ssec:LE-logics}
The framework of (normal) {\em LE-logics} (those logics canonically associated with varieties of normal  general -- i.e.~not necessarily distributive -- lattice expansions, cf.~\cite{gehrke2001bounded, CoPa:non-dist}, and therefore sometimes referred to as {\em non-distributive logics}) has been introduced as an overarching environment for developing duality, correspondence, and canonicity results for a wide class of logical systems covering the best known and most widely applied  nonclassical logics in connection with one another (cf.~Example \ref{example:LE languages} below); however, the resulting framework turned out to be very important also for studying the proof theory of these logics \cite{greco2018unified,chen2021syntactic}, precisely thanks to the systematic connections between proof-theoretic properties and correspondence-theoretic properties (more about this in the conclusions). The exposition in the present subsection if based on \cite[Section 1]{CoPa:non-dist},  \cite[Section 3]{conradie2020non},
\cite[Sections 2 and 3 ]{greco2018algebraic}, to which we refer the interested reader for more detail. 

An {\em order-type} over $n\in \mathbb{N}$ is an $n$-tuple $\epsilon\in \{1, \partial\}^n$. For every order-type $\epsilon$, we denote its {\em opposite} order-type by $\epsilon^\partial$, that is, $\epsilon^\partial_i = 1$ iff $\epsilon_i=\partial$ for every $1 \leq i \leq n$. For any lattice $\bba$, we let $\bba^1: = \bba$ and $\bba^\partial$ be the dual lattice, that is, the lattice associated with the converse partial order of $\bba$. For any order-type $\varepsilon$ over $n$, we let $\bba^\varepsilon: = \Pi_{i = 1}^n \bba^{\varepsilon_i}$.
	
	The language $\mathcal{L}_\mathrm{LE}(\mathcal{F}, \mathcal{G})$ (sometimes abbreviated as $\mathcal{L}_\mathrm{LE}$) takes as parameters: 1) a denumerable set $\Prop$ of proposition letters, elements of which are denoted $p,q,r$, possibly with indexes; 2) disjoint sets of connectives $\mathcal{F}$ and $\mathcal{G}$.\footnote{This partition of connectives  marks an important distinction in the order-theoretic properties of their interpretations: the interpretations of $\mathcal{F}$-connectives preserve (reverse) all finite joins (meets) coordinate-wise as dictated by the their order-types, and dually for the $\mathcal{G}$-connectives. This distinction turns out to be crucial for the general definition of inductive and Sahlqvist inequalities (see Definition \ref{Inducive:Ineq:Def}) and for the soundness of the rules of non-distributive ALBA (cf.~\cite[Section 4]{CoPa:non-dist}).}  Each $f\in \mathcal{F}$ and $g\in \mathcal{G}$ has arity $n_f\in \mathbb{N}$ (resp.\ $n_g\in \mathbb{N}$) and is associated with some order-type $\varepsilon_f$ over $n_f$ (resp.\ $\varepsilon_g$ over $n_g$). Unary $f$ (resp.\ $g$) connectives are typically denoted  $\Diamond$ (resp.\ $\Box$) if their order-type is 1, and $\lhd$ (resp.\ $\rhd$) if their order-type is $\partial$. The terms (formulas) of $\mathcal{L}_\mathrm{LE}$ are defined recursively as follows:
	\[
	\phi ::= p \mid \bot \mid \top \mid (\phi \wedge \phi) \mid (\phi \vee \phi) \mid f(\overline{\phi}) \mid g(\overline{\phi})
	\]
	where $p \in \Prop$, $f \in \mathcal{F}$, $g \in \mathcal{G}$. We will follow the standard rules for the elimination of
parentheses. Terms in $\mathcal{L}_\mathrm{LE}$ will be denoted either by $s,t$, or by lowercase Greek letters such as $\varphi, \psi, \gamma$ etc. 
The purpose of grouping LE-connectives in the families $\mathcal{F}$ and $\mathcal{G}$ is to identify -- and refer to -- the two types of order-theoretic behaviour relevant for the development of this theory. Roughly speaking, connectives in $\mathcal{F}$  and in $\mathcal{G}$ can be thought of as the logical counterparts of generalized operators, and of generalized dual operators, respectively.

	\begin{definition}
		\label{def:DLE:logic:general}
		For any language $\mathcal{L}_\mathrm{LE} = \mathcal{L}_\mathrm{LE}(\mathcal{F}, \mathcal{G})$, an $\mathcal{L}_\mathrm{LE}$-{\em logic} is a set of sequents $\phi\vdash\psi$, with $\phi,\psi\in\mathcal{L}_\mathrm{LE}$, which contains the following axioms:
		\begin{itemize}
			\item Sequents for lattice operations:
			\begin{align*}
				&p\vdash p, && \bot\vdash p, && p\vdash \top, & &  &\\
				&p\vdash p\vee q, && q\vdash p\vee q, && p\wedge q\vdash p, && p\wedge q\vdash q, &
			\end{align*}
			\item Sequents for each connective $f \in \mathcal{F}$ and $g \in \mathcal{G}$ with $n_f, n_g \geq 1$:
			\begin{align*}
				& f(p_1,\ldots, \bot,\ldots,p_{n_f}) \vdash \bot,~\mathrm{for}~ \varepsilon_f(i) = 1, \quad \quad f(p_1,\ldots, \top,\ldots,p_{n_f}) \vdash \bot,~\mathrm{for}~ \varepsilon_f(i) = \partial,\\
				&\top\vdash g(p_1,\ldots, \top,\ldots,p_{n_g}),~\mathrm{for}~ \varepsilon_g(i) = 1, \quad \quad \top\vdash g(p_1,\ldots, \bot,\ldots,p_{n_g}),~\mathrm{for}~ \varepsilon_g(i) = \partial,\\
				&f(p_1,\ldots, p\vee q,\ldots,p_{n_f}) \vdash f(p_1,\ldots, p,\ldots,p_{n_f})\vee f(p_1,\ldots, q,\ldots,p_{n_f}),~\mathrm{for}~ \varepsilon_f(i) = 1,\\
				&f(p_1,\ldots, p\wedge q,\ldots,p_{n_f}) \vdash f(p_1,\ldots, p,\ldots,p_{n_f})\vee f(p_1,\ldots, q,\ldots,p_{n_f}),~\mathrm{for}~ \varepsilon_f(i) = \partial,\\
				& g(p_1,\ldots, p,\ldots,p_{n_g})\wedge g(p_1,\ldots, q,\ldots,p_{n_g})\vdash g(p_1,\ldots, p\wedge q,\ldots,p_{n_g}),~\mathrm{for}~ \varepsilon_g(i) = 1,\\
				& g(p_1,\ldots, p,\ldots,p_{n_g})\wedge g(p_1,\ldots, q,\ldots,p_{n_g})\vdash g(p_1,\ldots, p\vee q,\ldots,p_{n_g}),~\mathrm{for}~ \varepsilon_g(i) = \partial,
			\end{align*}
		\end{itemize}
		and is closed under the following inference rules:
		\begin{displaymath}
			\frac{\phi\vdash \chi\quad \chi\vdash \psi}{\phi\vdash \psi}
			\quad
			\frac{\phi\vdash \psi}{\phi(\chi/p)\vdash\psi(\chi/p)}
			\quad
			\frac{\chi\vdash\phi\quad \chi\vdash\psi}{\chi\vdash \phi\wedge\psi}
			\quad
			\frac{\phi\vdash\chi\quad \psi\vdash\chi}{\phi\vee\psi\vdash\chi}
		\end{displaymath}
		where $\phi(\chi/p)$ denotes uniform substitution of $\chi$ for $p$ in $\phi$,  and for each connective $f \in \mathcal{F}$ and $g \in \mathcal{G}$,
		{\small
		\begin{displaymath}
		\begin{array}{ll}
			 \dfrac{\phi\vdash\psi}{f(\phi_1,\ldots,\phi,\ldots,\phi_{n_f})\vdash f(\phi_1,\ldots,\psi,\ldots,\phi_{n_f})}{\footnotesize~(\varepsilon_f(i) = 1)} \quad\quad
		&
			 \dfrac{\phi\vdash\psi}{f(\phi_1,\ldots,\psi,\ldots,\phi_{n_f})\vdash f(\phi_1,\ldots,\phi,\ldots,\phi_{n_f})}{\footnotesize~(\varepsilon_f(i) = \partial)}
		\\
			 \dfrac{\phi\vdash\psi}{g(\phi_1,\ldots,\phi,\ldots,\phi_{n_g})\vdash g(\phi_1,\ldots,\psi,\ldots,\phi_{n_g})}{\footnotesize~(\varepsilon_g(i) = 1)}
		&
			 \dfrac{\phi\vdash\psi}{g(\phi_1,\ldots,\psi,\ldots,\phi_{n_g})\vdash g(\phi_1,\ldots,\phi,\ldots,\phi_{n_g})}{\footnotesize~(\varepsilon_g(i) = \partial)}.
	    \end{array}
		\end{displaymath}
		}
		The minimal $\mathcal{L}_{\mathrm{LE}}(\mathcal{F}, \mathcal{G})$-logic is denoted by $\mathbf{L}_\mathrm{LE}(\mathcal{F}, \mathcal{G})$, or simply by $\mathbf{L}_\mathrm{LE}$ when $\mathcal{F}$ and $\mathcal{G}$ are clear from the context. 
	\end{definition}


As mentioned at beginning of this subsection, the framework of LE-logics is intended to encompass and simultaneously study a wide family of logics which includes the best known non-classical propositional logics. In the following example, we discuss a diverse selection of such logics, and show how they fit the LE-pattern. Notice that the range of examples includes both logics for which the distributive laws $\phi\wedge (\psi\vee \chi)\vdash (\phi\wedge\psi)\vee(\phi\wedge \chi)$ and $ (\phi\vee\psi)\wedge(\phi\vee \chi)\vdash \phi\vee (\psi\wedge \chi)$   hold and ones for which these laws fail. Notice also that some of these languages include connectives that capture some of the properties of (classical and intuitionistic) negation, and that can hence be understood as weak forms of negation.       

\begin{example}\label{example:LE languages}
The language $\mathcal{L}_{\mathrm{Bi}}$ of \emph{bi-intuitionistic logic} \cite{rauszer1974formalization} coincides with the LE-language associated with the following parameters: $\mathcal{F}: = \{{\pdra}\}$ and $\mathcal{G}: = \{\rightarrow\}$ with $n_{\footnotesize{\pdra}} = n_\rightarrow = 2$ and $\varepsilon_{\footnotesize{\pdra}} = \varepsilon_\rightarrow = (\partial, 1)$. The language $\mathcal{L}_{\mathrm{IL}}$ of \emph{intuitionistic logic} is the $\{{\pdra}\}$-free fragment of $\mathcal{L}_{\mathrm{Bi}}$. The language $\mathcal{L}_{\mathrm{DML}}$ of \emph{distributive modal logic} (cf.\  \cite{GNV,conradie2012algorithmic}) is obtained by instantiating $\mathcal{F}: = \{\Diamond, {\lhd}\}$ with $n_\Diamond = n_\lhd = 1$, $\varepsilon_\Diamond = 1$  and $\varepsilon_\lhd = \partial$, and   $\mathcal{G} = \{\Box, {\rhd}\}$ with $n_\Box = n_\rhd = 1$, $\varepsilon_\Box = 1$  and $\varepsilon_\rhd = \partial$. The language $\mathcal{L}_{\mathrm{PML}}$ of  \emph{positive modal logic} \cite{Dunn:Pos:ML} is the $\{{\lhd}, {\rhd} \}$-free fragment of $\mathcal{L}_{\mathrm{DML}}$. The language $\mathcal{L}_{\mathrm{BiML}}$ of \emph{bi-intuitionistic modal logic} \cite{wolter1998CoImplication} is obtained by instantiating $\mathcal{F}: = \{{\pdra}, \Diamond\}$ and $\mathcal{G}: = \{\rightarrow, \Box\}$ with $n_{\footnotesize{\pdra}} = n_\rightarrow = 2$, $n_\Diamond = n_\Box = 1$ and $\varepsilon_{\footnotesize{\pdra}} = \varepsilon_\rightarrow = (\partial, 1)$, $\epsilon_{\Diamond} = \epsilon_{\Box} = 1$. The language $\mathcal{L}_{\mathrm{IML}}$ of \emph{intuitionistic modal logic} \cite{fischerservi1977modal} is the $\{{\pdra}\}$-free fragment of $\mathcal{L}_{\mathrm{BiML}}$. The language $\mathcal{L}_{\mathrm{SDM}}$ of  \emph{semi-De Morgan logic}  \cite{sankappanavar1987semi} coincides with the language of \emph{orthologic} \cite{Goldblatt:Ortho:74} and is the $\{\Diamond, \Box, {\lhd}\}$-free fragment of $\mathcal{L}_{\mathrm{DML}}$. The language $\mathcal{L}_{\mathrm{FL}}$ of the \emph{full Lambek calculus} \cite{la61, GaJiKoOn07} is obtained by instantiating $\mathcal{F}: = \{e, \circ\}$ with $n_e = 0$, $n_\circ  = 2$, $\varepsilon_\circ = (1, 1)$   and   $\mathcal{G}: = \{\backslash, /\}$ with $n_\backslash = n_/ = 2$, $\varepsilon_\backslash = (\partial, 1)$  and $\varepsilon_/ = (1, \partial)$. The language $\mathcal{L}_{\mathrm{LG}}$ of the \emph{ Lambek-Grishin calculus} (cf.\ \cite{MoortgatSymmCatGramm}) is  obtained by instantiating $\mathcal{F}: = \{e, \circ, \starfor, \starback \}$ with $n_{e} = 0$, $n_\circ = n_{\starback} = n_{\starfor} = 2$, $\varepsilon_\circ = (1, 1)$, $\varepsilon_{\starback} = (\partial, 1)$, $\varepsilon_{\starfor} = (1, \partial)$   and   $\mathcal{G} := \{\upsE, \star, \circfor, \circback\}$ with $n_{\footnotesize{\upsE}} = 0$,  $n_\star = n_{\circfor} = n_{\circback} = 2$, $\varepsilon_\star = (1, 1)$,  $\varepsilon_{\circback} = (\partial, 1)$, $\varepsilon_{\circfor} = (1, \partial)$. The language $\mathcal{L}_{\mathrm{MALL}}$ of the  {\em multiplicative-additive fragment of linear logic} (cf.\ \cite{GaJiKoOn07}) is the $\{\star, \starfor, \starback, \circfor\}$-free fragment of $\mathcal{L}_{\mathrm{LG}}$.
\end{example}

\paragraph{Polarity-based frames and their complex algebras.} 
Several complete relational semantics have been introduced for LE-logics, among which, the {\em polarity-based} \cite{conradie2016categories,Tark1} and the {\em graph-based} \cite{graph-based-wollic} give rise to two -- very different but equally compelling and mathematically well grounded --   intuitive interpretations of these logics \cite{conradie2020non}. In particular, the interpretation of LE-logics supported by the polarity-based semantics is that LE-logics are the logics of {\em formal concepts} in the sense of Formal Concept Analysis (FCA) \cite{ganter2012formal}, which in turn are mathematical representations of  {\em categories}. Thanks to this interpretation, in \cite{conradie2021rough}, the polarity-based semantics of basic normal lattice-based modal logic was taken as a unifying mathematical environment for FCA and RST, and in \cite{conradie2019logic}, this environment was generalized from crisp to many-valued.

A {\em formal context} or {\em polarity} \cite{ganter2012formal} is a structure $\mathbb{P} = (A, X, I)$ s.t.~$A$ and $X$ are sets and $I\subseteq A\times X$ is a binary relation. 
For every polarity $\mathbb{P}$, maps $(\cdot)^\uparrow: \mathcal{P}(A)\to \mathcal{P}(X)$ and $(\cdot)^\downarrow: \mathcal{P}(X)\to \mathcal{P}(A)$ can be defined as follows:
$B^\uparrow: = \{x\in X \mid\forall a \in A(a \in B\to aIx)\}$ and $Y^\downarrow: = \{a\in A\mid \forall x \in X(x \in Y \to aIx)\}$. The maps $(\cdot)^\uparrow$ and $(\cdot)^\downarrow$ form a \textit{Galois connection} between $(\mathcal{P}(A), \subseteq)$ and $(\mathcal{P}(X), \subseteq)$, i.e. $Y \subseteq B^\uparrow$ iff $B\subseteq Y^\downarrow$
for all $B \in \mathcal{P}(A)$ and $Y\in \mathcal{P}(X)$. 
A {\em formal concept} of $\mathbb{P}$ is a pair 
$c = (\val{c}, \descr{c})$ s.t.~$\val{c}\subseteq A$ and $\descr{c}\subseteq X$, and  $\val{c}^{\uparrow} = \descr{c}$ and $\descr{c}^{\downarrow} = \val{c}$. The set $\val{c}$ is the {\em extension} of $c$, while $\descr{c}$ is its {\em intension}. It  immediately follows from this definition that if $(\val{c}, \descr{c})$ is a formal concept, then $\val{c}^{\uparrow\downarrow} = \val{c}$ and $\descr{c}^{\downarrow\uparrow} = \descr{c}$. In this case, the sets $\val{c}$ and $\descr{c}$ are referred to as being \textit{Galois-stable}.  The set $\mathbb{L}(\mathbb{P})$  of the formal concepts of $\mathbb{P}$ can be partially ordered as follows: for any $c, d\in \mathbb{L}(\mathbb{P})$, \[c\leq d\quad \mbox{ iff }\quad \val{c}\subseteq \val{d} \quad \mbox{ iff }\quad \descr{d}\subseteq \descr{c}.\]
With this order, $\mathbb{L}(\mathbb{P})$ is a complete lattice such that, for any $\mathcal{X}\subseteq  \mathbb{L}(\mathbb{P})$,
\begin{eqnarray*}
\bigwedge \mathcal{X}& =  &(\bigcap \{\val{c}\mid c\in \mathcal{X}\}, (\bigcap \{\val{c}\mid c\in \mathcal{X}\})^\uparrow)\\
 \bigvee \mathcal{X} & =  &( (\bigcap \{\descr{c}\mid c\in \mathcal{X}\})^\downarrow, \bigcap \{\descr{c}\mid c\in \mathcal{X}\}). 
\end{eqnarray*}
This complete lattice is referred to as the {\em concept lattice} $\mathbb{P}^+$ of $\mathbb{P}$. Moreover, 
\begin{proposition}[\cite{conradie2020non}, Proposition 3.1]
\label{prop:join and meet generators}
For any polarity $\mathbb{P} = (A, X, I)$, the complete lattice $\mathbb{P}^+$ is completely join-generated by the set $\{\mathbf{a}: = (a^{\uparrow\downarrow}, a^{\uparrow})\mid a\in A\}$ and is completely meet-generated by the set $\{\mathbf{x}: =(x^{\downarrow},x^{\downarrow\uparrow})\mid x\in X\}$.
\end{proposition}
The elements $c = (\val{c}, \descr{c})$ of the concept lattice of a polarity will be used to interpret the formulas of the associated LE-language. Specifically, via valuations $V: \mathcal{L}_{\mathrm{LE}}\to \mathbb{P}^+$, formulas $\phi$ will be interpreted as names for formal concepts, of which $\val{V(\phi)}\subseteq A$ will be the {\em extension}, and $\descr{V(\phi)}\subseteq X$ will be the {\em intension} (the recursive definitions will be detailed below). 

The following definition is based on \cite[Definition 17]{greco2018algebraic} \cite[Definition A.1]{conradie2020non}.
\begin{definition}\label{def:polarity:based:frm}
	A {\em polarity-based $\mathcal{L}_{\text{LE}}$-frame}  is a tuple $\mathbb{F} = (\mathbb{P}, \mathcal{R}_{\mathcal{F}}, \mathcal{R}_{\mathcal{G}})$, where  $\mathbb{P} = (A, X,  I)$ is a polarity, $\mathcal{R}_{\mathcal{F}} = \{R_f\mid f\in \mathcal{F}\}$, and $\mathcal{R}_{\mathcal{G}} = \{R_g\mid g\in \mathcal{G}\}$, such that  for each $f\in \mathcal{F}$ and $g\in \mathcal{G}$, the symbols $R_f$ and  $R_g$ respectively denote $(n_f+1)$-ary and $(n_g+1)$-ary relations 
	\begin{equation*}
	R_f \subseteq X \times A^{\epsilon_{f}}    \ \mbox{ and }\ R_g \subseteq A \times X^{\epsilon_{g}},
	\end{equation*}
	where $A^{\epsilon_{f}}$ denotes the $n_f$-fold cartesian product of $A$ and $X$ such that, for each $1\leq i\leq n_f$, the $i$th projection of $A^{\epsilon_{f}}$ is $A$ if $\epsilon_f(i ) = 1$ and is $X$ if $\epsilon_f(i ) = \partial$, and $X^{\epsilon_g}$ denotes the $n_g$-fold cartesian product of $A$ and $X$ such that for each $1\leq i\leq n_g$ the $i$th projection of $X^{\epsilon_{g}}$ is $X$ if $\epsilon_g(i ) = 1$ and is $A$ if $\epsilon_g(i ) = \partial$.
	In addition, all relations $R_f$ and $R_g$ are required to be {\em $I$-compatible}, i.e.\ the following sets are assumed to be  Galois-stable for all $a \in A$, $x \in X $, $\overline{a} \in A^{\epsilon_f}$, and $\overline{x} \in X^{\epsilon_g}$:
	\[
	R_f^{(0)}[\overline{a}]\text{ and }R_f^{(i)}[x, \overline{a}^{\, i}] \quad\quad R_g^{(0)}[\overline{x}]\text{ and }R_g^{(i)}[a, \overline{x}^{\,i}].
	\]
	
	Thus, when applying this definition to all unary connectives  $\Diamond, {\lhd}\in \mathcal{F}$ and $\Box, {\rhd}\in \mathcal{G}$, their associated relations are of the following types: 
	\begin{equation*}
	R_\Diamond \subseteq X \times A   \quad R_\lhd \subseteq X \times X \quad R_\Box \subseteq A \times X \quad R_\rhd \subseteq A \times A,
	\end{equation*}
	and the $I$-compatibility condition requires the following inclusions to hold  for all $a \in A$ and $x \in X $:
\begin{align*}
	(R_\Diamond^{(0)}[a])^{\downarrow\uparrow}\subseteq R_\Diamond^{(0)}[a] \quad (R_\Diamond^{(1)}[x])^{\uparrow\downarrow}\subseteq R_\Diamond^{(1)}[x]\quad\quad (R_\Box^{(0)}[x])^{\uparrow\downarrow}\subseteq R_\Box^{(0)}[x] \quad (R_\Box^{(1)}[a])^{\downarrow \uparrow}\subseteq R_\Box^{(1)}[a]\\
	(R_\lhd^{(0)}[x])^{\downarrow\uparrow}\subseteq R_\lhd^{(0)}[x] \quad (R_\lhd^{(1)}[x])^{\downarrow\uparrow}\subseteq R_\lhd^{(1)}[x]\quad\quad (R_\rhd^{(0)}[a])^{\uparrow\downarrow}\subseteq R_\rhd^{(0)}[a] \quad (R_\rhd^{(1)}[a])^{\uparrow\downarrow }\subseteq R_\rhd^{(1)}[a].\\
	\end{align*}
\end{definition}
For any polarity $\mathbb{P} = (A, X, I)$, examples of $I$-compatible relations are $I$ itself (as a consequence of $B^{\uparrow}$ and $Y^{\downarrow}$ being Galois-stable for any $B\subseteq A$ and $Y\subseteq X$), and the full relation $R:=A\times X$ (since $R^{(0)}[x] =A$ and $R^{(1)}[a] = X$ for all $a\in A$ and $x\in X$). Moreover, if $I$ is such that $\{a\in A\mid \forall x (x\in X\Rightarrow aIx)\} = \varnothing = \{x\in X\mid \forall a (a\in A\Rightarrow aIx)\}$, then the empty  relation $R:=\varnothing$ is $I$-compatible. Finally, if $R_{\Box}\subseteq A\times X$ (resp.~$R_{\Diamond}\subseteq X\times A$) is $I$-compatible, so is $R_{\Diamondblack}\subseteq X\times A$ (resp.~$R_{\blacksquare}\subseteq A\times X$), defined as $x R_{\Diamondblack} a$ iff $aR_{\Box}x$ (resp.~$a R_{\blacksquare} x$ iff $xR_{\Diamond}a$). In view of the definition above, throughout the remainder of the paper, we will sometimes refer to e.g.~($I$-compatible) relations $R\subseteq A\times X$ as `box-type relations', and to $R\subseteq X\times A$ as `diamond-type relations'.
\begin{definition}
\label{def:complex algebra of LE-frame}
The {\em complex algebra} of  a polarity-based $\mathcal{L}_{\text{LE}}$-frame $\mathbb{F} = (\mathbb{P}, \mathcal{R}_\mathcal{F}, \mathcal{R}_\mathcal{G})$  is the $\mathcal{L}_{\text{LE}}$-algebra
$$\mathbb{F}^+ = (\mathbb{L}, \{f_{R_f}\mid f\in \mathcal{F}\}, \{g_{R_g}\mid g\in \mathcal{G}\}),$$
where $\mathbb{L}: = \mathbb{P}^+$, and for all $f \in \mathcal{F}$ and $g \in \mathcal{G}$,

\begin{enumerate}
	\setlength{\itemsep}{0.2pt}
        \setlength{\parskip}{0pt}
        \setlength{\parsep}{0pt}
	\item $ f_{R_f}: \mathbb{L}^{n_f}\to \mathbb{L}$ is defined by the assignment $f_{R_f}(\overline{c}) = \left(\left(R_f^{(0)}[\overline{\val{c}}^{\epsilon_f}]\right)^{\downarrow}, R_f^{(0)}[\overline{\val{c}}^{\epsilon_f}]\right)$;
	
	\item $g_{R_g}: \mathbb{L}^{n_g}\to \mathbb{L}$ is defined by the assignment  $g_{R_g}(\overline{c}) = \left(R_g^{(0)}[\overline{\descr{c}}^{\epsilon_g}], \left(R_g^{(0)}[\overline{\descr{c}}^{\epsilon_g}]\right)^{\uparrow}\right)$. 
	
\end{enumerate}
\noindent
Here, for every $\overline{c}\in \mathbb{L}^{n_f}$, the tuple $\overline{\val{c}}^{\epsilon_f}$ is such that for
each $1 \leq i \leq n_f$,  the $i$-th coordinate of $\overline{\val{c}}^{\epsilon_f}$ is $\val{c_i}$ if $\epsilon_f(i) = 1$ and is $\descr{c_i}$ if $\epsilon_f(i) = \partial$, and $\overline{\descr{c}}^{\epsilon_g}$ is such that for
each $1 \leq i \leq n_g$  the $i$-th coordinate of $\overline{\descr{c}}^{\epsilon_g}$ is $\descr{c_i}$ if $\epsilon_g(i) = 1$, and is  $\val{c_i}$ if $\epsilon_g(i) = \partial$. 
Thus, for all $\Diamond, {\lhd}\in \mathcal{F}$ and all $\Box, {\rhd}\in \mathcal{G}$, 
\begin{enumerate}
	\setlength{\itemsep}{0.2pt}
        \setlength{\parskip}{0pt}
        \setlength{\parsep}{0pt}
	\item $ \langle R_\Diamond\rangle, \langle R_\lhd]: \mathbb{L}\to \mathbb{L}$ are respectively defined by the assignments \footnote{The symbols $\langle R_\Diamond \rangle$, $[ R_\Box ]$, $\langle R_{\lhd} ]$, $[ R_\rhd \rangle$ were introduced in \cite{GNV}  in the context of the relational semantics for distributive modal logic.}
	\[
	c\mapsto \left(\left(R_\Diamond^{(0)}[\val{c}]\right)^{\downarrow}, R_\Diamond^{(0)}[\val{c}]\right) \quad\text{ and }\quad c\mapsto \left(\left(R_\lhd^{(0)}[\descr{c}]\right)^{\downarrow}, R_\lhd^{(0)}[\descr{c}]\right);\]
	
	\item $[R_\Box], [R_\rhd\rangle : \mathbb{L}\to \mathbb{L}$ are respectively defined by the assignments 
  \[ c \mapsto \left(R_\Box^{(0)}[\descr{c}], \left(R_\Box^{(0)}[\descr{c}]\right)^{\uparrow}\right) \quad\text{ and }\quad c\mapsto \left(R_\rhd^{(0)}[\val{c}], \left(R_\lhd^{(0)}[\val{c}]\right)^{\uparrow} \right).\]

\end{enumerate}
\end{definition}

\begin{proposition}[cf.\ \cite{greco2018algebraic} Proposition 21]\label{prop:F plus is complete LE}
	If $\mathbb{F} = (\mathbb{P}, \mathcal{R}_\mathcal{F}, \mathcal{R}_\mathcal{G})$ is a polarity-based $\mathcal{L}_{\mathrm{LE}}$-frame, then $\mathbb{F}^+ = (\mathbb{L}, \{f_{R_f}\mid f\in \mathcal{F}\}, \{g_{R_g}\mid g\in \mathcal{G}\})$ is a complete normal $\mathcal{L}_{\mathrm{LE}}$-algebra (cf.~\cite[Definition 1.3]{CoPa:non-dist}). 
\end{proposition}

\paragraph{Interpretation of LE-languages on polarity-based frames.}
For any polarity-based frame $\mathbb{F}=(\mathbb{P}, \mathcal{R}_\mathcal{F}, \mathcal{R}_\mathcal{G})$ with $\mathbb{P} = (A, X, I)$, a {\em valuation} on $\mathbb{F}$ is a map $V:\Prop\to \mathbb{P}^+$. For every  $p\in \Prop$, we let  $\val{p}: = \val{V(p)}$ (resp.~$\descr{p}: = \descr{V(p)}$) denote the extension (resp.~the intension) of the interpretation of $p$ under $V$.  The elements (objects) of $\val{p}$ are the {\em members} of concept $p$ under  $V$; the elements (features) of $\descr{p}$ {\em describe}  concept $p$ under $V$. Any valuation $V$ on $\mathbb{F}$ extends homomorphically to a unique interpretation map of $\mathcal{L}$-formulas, which, abusing notation, we also denote $V$, and which is defined as follows:
\smallskip

{{\centering
\begin{tabular}{rcl c rcl}
$V(p)$ & $ = $ & $(\val{p}, \descr{p})$\\
$V(\top)$ & $ = $ & $(A, A^{\uparrow})$ & \quad\: &
 $V(\bot)$ & $ = $ & $(X^{\downarrow}, X)$\\
$V(\phi\wedge\psi)$ & $ = $ & $(\val{\phi}\cap \val{\psi}, (\val{\phi}\cap \val{\psi})^{\uparrow})$ &&
$V(\phi\vee\psi)$ & $ = $ & $((\descr{\phi}\cap \descr{\psi})^{\downarrow}, \descr{\phi}\cap \descr{\psi})$\\
$V(f(\overline{\phi}))$ & $ = $ & $\left(\left(R_f^{(0)}[\overline{\val{\phi}}^{\epsilon_f}]\right)^{\downarrow}, R_f^{(0)}[\overline{\val{\phi}}^{\epsilon_f}]\right)$ &&
$V(g(\overline{\phi}))$ & $ = $ & $\left(R_g^{(0)}[\overline{\descr{\phi}}^{\epsilon_g}], \left(R_g^{(0)}[\overline{\descr{\phi}}^{\epsilon_g}]\right)^{\uparrow}\right)$
\end{tabular}
\par}}

Here, for every $\overline{\phi}\in \mathcal{L}^{n_f}$  (resp.~$\overline{\phi}\in \mathcal{L}^{n_g}$), the tuple $\overline{\val{\phi}}^{\epsilon_f}$ is such that for
each $1 \leq i \leq n_f$,  the $i$-th coordinate of $\overline{\val{\phi}}^{\epsilon_f}$ is $\val{\phi_i}$ if $\epsilon_f(i) = 1$, and is $\descr{\phi_i}$ if $\epsilon_f(i) = \partial$, and $\overline{\descr{\phi}}^{\epsilon_g}$ is such that for
each $1 \leq i \leq n_g$  the $i$-th coordinate of $\overline{\descr{\phi}}^{\epsilon_g}$ is $\descr{\phi_i}$ if $\epsilon_g(i) = 1$, and is  $\val{\phi_i}$ if $\epsilon_g(i) = \partial$. Thus, for all $\Diamond, {\lhd}\in \mathcal{F}$ and  $\Box, {\rhd}\in \mathcal{G}$,

{{\centering
\begin{tabular}{rcl c rcl}
$V(\Box\phi)$ & $ = $ & $(R_{\Box}^{(0)}[\descr{\phi}], (R_{\Box}^{(0)}[\descr{\phi}])^{\uparrow})$ & \quad\quad &
$V(\Diamond\phi)$ & $ = $ & $((R_{\Diamond}^{(0)}[\val{\phi}])^\downarrow, R_{\Diamond}^{(0)}[\val{\phi}])$\\
$V({\rhd}\phi)$ & $ = $ & ($R_\rhd^{(0)}[\val{\phi}], (R_\rhd^{(0)}[\val{\phi}])^\uparrow)$ &&
$V({\lhd}\phi)$ & $ = $ & $((R_\lhd^{(0)}[\descr{\phi}])^\downarrow, R_\lhd^{(0)}[\descr{\phi}])$.\\
\end{tabular}
\par}} 
\smallskip

\noindent
A {\em model}  $\mathbb{M} = (\mathbb{F}, V)$ is  such that $\mathbb{F} $ is a polarity-based frame and $V$ is a  valuation on $\mathbb{F}$.  For any $\phi\in \mathcal{L}$, any $a \in A$ and $x \in X$, we let:

{{\centering
\begin{tabular}{l}
$\mathbb{M}, a \Vdash \phi$  \quad iff \quad  $a\in \val{\phi}_{\mathbb{M}}$  \quad \quad \quad
$\mathbb{M}, x \succ \phi$  \quad iff \quad $x\in \descr{\phi}_{\mathbb{M}}$  \\
\end{tabular}
\par}} 
\smallskip

\noindent
In line with the idea,  discussed early on, that the polarity-based semantics supports an intuitive interpretation of LE-logics as the logics of formal concepts in the sense of FCA, we read $\mathbb{M}, a \Vdash \phi$ as `` object $a$ is a member of category $\phi$'', and $\mathbb{M}, x \succ \phi$ as ``feature $x$ describes category $\phi$''. 
%
Spelling out the definition above, we can equivalently rewrite it in the following recursive form: 
\smallskip

{{\centering
\begin{tabular}{rcl c rcl}
$\mathbb{M}, a \Vdash p$ & iff & $a\in \val{p}_{\mathbb{M}}$ &\quad&
$\mathbb{M}, x \succ p$ & iff & $x\in \descr{p}_{\mathbb{M}}$ \\
$\mathbb{M}, a \Vdash\top$ &  & always &&
$\mathbb{M}, x \succ \top$ & iff &   $a I x$ for all $a\in A$\\
$\mathbb{M}, x \succ  \bot$ &  & always &&
$\mathbb{M}, a \Vdash \bot $ & iff & $a I x$ for all $x\in X$\\
$\mathbb{M}, a \Vdash \phi\wedge \psi$ & iff & $\mathbb{M}, a \Vdash \phi$ and $\mathbb{M}, a \Vdash  \psi$ & &
$\mathbb{M}, x \succ \phi\wedge \psi$ & iff & $\forall a\in A(\mathbb{M}, a \Vdash \phi\wedge \psi \Rightarrow a I x)$\\
$\mathbb{M}, x \succ \phi\vee \psi$ & iff &  $\mathbb{M}, x \succ \phi$ and $\mathbb{M}, x \succ  \psi$ &&
$\mathbb{M}, a \Vdash \phi\vee \psi$ & iff & $\forall x\in X(\mathbb{M}, x \succ \phi\vee \psi \Rightarrow a I x)$  \\
$\mathbb{M}, a \Vdash g(\overline{\phi})$ & iff & $\forall \overline{x}\in \overline{X}^{\epsilon_g}(\mathbb{M},\overline{x} \succ^{\epsilon_g}\overline{ \phi} \Rightarrow R_{g} (a, \overline{x}))$ &&
$\mathbb{M}, x \succ g(\overline{\phi})$ & iff & $\forall a\in A(\mathbb{M}, a \Vdash g(\overline{\phi}) \Rightarrow a I x)$\\
$\mathbb{M}, x \succ f(\overline{\phi})$ & iff & $\forall \overline{a}\in \overline{A}^{\epsilon_f}(\mathbb{M},\overline{a} \Vdash^{\epsilon_f}\overline{ \phi}\Rightarrow R_{f} (x, \overline{a}))$ &&
$\mathbb{M}, a \Vdash f(\overline{\phi})$ & iff &  $\forall x\in X(\mathbb{M}, x\succ g(\overline{\phi}) \Rightarrow a I x)$,\\
\end{tabular}
\par}} 
\smallskip 

\noindent
where, if  $\overline{a}\in A^{\epsilon_f}$,\footnote{That is, the $i$th coordinate of $\overline{a}$ is $a_i\in A$ if $\epsilon_f(i) = 1$ and is $x_i\in X$ if $\epsilon_f(i) = \partial$, for each $1\leq i\leq n_f$, and similarly for $\overline{x}\in X^{\epsilon_g}$.} the notation $\mathbb{M}, \overline{a} \Vdash^{\epsilon_f}\overline{ \phi}$  refers to the conjunction over $1\leq i\leq n_f$ of statements of the form $\mathbb{M}, a_i\Vdash\phi_i$, if $\epsilon_f(i) = 1$, or $\mathbb{M}, x_i\succ \phi_i$, if $\epsilon_f(i) = \partial$, whereas if  $\overline{x}\in X^{\epsilon_g}$, the notation $\mathbb{M}, \overline{x} \succ^{\epsilon_g}\overline{ \phi}$  refers to the conjunction over $1\leq i\leq n_g$ of statements of the form  $\mathbb{M}, x_i\succ \phi_i$, if $\epsilon_g(i) = 1$, or $\mathbb{M}, a_i\Vdash\phi_i$, if $\epsilon_g(i) = \partial$. 
Specializing the last four clauses above to the interpretation of $f$-formulas and $g$-formulas when $n_f = n_g = 1$: \\[1mm]
{{
\small
\begin{tabular}{llll c llll}
$\mathbb{M}, a \Vdash \Box\phi$ & iff & for all $x\in X$, if $\mathbb{M}, x \succ \phi$, then $a R_\Box x$& 
$\mathbb{M}, x \succ \Box\phi$ & iff & for all $a\in A$, if $\mathbb{M}, a \Vdash \Box\phi$, then $a I x$.\\
$\mathbb{M}, x \succ \Diamond\phi$ & iff &  for all $a\in A$, if $\mathbb{M}, a \Vdash \phi$, then $x R_\Diamond a$ & 
$\mathbb{M}, a \Vdash \Diamond\phi$ & iff & for all $x\in X$, if $\mathbb{M}, x \succ \Diamond\phi$, then $a I x$   &\\
$\mathbb{M}, a \Vdash {\rhd}\phi$ & iff & for all $b\in A$, if $\mathbb{M}, b \Vdash \phi$, then $a R_\Box b$&  
$\mathbb{M}, x \succ {\rhd}\phi$ & iff & for all $a\in A$, if $\mathbb{M}, a \Vdash {\rhd}\phi$, then $a I x$.\\
$\mathbb{M}, x \succ {\lhd}\phi$ & iff &  for all $y\in X$, if $\mathbb{M}, y \Vdash \phi$, then $x R_\lhd y$ &
$\mathbb{M}, a \Vdash {\lhd}\phi$ & iff & for all $x\in X$, if $\mathbb{M}, x \succ \lhd\phi$, then $a I x$.  &\\
\end{tabular}
}}

\noindent
Finally, we define  the interpretation of sequents in the following, equivalent, ways:
\smallskip 

{{\centering
\begin{tabular}{llll}
$\mathbb{M}\models \phi\vdash \psi$ & iff & for all $a \in A$, $\mbox{if } \mathbb{M}, a \Vdash \phi, \mbox{ then } \mathbb{M}, a \Vdash \psi$&\\
                                    & iff & for all $x \in X$, $\mbox{if } \mathbb{M}, x \succ \psi, \mbox{ then } \mathbb{M}, x \succ \phi$.&
\end{tabular}
\par}}


A sequent $\phi\vdash \psi$ is {\em valid} on an enriched formal context $\mathbb{F}$ (in symbols: $\mathbb{F}\models \phi\vdash \psi$) if $\mathbb{M}\models \phi\vdash \psi$  for every model $\mathbb{M}$ based on $\mathbb{F}$. As discussed in \cite[Section 2]{CoPa:non-dist}, the basic $\mathcal{L}_{\mathrm{LE}}$-logic of formal concepts is complete w.r.t.~its associated class of polarity-based frames defined as in Definition \ref{def:polarity:based:frm}.
\medskip

The reader familiar with modal logic would have noticed that the interpretation clauses  for the modal connectives given above diverge from those adopted in the well-known Kripke semantics. In particular, in the interpretation of the diamond, one notices that there are no existential clauses, and in the case of the box, the relational restrictor of the quantification appears in the consequent, rather than in the antecedent of the implication.
In \cite{conradie2016categories,Tark1, conradie2021rough}, these modal operators and ensuing semantic clauses were given epistemic/evidentialist interpretations, as follows. If    $R_{\Box}\subseteq A\times X$ is understood as the set of attributions of features to objects {\em as perceived by, or known by, or according to a given agent}, or the set of attributions of features to objects {\em for which there is positive evidence}, then the intended meaning of $\Box \phi$  is `category $\phi$ {\em according to the agent}' or `the category of objects which are {\em demonstrably} $\phi$-objects', and $\mathbb{M}, a \Vdash \Box\phi$ (which, as discussed above, reads `object $a$ is a member of $\Box \phi$') exactly when the agent attributes to $a$ every feature $x$ describing $\phi$, or there is positive evidence of $a$ having each feature $x$ describing $\phi$. Likewise, interpreting $R_{\Diamond}\subseteq X\times A$ in a similar way as $R_{\Box}$, the intended meaning of $\mathbb{M}, a \Vdash \Diamond\phi$, as is read off from the interpretation  clauses for $\Diamond$-formulas above, is that object $a$ has every feature that can demonstrably be associated with (or that the agent attributes to) every member of $\phi$. Notice that, when $R_{\Box}\subseteq I$ (resp.~$R_{\Diamond}\subseteq J$, where $J\subseteq X\times A$ and $xJa$ iff $aIx$)
these requirements translate into a  {\em more restrictive} (resp.~{\em laxer}) threshold of acceptance of objects as members of $\Box \phi$ (resp.~$\Diamond\phi$) than as members of $\phi$. This is the first motivating idea for the framework of \cite{conradie2021rough} unifying RST and FCA: that is, under the epistemic/evidentialist interpretations of $R_{\Box}$ and $R_{\Diamond}$, their associated modal operators $\Box$ and $\Diamond$ would provide the {\em lower} and {\em upper} approximation of (formal) concepts, respectively.

Notice that, in contrast with classical epistemic  semantics,  this epistemic/evidentialist interpretation of $R_{\Box}$ and $R_{\Diamond}$  does not take the negative aspect of the notion being modelled (i.e.~ignorance, uncertainty, or indiscernibility, lack of evidence) as primary, but rather encodes  the positive side, that is, the (possibly partial, or incomplete, or flawed) knowledge or perception of an agent about which objects have which features, or which attributions are grounded in evidence. This observation suggested the second motivating idea for the framework of \cite{conradie2021rough}, namely that the `lift' from, or embedding of,  the setting of approximation spaces in RST to the more general setting of polarities in FCA involves representing (indiscernibility) relations in RST by way of  their set-theoretic complements, as will be discussed technically in Section \ref{ssec: lifting}. 
The third motivating idea, on which the main definition in \cite{conradie2021rough} is grounded (namely, that of {\em conceptual approximation spaces}, cf.~\cite[Definition 4.1]{conradie2021rough}), stems from the observation that also the characteristic properties (such as reflexivity or transitivity) of the indiscernibility relation in RST would `lift', or be preserved, along the embedding of the RST setting into the FCA setting, and even more interestingly, when each of these properties is captured by (i.e., more technically, is the first-order correspondent of) a modal axiom in the RST setting, the {\em same} modal axiom would capture (i.e.~have as first-order correspondent on polarity-based frames) the `lifted' property in the FCA setting (cf.~\cite[Proposition 4.3]{conradie2021rough}). Besides providing the mathematical principles justifying the introduction of conceptual approximation spaces in the specific way they are defined, \cite[Proposition 4.3]{conradie2021rough} naturally elicited the question of whether the phenomenon it captures concerning a finite number of instances can be generalized to classes of modal axioms, and whether this technical result underlies a more conceptual understanding of how the meaning of logical axioms can transfer or be retrieved across different but related semantic contexts. These questions  are the main initial drive of the present paper, which will go even further; namely, we  consider this transfer phenomenon across four different contexts: crisp and many-valued Kripke semantics, and crisp and many-valued polarity-based semantics. We will show that the interpretation of inductive modal reduction principles seamlessly  transfers across these four environments; moreover, in Section \ref{sec:conclusions} we will discuss how these results can be extended further,  to encompass an even wider range of semantic environments. 

\subsection{Inductive modal reduction principles and their ALBA-runs}
\label{sec:mrp}

Let us fix an LE-language $\mathcal{L} = \mathcal{L}(\mathcal{F}, \mathcal{G})$. A {\em modal reduction principle (MRP)} of $\mathcal{L}$ (cf.~\cite[Definition 4.18]{vBPhD}) is an inequality $s(p)\leq t(p)$ such that both $s = s(p)$ and $t = t(p)$ are $\mathcal{L}$-terms generated as follows:
\[s :: =  p\mid f(s)\mid g(s),\]
with $p\in \Prop$, $f\in \mathcal{F}$ and $g\in \mathcal{G}$ such that  $n_f = n_g = 1$ and $\epsilon_f(1) = \epsilon_g(1) = 1$.\footnote{This definition straightforwardly adapts the definition of modal reduction principle in classical modal logic to the context of LE-logics.  We could have included in it also unary connectives $f$ and $g$ such that $\epsilon_f(1) = \epsilon_g(1) = \partial$; however, we chose to not do so, since this would have generated a combinatorial proliferation of cases without significantly adding to the conceptual depth.} 
For the sake of the development of the present theory, we will find it convenient to represent the first-order correspondents of inductive MRPs in a certain compact shape\footnote{Namely, as a pure inequality in the expanded language  $\mathcal{L}_\mathrm{LE}^+$, defined below.} which, even more importantly for our purposes, turns out to be {\em invariant} across different semantic contexts, thanks to duality-theoretic considerations.\footnote{We refer the reader to \cite[Section 2]{CoPa:non-dist} for an extensive discussion on the invariance of this representation.} Therefore, below, we preliminarily introduce the   language-expansion which is used to perform the effective computation, and compact representation, of the first-order correspondents of inductive MRPs; then we will proceed to define inductive MRPs and provide the effective computation and compact/invariant representation of their first-order correspondents.

\paragraph{The expanded language $\mathcal{L}_\mathrm{LE}^+$.} 	Any  language $\mathcal{L}_\mathrm{LE} = \mathcal{L}_\mathrm{LE}(\mathcal{F}, \mathcal{G})$ can be associated with the language $\mathcal{L}_\mathrm{LE}^* = \mathcal{L}_\mathrm{LE}(\mathcal{F}^*, \mathcal{G}^*)$, where $\mathcal{F}^*\supseteq \mathcal{F}$ and $\mathcal{G}^*\supseteq \mathcal{G}$ are obtained by expanding $\mathcal{L}_\mathrm{LE}$ with the following connectives:
	\begin{enumerate}
	\setlength{\itemsep}{0.2pt}
        \setlength{\parskip}{0pt}
        \setlength{\parsep}{0pt}
		\item the $n_f$-ary connective $f^\sharp_i$ for $1\leq i\leq n_f$, the intended interpretation of which is the right residual of $f\in\mathcal{F}$ in its $i$th coordinate if $\varepsilon_f(i) = 1$ (resp.\ its Galois-adjoint\footnote{With \textit{Galois-adjoint} we mean a dual residual. } if $\varepsilon_f(i) = \partial$);
		\item the $n_g$-ary connective $g^\flat_i$ for $1\leq i\leq n_g$, the intended interpretation of which is the left residual of $g\in\mathcal{G}$ in its $i$th coordinate if $\varepsilon_g(i) = 1$ (resp.\ its Galois-adjoint if $\varepsilon_g(i) = \partial$).
		\footnote{The adjoints of the unary connectives $\Box$, $\Diamond$, $\lhd$ and $\rhd$ are denoted $\Diamondblack$, $\blacksquare$, $\blhd$ and $\brhd$, respectively.}
	\end{enumerate}
	We stipulate that
	$f^\sharp_i\in\mathcal{G}^*$ if $\varepsilon_f(i) = 1$, and $f^\sharp_i\in\mathcal{F}^*$ if $\varepsilon_f(i) = \partial$. Dually, $g^\flat_i\in\mathcal{F}^*$ if $\varepsilon_g(i) = 1$, and $g^\flat_i\in\mathcal{G}^*$ if $\varepsilon_g(i) = \partial$. The order-type assigned to the additional connectives is predicated on the order-type of their intended interpretations. That is, for any $f\in \mathcal{F}$ and $g\in\mathcal{G}$,
	
	\begin{tabular}{ll}
	1. &	if $\epsilon_f(i) = 1$, then $\epsilon_{f_i^\sharp}(i) = 1$ and $\epsilon_{f_i^\sharp}(j) = \epsilon_f^\partial(j)$ for any $j\neq i$.\\
		2. & if $\epsilon_f(i) = \partial$, then $\epsilon_{f_i^\sharp}(i) = \partial$ and $\epsilon_{f_i^\sharp}(j) = \epsilon_f(j)$ for any $j\neq i$.\\
		3. & if $\epsilon_g(i) = 1$, then $\epsilon_{g_i^\flat}(i) = 1$ and $\epsilon_{g_i^\flat}(j) = \epsilon_g^\partial(j)$ for any $j\neq i$.\\
		4. & if $\epsilon_g(i) = \partial$, then $\epsilon_{g_i^\flat}(i) = \partial$ and $\epsilon_{g_i^\flat}(j) = \epsilon_g(j)$ for any $j\neq i$.
	\end{tabular}
	
The expanded language of ALBA will include the connectives corresponding to all the residuals of the original connectives, as well as a denumerably infinite set of sorted variables $\mathsf{NOM}$ called {\em nominals},
ranging over a join-generating set of elements of the complex algebra of a given polarity-based frame (e.g., the completely join-irreducible elements of perfect LEs, or, constructively, the closed elements of the constructive canonical extensions, as in \cite{palmigiano2020constructive}), and a denumerably infinite set of
sorted variables $\mathsf{CO\text{-}NOM}$, called {\em co-nominals}, ranging over a meet-generating set of elements of the complex algebra of a given polarity-based frame (e.g., the completely meet-irreducible elements of perfect LEs, or, constructively, on the open elements of the constructive canonical extensions). The elements of $\mathsf{NOM}$ will be denoted with $\nomi, \nomj$, possibly indexed, and those of
$\mathsf{CO\text{-}NOM}$ with $\cnomm, \cnomn$, possibly indexed.
Let us introduce the expanded language formally. The formulas $\phi$ of $\mathcal{L}_\mathrm{LE}^{+}$ are given by the following recursive definition:
\begin{center}
\begin{tabular}{r c |c|c|c|c|c|c c c c c c c}
$\phi ::= $ &$\nomj$ & $\cnomm$ & $\psi$ & $(\phi\wedge\phi)$ & $(\phi\vee\phi)$ & $f(\overline{\phi})$ &$g(\overline{\phi})$
\end{tabular}
\end{center}
with $\psi  \in \mathcal{L}_\mathrm{LE}$, $\nomj \in \mathsf{NOM}$ and $\cnomm \in \mathsf{CO\text{-}NOM}$,  $f\in \mathcal{F}^*$ and $g\in \mathcal{G}^*$. We will follow the standard rules for the elimination of
parentheses.
An $\mathcal{L}_\mathrm{LE}^{+}$-formula (resp.\ inequality) is called {\em pure} if it contains no variables in $\Prop$.

\paragraph{Inductive MRPs.} A MRP is {\em inductive}\footnote{The general definition of inductive  $\mathcal{L}$-inequalities can be found in Appendix \ref{Inductive:Fmls:Section}. The two clauses given here are the result of specializing this general definition to MRPs.} if it is of one of the following forms:

\begin{itemize}
\setlength{\itemsep}{0.2pt}
        \setlength{\parskip}{0pt}
        \setlength{\parsep}{0pt}
\item[(a)] $\phi [\alpha (p)/!y]\leq \psi [\chi (p)/!x]$, where $\psi (!x)$ (resp.~$\phi(!y)$) is a finite (possibly empty) concatenation of box-type connectives in $\mathcal{G}$ (resp.~diamond-type connectives in $\mathcal{F}$),  and $\alpha(p)$ and $\chi (p)$ are terms built up out of the same variable $p\in \mathsf{Prop}$ by finite (possibly empty) concatenations of  box-type connectives  in $\mathcal{G}$, in the case of $\alpha(p)$, and of box- and diamond-type connectives  in the case of $\chi(p)$;
\item[(b)] $\phi [\zeta (p)/!y]\leq \psi [\delta (p)/!x]$, where $\psi (!x)$ (resp.~$\phi(!y)$) is a finite (possibly empty) concatenation of box-type connectives in $\mathcal{G}$ (resp.~diamond-type connectives in $\mathcal{F}$),  and $\delta(p)$ and $\zeta (p)$ are terms built up out of the same variable $p\in \mathsf{Prop}$ by finite (possibly empty) concatenations of  diamond-type connectives  in $\mathcal{F}$, in the case of $\delta(p)$, and of box- and diamond-type connectives  in the case of $\zeta(p)$,
 \end{itemize}

{\em Analytic inductive  modal reduction principles} are of the form $\phi [\alpha (p)/!y]\leq \psi [\delta (p)/!x]$, where $\psi (!x)$, $\phi(!y)$, $\delta(p)$ and $\alpha (p)$ are as above, and thus they are of both shapes (a) and (b).

For example, the inequality $\Diamond \Box p \leq \Diamond \Box \Diamond p$ is an inductive MRP of type (a) with $\phi(!y) := \Diamond y$, $\alpha(p) := \Box p$, $\psi(!x) := x$ and $\chi(p) := \Diamond \Box \Diamond p$.  Note that it is not of type (b). The inequality $\Diamond \Box \Diamond p \leq \Box  p$  is an inductive MRP of type (b) with $\phi(!y) := \Diamond y$, $\zeta(p) := \Box \Diamond p$, $\psi(!x) := \Box x$ and $\delta(p) := p$. Note that it is not of type (a).  Lastly, the inequality $ \Diamond \Box p \leq \Box \Diamond p$  is an analytic inductive MRP with $\phi(!y) := \Diamond y$, $\alpha(p) := \Box p$, $\psi(!x) := \Box x$ and $\delta(p) := \Diamond p$.


\begin{proposition}
\label{prop:albaoutput}
Every inductive MRP can be reduced to an equivalent pure inequality in the expanded language $\mathcal{L}_\mathrm{LE}^+$ as follows: 

\begin{tabular}{rl}
    1. & $\forall p [\phi [\alpha(p)/!y]\leq \psi [\chi (p)/!x]]$ \quad iff \quad $\forall \nomj[\mathsf{LA}(\psi)[\phi[\nomj/!y]/!u] \leq \chi[\mathsf{LA}(\alpha)[\nomj/!u]/p]]$,\\
    2. & $\forall p [\phi [\zeta (p)/!y]\leq \psi [\delta (p)/!x]]$ \quad iff \quad $\forall \cnomm[\zeta[\mathsf{RA}(\delta)[\cnomm/!u]/p] \leq \mathsf{RA}(\phi)[\psi[\cnomm/!y]/!u]$.
\end{tabular}
\end{proposition}
\begin{proof}
In what follows, we apply the manipulation rules of the algorithm ALBA (cf.~\cite{CoPa:non-dist}) in  the following chains of equivalences:
\smallskip

{{\centering
\begin{tabular}{c l l}
& $\forall p [\phi [\alpha(p)/!y]\leq \psi [\chi (p)/!x]]$\\
iff & $\forall p \forall \nomj\forall \cnomm[(\nomj\leq \alpha(p)\ \& \ \chi (p)\leq\cnomm)\Rightarrow \phi[\nomj/!y]\leq \psi [\cnomm/!x]]$ & first approx.\\
iff & $\forall p \forall \nomj\forall \cnomm\left [(\mathsf{LA}(\alpha)[\nomj/!u]\leq p\ \& \  \chi(p)\leq \cnomm)\Rightarrow  \mathsf{LA}(\psi)[\phi[\nomj/!y]/!u]\leq \cnomm \right ] $ & adjunction\\
iff & $\forall \nomj\forall \cnomm[\chi[\mathsf{LA}(\alpha)[\nomj/!u]/p] \leq \cnomm\Rightarrow \mathsf{LA}(\psi)[\phi[\nomj/!y]/!u]\leq \cnomm]$ & Ackermann\\
iff & $\forall \nomj[\mathsf{LA}(\psi)[\phi[\nomj/!y]/!u] \leq \chi[\mathsf{LA}(\alpha)[\nomj/!u]/p]]$ & Ackermann\\[1.5mm]
& $\forall p [\phi [\zeta (p)/!y]\leq \psi [\delta (p)/!x]]$\\
iff & $\forall p \forall \nomj\forall \cnomm[(\nomj\leq \zeta(p)\ \& \ \delta (p)\leq\cnomm)\Rightarrow \phi[\nomj/!y]\leq \psi [\cnomm/!x]]$ & first approx.\\
iff & $\forall p \forall \nomj\forall \cnomm\left [(\nomj\leq \zeta(p)\ \& \ p\leq\mathsf{RA}(\delta)[\cnomm/!u])\Rightarrow \nomj\leq \mathsf{RA}(\phi)[\psi[\cnomm/!y]/!u] \right ] $ & adjunction\\
iff & $\forall \nomj\forall \cnomm[\nomj\leq \zeta[\mathsf{RA}(\delta)[\cnomm/!u]/p] \Rightarrow \nomj\leq \mathsf{RA}(\phi)[\psi[\cnomm/!y]/!u]$ & Ackermann\\
iff & $\forall \cnomm[\zeta[\mathsf{RA}(\delta)[\cnomm/!u]/p] \leq \mathsf{RA}(\phi)[\psi[\cnomm/!y]/!u]$, & Ackermann\\
\end{tabular}
\par}}

\noindent
where if $\phi(!y) := \Diamond_{i_1}\cdots \Diamond_{i_n}y$ (resp.\ $\delta(p) := \Diamond_{i_1}\cdots \Diamond_{i_n} p$) and $\psi(!x): = \Box_{i_1}\cdots \Box_{i_n}x$ (resp.\ $\alpha(p) := \Box_{i_1}\cdots \Box_{i_n} p$), then $\mathsf{RA}(\phi)(!u) =\mathsf{RA}(\delta)(!u) = \blacksquare_{i_n}\cdots \blacksquare_{i_1} u$  and $\mathsf{LA}(\psi)(!u) = \mathsf{LA}(\alpha)(!u) = \Diamondblack_{i_n}\cdots \Diamondblack_{i_1} u$.\footnote{
$\mathsf{LA}(\alpha)(!u)$, $\mathsf{RA}(\delta)(!u)$, $\mathsf{LA}(\psi)(!u)$, and $\mathsf{RA}(\phi)(!u)$  are defined more generally in Definition \ref{def: RA and LA}, however, for the purpose of treating MRPs, the simplified version given here suffices.}
\end{proof}

Let us illustrate the computations of the proof of the proposition above on the MRP 
$\Diamond \Box p \leq \Diamond \Box \Diamond p$, which, as discussed above, is inductive of type (a) with $\phi(!y) := \Diamond y$, $\alpha(p) := \Box p$, $\psi(!x) := x$ and $\chi(p) := \Diamond \Box \Diamond p$:
\smallskip

{{\centering
\begin{tabular}{r c l l}
 $\forall p (\Diamond \Box p \leq \Diamond \Box \Diamond p)$
&iff & $\forall p \forall \nomj\forall \cnomm((\nomj\leq \Box p\ \& \ \Diamond \Box \Diamond p \leq\cnomm)\Rightarrow \Diamond \nomj \leq  \cnomm)$ & first approx.\\
&iff & $\forall p \forall \nomj\forall \cnomm((\Diamondblack \nomj \leq p\ \& \ \Diamond \Box \Diamond p \leq\cnomm)\Rightarrow \Diamond \nomj \leq  \cnomm)$ & adjunction\\
&iff & $\forall \nomj\forall \cnomm(\Diamond \Box \Diamond \Diamondblack \nomj \leq\cnomm\Rightarrow \Diamond \nomj \leq  \cnomm)$ & Ackermann\\
&iff & $ \forall \nomj(\Diamond \nomj\leq \Diamond \Box \Diamond \Diamondblack \nomj).$ & Ackermann\\
\end{tabular}
\par}}
The justification of these computations rests on the order-theoretic properties of the complex algebras of Kripke frames; for instance, the first approximation step holds thanks to the fact that nominals and co-nominals range over join-generators (atoms) and meet-generators (co-atoms) of complex algebras (powerset algebras with operators), respectively, and that the interpretations of $\Box$ and $\Diamond$  are completely meet-preserving and completely join-preserving, respectively, while the justification of the `Ackermann' steps rests on applications of the well known Ackermann lemma (cf.~\cite{Ackermann}, see also \cite[Lemma 6.3, Lemma 6.4]{CoPa:non-dist} for the specific applications to the present context). Interestingly, not only are  these computations sound w.r.t.~the complex algebras of Kripke frames, but their soundness is preserved also in the much more general environment of the complex algebras of polarity-based frames, since the order-theoretic properties on which they rest are maintained.\footnote{The computations above are in fact instances of runs of the algorithm ALBA \cite{CoPa:non-dist}, which is shown to succesfully transform every inductive inequality to a set of pure inequalities in the expanded language $\mathcal{L}^+$. Because in the present paper we focus on MRPs, the full generality of ALBA is not required, and the computations described above suffice.}

\subsection{Lifting Kripke frames to polarity-based frames}
\label{ssec: lifting}
As discussed  in the previous subsections, the present paper aims to extend results achieved in \cite{conradie2021rough}, establishing systematic connections between the first-order correspondents of  a finite number of well known inductive MRPs, to the whole class of inductive MRPs. These systematic connections are established on the basis of an embedding mapping any set $S$ to its associated polarity $\mathbb{P}_{S}$; this embedding  can be expanded to an embedding mapping any  Kripke frame $\mathbb{X}$ to its associated polarity-based frame $\mathbb{F}_{\mathbb{X}}$. In the present section, which is based on \cite[Sections 2 and 3]{conradie2021rough},  we partly recall  the definition of this embedding, and partly extend it.

Throughout this section, for every set $S$, we let $\Delta_S: = \{(s, s)\mid s\in S\}$, and we typically drop the subscript when it does not cause ambiguities. Hence we write e.g.~$\Delta^c = \{(s, s')\mid s, s'\in S\mbox{ and }s\neq s'\}$. We let $S_A$ and $S_X$ be  copies of $S$, and for every $P\subseteq S$, we let $P_A\subseteq S_A$ and $P_X\subseteq S_X$ denote the corresponding copies of $P$ in $S_A$ and $S_X$, respectively. Then $P^c_X$ (resp.~$P^c_A$) stands both for $(P^c)_X$ (resp.~$(P^c)_A$) and $(P_X)^c$ (resp.~$(P_A)^c$). For the sake of a more manageable notation, we will use $a$ and $b$ (resp.~$x$ and $y$)  to indicate both elements of $A$ (resp.~$X$) and their corresponding elements in $S_A$  (resp.~$S_X$), relying on the types of the relations for disambiguation. In what follows, for any order-type $\epsilon$ on $n$, the symbol $S_X^{\epsilon}$ denotes the $n$-fold cartesian product of $S_A$ and $S_X$ such that for any $1\leq i\leq n$, its $i$th projection is $S_X$ if $\epsilon(i) = 1$ and is $S_A$ if $\epsilon(i) = \partial$; similarly, the symbol $S_A^{\epsilon}$ denotes the $n$-fold cartesian product of $S_A$ and $S_X$ such that for any $1\leq i\leq n$, its $i$th projection is $S_A$ if $\epsilon(i) = 1$ and is $S_X$ if $\epsilon(i) = \partial$.
\begin{definition}
\label{def:liftings relations-n plus one}
 For every $R\subseteq S^{n+1}$ and any order-type $\epsilon$ on $n$, we let
 \smallskip
 
{{\centering
$I_{R}^{\epsilon}\subseteq S_A\times S_X^{\epsilon}\; $ such that $\; a I_{R}^\varepsilon \overline{x}$ iff $a R\overline{x}$;
 $\quad\quad J_{R}^{\epsilon}\subseteq S_X\times S_A^\epsilon\; $ such that $\; x J_{R}^\varepsilon \overline{a}$ iff $x R\overline{a}$.
\par}}
\end{definition}
In what follows, we will apply the definition above to relations associated with logical connectives $f$ and $g$. In this case, the order-type $\epsilon$ will be instantiated to the one associated with a given connective. Since we will focus on unary connectives in the remainder of the paper, the following definition specializes the definition above to binary relations (i.e.~to potential interpretants of unary connectives).
\begin{definition}[cf.~\cite{conradie2021rough} Definition 3.1]
\label{def:liftings relations}
 For every $R\subseteq S\times S$, we let
 \smallskip
 
{{\centering
\begin{tabular}{ll}
$I_{R}\subseteq S_A\times S_X$ such that $a I_{R} x$ iff $a Rx$;
\quad\quad & 
$J_{R}\subseteq S_X\times S_A$ such that $x J_{R} a$ iff $x Ra$;
\\
$H_{R}\subseteq S_A\times S_A$ such that $a H_{R} b$ iff $a Rb$;
&
$K_{R}\subseteq S_X\times S_X$ such that $xK_{R} y$ iff $x Ry$.
\end{tabular}
\par}}
\end{definition}

\begin{lemma}[cf.~\cite{conradie2021rough}, Lemma 3.2]
\label{lemma:liftings and converses}
For every $R\subseteq S\times S$,
\begin{equation}
\label{eq: converses and liftings}
(J_R)^{-1} = I_{R^{-1}}\quad (I_R)^{-1} = J_{R^{-1}}\quad (H_R)^{-1} = H_{R^{-1}}\quad (K_R)^{-1} = K_{R^{-1}}.
\end{equation}
\end{lemma}
\begin{definition}[Kripke $\mathcal{L}$-frame]
\label{def:kripkelframe}
  For any LE-language $\mathcal{L} = \mathcal{L}(\mathcal{F}, \mathcal{G})$, a {\em Kripke} $\mathcal{L}$-{\em frame} is a tuple  $\mathbb{X}= (S,\mathcal{R}_{\mathcal{F}}, \mathcal{R}_{\mathcal{G}})$ such that $\mathcal{R}_{\mathcal{F}} = \{R_f\mid f\in \mathcal{F}\}$ and  $\mathcal{R}_{\mathcal{G}} = \{R_g\mid g\in \mathcal{G}\}$, and $R_f\subseteq S^{n_f +1}$ and $R_g\subseteq S^{n_g+1}$  for all $f\in \mathcal{F}$ and  $g\in \mathcal{G}$. 
  
  The {\em complex algebra} of $\mathbb{X}$ is the Boolean algebra with operators 
  \[\mathbb{X}^+ = (\mathcal{P}(S), \{f_{R_f}\mid f\in \mathcal{F}\} ,\{ g_{R_g}\mid g\in \mathcal{G}\}),
  \]
where, for every $f\in \mathcal{F}$ and $g\in \mathcal{G}$, the symbols $f_{R_f}$ and $g_{R_g}$ respectively denote an $n_f$-ary and an $n_g$-ary operation on $\mathcal{P}(S)$, defined by the assignments $\overline{X}\mapsto \{x\in S\mid \exists \overline{y}(R_f(x, \overline{y})\ \&\  \overline{y}\in \overline{X}^{\epsilon_f})\}$ and $\overline{X}\mapsto \{x\in S\mid \forall \overline{y}( R_g(x, \overline{y}) \Rightarrow \overline{y}\in \overline{X}^{\epsilon_g}) \}$, respectively, where, for  any order-type $\epsilon$ on $n$, the symbol $\overline{X}^{\epsilon}$ denotes the element of $(\mathcal{P}(S))^n$  whose $i$th coordinate for each $1\leq i\leq n$ is $X_i$ if $\varepsilon(i) = 1$ and is $X_i^{c}$ if $\varepsilon(i) = \partial$. 
 \end{definition}
  \begin{definition}[Lifting of Kripke $\mathcal{L}$-frame]
  \label{def:lifting kripke frames}
 For any Kripke $\mathcal{L}$-frame $\mathbb{X}$ as above,  we let $\mathbb{F}_{\mathbb{X}}: = (\mathbb{P}_{S}, \mathcal{J}_{\mathcal{R}_{\mathcal{F}}^c}, \mathcal{I}_{\mathcal{R}_{\mathcal{G}}^c})$, where  $\mathbb{P}_{S} = (S_A, S_X, I_{\Delta^c})$ with $I_{\Delta^c}$ such as in Def.\ \ref{def:liftings relations}, and $\mathcal{J}_{\mathcal{R}_{\mathcal{F}}^c} = \{J_{R_f^c}^{\epsilon_f}\mid f\in \mathcal{F}\}$ and $\mathcal{I}_{\mathcal{R}_{\mathcal{G}}^c} = \{I_{R_g^c}^{\epsilon_g}\mid g\in \mathcal{G}\}$.
 \end{definition}
 Since the concept lattice of $\mathbb{P}_S$ is isomorphic\footnote{As for all $W\subseteq S$, $I_{\Delta^c}^{(0)}[W] = W^c = I_{\Delta^c}^{(1)}$; hence for all $B\subseteq S_A$, $Y \subseteq S_X$, we have $B = I_{\Delta^c}^{(0)}[I_{\Delta^c}^{(1)}[B]]$ and $Y = I_{\Delta^c}^{(1)}[I_{\Delta^c}^{(0)}[Y]]$.} to $\mathcal{P}(S)$, the relations  $I_{R_g^c}^{\epsilon_g}\subseteq S_A\times S_X^{\epsilon_g}$ and  $J_{R_f^c}^{\epsilon_f}\subseteq S_X\times S_A^{\epsilon_f}$ are trivially $I_{\Delta^c}$-compatible  for all $f\in \mathcal{F}$ and  $g\in \mathcal{G}$, hence $\mathbb{F}_{\mathbb{X}}$ is a polarity-based $\mathcal{L}$-frame (cf.~Definition \ref{def:polarity:based:frm}).
%
%
The following proposition generalizes \cite[Proposition 3.7]{conradie2021rough}.
\begin{proposition}
\label{prop:from Kripke frames to enriched polarities}
  For any LE-language $\mathcal{L} = \mathcal{L}(\mathcal{F}, \mathcal{G})$, if $\mathbb{X}$ is a Kripke $\mathcal{L}$-frame, then $\mathbb{F}_{\mathbb{X}}^+ \cong \mathbb{X}^+$.
  
\end{proposition}
\begin{proof}
Let  $\mathbb{X}=(S, \mathcal{R}_{ \mathcal{F}}, \mathcal{R}_{ \mathcal{G}})$ be an $\mathcal{L}$-Kripke frame,  let   $\mathbb{X}^+= (\mathcal{P}(S), \{f_{R_f}\mid f \in \mathcal{F} \},  \{g_{R_g}\mid g \in \mathcal{G} \})$ be its complex algebra as in Definition \ref{def:kripkelframe}. Moreover, let $\mathbb{F}_{\mathbb{X}} = (\mathbb{P}_{S}, \mathcal{J}_{\mathcal{R}_{\mathcal{F}}^c}, \mathcal{I}_{\mathcal{R}_{\mathcal{G}}^c})$ as in Definition \ref{def:lifting kripke frames}, and $\mathbb{F}_{\mathbb{X}}^+ = (\mathbb{P}_{S}^+,\{ f_{J_{R_f^c}}\mid f\in \mathcal{F}\}, \{g_{I_{R_g^c}}\mid g\in \mathcal{G}\} )$ as in Definition \ref{def:complex algebra of LE-frame}. By  \cite[Proposition 3.4]{conradie2021rough},  $\mathcal{P}(S)\cong \mathbb{P}_S^+$. 
Thus, it is enough to show that 
for any  $f \in \mathcal{F}$ and $ g\in  \mathcal{G}$,
\[
(f_{R_f}(\overline{\val{c}}^{\epsilon_{f}}))_A= \val{f_{J_{{R_f^c}}}(\overline{c})} \quad \quad \text{and} \quad \quad ( g_{R_g}(\overline{\descr{c}}^{\epsilon_{g}}))_A= \val{g_{I_{{R_g^c}}}(\overline{c})}
\]
where the $i$th coordinate\footnote{When writing $f_{R_f}(\overline{\val{c}}^{\epsilon_{f}})$ and $g_{R_g}(\overline{\descr{c}}^{\epsilon_{g}})$, we are slightly abusing notation, since any coordinate of $\overline{\val{c}}^{\epsilon_{f}}$ and $\overline{\descr{c}}^{\epsilon_{g}}$ is a subset of $S_A$ or of $S_X$, and not, strictly speaking, of $S$. However, as $S_A$, $S_X$ and $S$ are clearly isomorphic,  each coordinate of $\overline{\val{c}}^{\epsilon_{f}}$ and $\overline{\descr{c}}^{\epsilon_{g}}$ uniquely identifies a subset of $S$, and hence the vectors $\overline{\val{c}}^{\epsilon_{f}}$ and $\overline{\descr{c}}^{\epsilon_{g}}$ uniquely identify the arguments of $f_{R_f}$ and $g_{R_g}$.} in $\overline{\val{c}}^{\epsilon}$ (resp.~$\overline{\descr{c}}^{\epsilon}$) is $\val{c_i}$ (resp.~$\descr{c_i} =  \val{c_i}^c$) if $\varepsilon(i) = 1$ and is $\descr{c_i} = \val{c_i}^c$ (resp.~$\val{c_i}$) if $\varepsilon(i) = \partial$.
\smallskip

{{\centering
\begin{tabular}{cc}
\begin{tabular}{rcl}
& & $ \val{f_{J_{R_f^c}}(\overline{c} )}$   \\                    & = & $ \left(J_{R_f^c}^{(0)}[\overline{\val{c}}^{\epsilon_{f}}]\right)^{\downarrow}$\\
       & = & $(\{x \in S_X \mid \forall \overline{b} (  \overline{b} \in \overline{\val{c}}^{\epsilon_{f}} \Rightarrow x  J_{R_f^c} \overline{b})\})^\downarrow$\\

	     & = & $((\{x \in S \mid \forall \overline{b} (  \overline{b} \in \overline{\val{c}}^{\epsilon_{f}} \Rightarrow x  R_f^c \overline{b})\})^c)_A$\\
	     & = & $(\{x \in S \mid \exists \overline{b} (  \overline{b} \in \overline{\val{c}}^{\epsilon_{f}}\ \&\;  x  R_f \overline{b})\})_A$\\
	      &= & $(f_{R_f}(\overline{\val{c}}^{\epsilon_f}))_A$\\

 \end{tabular}          
&
\begin{tabular}{rcl}
&& $ \val{g_{I_{R_g^c}}(\overline{c} )}$  \\                    & = & $ I_{R_f^c}^{(0)}[\overline{\descr{c}}^{\epsilon_{g}}]$\\
& = & $\{a \in S_A \mid \forall \overline{y} (  \overline{y} \in \overline{\descr{c}}^{\epsilon_{g}} \Rightarrow a  I_{R_g^c} \overline{y})\}$\\
 & = & $(\{a \in S \mid \forall \overline{y} (  \overline{y} \in \overline{\descr{c}}^{\epsilon_{g}} \Rightarrow a  R_g^c \overline{y})\})_A$\\
  & = & $(\{a \in S \mid \forall \overline{y} (   a  R_g \overline{y}\Rightarrow \overline{y} \in \overline{\val{c}}^{\epsilon_{g}} )\})_A$\\

	      &= & $(g_{R_g}(\overline{\val{c}}^{\epsilon_g}))_A$.\\

 \end{tabular}   
 \\
 \end{tabular}
\par}}

\end{proof}


\subsection{Many-valued polarity-based semantics for LE-logics}
\label{ssec:MV-polarities}
Many-valued Kripke semantics for modal logics were introduced in \cite{fitting1991many} while many-valued polarities have been studied in \cite{belohlavek} on which basis many-valued polarity-based frames were introduced in \cite[Section 7]{conradie2021rough}. Just like many-valued accessibility relations in many-valued Kripke frames capture the extent to which two states are related, the incidence relation in many-valued polarities intuitively captures the extent to which an object has a feature, and, when interpreted e.g.\ epistemically, the additional relations on polarity-based frames capture the extent to which an agent attributes a feature to an object. In \cite{conradie2019logic}, many-valued polarity-based frames have been proposed as a semantic environment for the logic of `vague categories', formally represented as formal concepts arising from many valued polarities. Vague categories are those categories for which membership is a matter of degrees, and common examples are the category of blue objects or the category of tall individuals.   

Throughout this paper,  we let $\mathbf{A} = (D, 1, 0, \vee, \wedge, \to)$ denote an arbitrary but fixed 
complete frame-distributive and dually frame-distributive Heyting algebra  (understood as the algebra of truth-values). For every set $W$, an $\mathbf{A}$-{\em valued subset}  (or $\mathbf{A}$-{\em subset}) of $W$ is a map $u: W\to \mathbf{A}$.  We let $\mathbf{A}^W$ denote the set of all $\mathbf{A}$-subsets. Clearly, $\mathbf{A}^W$ inherits the algebraic structure of $\mathbf{A}$ by defining the operations and the order pointwise. The $\mathbf{A}$-{\em subsethood} relation between elements of $\mathbf{A}^W$ is the map $S_W:\mathbf{A}^W\times \mathbf{A}^W\to \mathbf{A}$ defined as $S_W(h, k) :=\bigwedge_{z\in W }(h(z)\rightarrow k(z)) $. For every $\alpha\in \mathbf{A}$, 
let $\{\alpha\slash w\}: W\to \mathbf{A}$ be defined by $v\mapsto \alpha$ if $v = w$ and $v\mapsto 0$ if $v\neq w$. Then, for all $h\in \mathbf{A}^W$,
\begin{equation}\label{eq:MV:join:generators}
h = \bigvee_{w\in W}\{h(w)\slash w\}.
\end{equation}
 When $h, k: W\to \mathbf{A}$ and $h\leq k$ w.r.t.~the pointwise order, we write $h\subseteq k$.
A binary $\mathbf{A}$-{\em valued relation} (or $\mathbf{A}$-{\em relation}) is a map $R: U \times W \rightarrow \mathbf{A}$. Two-valued relations can be regarded as  $\mathbf{A}$-relations. In particular for any set $Z$, we let $\Delta_Z: Z\times Z\to \mathbf{A}$ be defined by $\Delta_Z(z, z') = 1$ if $z = z'$ and $\Delta_Z(z, z') = 0$ if $z\neq z'$. Any $\mathbf{A}$-valued relation $R: U \times W \rightarrow \mathbf{A}$ induces  maps $R^{(0)}[-] : \mathbf{A}^W \rightarrow \mathbf{A}^U$ and $R^{(1)}[-] : \mathbf{A}^U \rightarrow \mathbf{A}^W$ defined as follows: for every $h: U \to \mathbf{A}$ and $u: W \to \mathbf{A}$,

{{\centering
	\begin{tabular}{r l r l }
$R^{(1)}[h]:$ & $ W\to \mathbf{A}$ & $R^{(0)}[u]: $ & $U\to \mathbf{A} $\\
		
		& $ x\mapsto \bigwedge_{a\in U}(h(a)\rightarrow R(a, x))$ \quad\quad\quad&
		& $a\mapsto \bigwedge_{x\in W}(u(x)\rightarrow R(a, x))$\\
	\end{tabular}
\par}} 

\noindent
The following lemma is the many valued counterpart of \cite[Lemma 2.2]{conradie2021rough}, and collects some basic properties  of this construction.
 \begin{lemma}\label{lemma: basic mv}
 For any $\mathbf{A}$-valued relation $R: U \times W \rightarrow \mathbf{A}$, any $h, k: U \to \mathbf{A}$, any $u, v:W \to \mathbf{A}$, any collections $\mathcal{H}$ (resp.\ $\mathcal{U}$) of $\mathbf{A}$-valued subsets of $U$ (resp.\ $W$),
 
\begin{tabular}{ll}
1. & if $h \leq k$ then $R^{(1)}[k]\leq R^{(1)}[h]$, and   if $u \leq v$ then $R^{(0)}[v]\leq R^{(0)}[u]$.\\
2. & $h \leq R^{(0)}[u]$ \ iff \  $u \leq R^{(1)}[h]$.\\
 3. & $h\leq R^{(0)}[R^{(1)}[h]]$ and $u\leq R^{(1)}[R^{(0)}[u]]$. \\
 4. & $R^{(1)}[h] = R^{(1)}[R^{(0)}[R^{(1)}[h]]]$ and $R^{(0)}[u] = R^{(0)}[R^{(1)}[R^{(0)}[u]]]$.\\
 5. & $R^{(0)}[\bigvee\mathcal{H}] = \bigwedge_{h\in \mathcal{H}}R^{(0)}[h]$ and $R^{(1)}[\bigvee\mathcal{U}] = \bigwedge_{u \in \mathcal{U}}R^{(1)}[u]$.\\
\end{tabular}
 \end{lemma}
\begin{proof}
1. For all $x \in W$,

{{\centering
\begin{tabular}{rclr}
    $R^{(1)}[k](x)$ & $=$ & $\bigwedge_{a \in U}(k(a) \rightarrow R(a,x))$ & Definition of $R^{(1)}[\cdot]$\\
    & $\leq$ & $\bigwedge_{a \in U}(h(a) \rightarrow R(a,x))$ & $h \leq k$ and antitonicity of $\rightarrow$ \\
    & $=$ & $R^{(1)}[h](x)$ & Definition of $R^{(1)}[\cdot]$
\end{tabular}
\par}}

\noindent
The second statement is proven similarly. As to item 2,

{{\centering
\begin{tabular}{r rlr}
 $h \leq R^{(0)}[u]$ & 
iff & $\forall a\left( h(a) \leq R^{(0)}[u](a) \right)$ & Definition of $\leq$ \\
&iff & $\forall a\left( h(a) \leq \bigwedge_{x\in W}(u(x) \rightarrow R(a,x)) \right)$ & Definition of $R^{(0)}[\cdot]$ \\
&iff & $\forall a \forall x \left( h(a) \leq u(x) \rightarrow R(a,x) \right)$ & Definition of $\wedge$ \\
&iff & $\forall a \forall x \left( u(x) \leq h(a) \rightarrow R(a,x) \right)$ & residuation \\
&iff & $\forall x \forall a \left( u(x) \leq h(a) \rightarrow R(a,x) \right)$ &  \\
&iff & $\forall x \left( u(x) \leq \bigwedge_{a\in U}(h(a) \rightarrow R(a,x)) \right)$ & definition of $\wedge$ \\
&iff & $\forall x \left( u(x) \leq R^{(1)}[h](x) \right)$ & definition of $R^{(1)}[\cdot]$ \\
&iff & $u \leq R^{(1)}[h]$ \\
\end{tabular}
\par}}

\noindent
Item 2 can be equivalently restated by saying that the maps $R^{(0)}[\cdot]$ and $R^{(1)}[\cdot]$ form a Galois connection of the partial orders $(\mathbf{A}^U, \leq)$ and $(\mathbf{A}^W, \leq)$, where $\leq$ is the pointwise order. Hence, items 3, 4, and 5 immediately follow from the basic properties of Galois connections.
\end{proof}

For any $n\geq 1$, an $(n+1)$-ary $\mathbf{A}$-{\em valued relation} (or $\mathbf{A}$-{\em relation}) is a map $R: \Pi_{0\leq i\leq n} W_i \rightarrow \mathbf{A}$.
Any such relation  induces  maps $R^{(i)}[-] : \Pi_{j\neq i} \mathbf{A}^{W_j} \rightarrow \mathbf{A}^{W_i}$, for every $0\leq i\leq n$,
defined as follows: for every $\overline{h}\in\Pi_{j\neq i} \mathbf{A}^{W_j}$,

{{\centering
	\begin{tabular}{r l }
$R^{(i)}[\overline{h}]:$ & $ W_i \to \mathbf{A}$\\
		
		& $ x\mapsto \bigwedge_{\overline{a}^i\in \Pi_{j\neq i} W_j}((\bigwedge_{j\neq i}h_j(a_j))\rightarrow R(\overline{a}_x^i))$,\\
	\end{tabular}
\par}}

where $\overline{a}\in \Pi_{0\leq j \leq n} W_j$.
For $R^{(i)}[-]$ a similar result as Lemma \ref{lemma: basic mv} holds:
 \begin{lemma}\label{lemma: almost basic mv}
 For any $n\geq 1$ and any $(n+1)$-ary $\mathbf{A}$-valued relation $R: \Pi_{0\leq i\leq n} W_i \rightarrow \mathbf{A}$, any $0\leq i,j\leq n$ and $\overline{h}\in\prod_{0\leq i\leq n}\mathbf{A}^{W_i}$, 
\[h_i \leq R^{(i)}[\overline{h}^i]\ \iff\ h_j \leq R^{(j)}[\overline{h}^j].\]
\end{lemma}
\begin{proof}
The proof is similar to item 2 of Lemma \ref{lemma: basic mv}:

{{\centering
\begin{tabular}{rrlr}
 $h_i \leq R^{(i)}[\overline{h}^i]$ & 
iff & $\forall a_i\left( h_i(a_i) \leq R^{(i)}[\overline{h}^i](a_i) \right)$ & Definition of $\leq$ \\
&iff & $\forall a_i\left( h_i(a_i) \leq \bigwedge_{\overline{a}^i\in \Pi_{k\neq i} W_k}((\bigwedge_{k\neq i}h_k(a_k))\rightarrow R(\overline{a}_{a_i}^i)) \right)$ & Definition of $R^{(i)}[-]$ \\
& iff & $\forall a_i \forall \overline{a}^i \left( h_i(a_i) \leq ((\bigwedge_{k\neq i}h_k(a_k))\rightarrow R(\overline{a}_{a_i}^i)) \right)$ & Definition of $\wedge$ \\
&iff & $\forall a_i \forall \overline{a}^i \left( h_j(a_j) \leq (h_i(a_i)\wedge \bigwedge_{k\neq i,j}h_k(a_k)  ) \rightarrow R(\overline{a}^j_{a_j}) \right)$ & residuation \\
&iff & $\forall a_j \forall \overline{a}^j \left( h_j(a_j) \leq ((\bigwedge_{k\neq j}h_k(a_k))\rightarrow R(\overline{a}_{a_j}^j)) \right)$ &  \\
&iff & $\forall a_j \left( h_j(a_j) \leq \bigwedge_{\overline{a}^j\in \Pi_{k\neq j} W_k}((\bigwedge_{k\neq j}h_k(a_k))\rightarrow R(\overline{a}_{a_j}^j)) \right)$ & definition of $\wedge$ \\
&iff & $\forall a_j \left( h_j(a_j) \leq R^{(j)}[\overline{h}^j](a_j) \right)$ & definition of $R^{(j)}[-]$ \\
&iff & $h_j \leq R^{(j)}[\overline{h}^j]$. \\
\end{tabular}
\par}}
\end{proof}
\begin{lemma}\label{lem:keytouniversemysteries}
For any $n\geq 1$, any $(n+1)$-ary $\mathbf{A}$-valued relation $R: \Pi_{0\leq i\leq n} W_i \rightarrow \mathbf{A}$, any  $0\leq i\leq n$, any $\overline{h}\in\prod_{0\leq i\leq n}\mathbf{A}^{W_i}$,
 and any $\overline{\alpha}\in\mathbf{A}^{n+1}$, \[(\bigwedge_{j\neq i}\alpha_j) \to R^{(i)}[\overline{h}_i]=R^{(i)}[\overline{(\alpha\land h)}_i].\]
\end{lemma}
\begin{proof}
For  any $w\in W_i$,

{{\centering
    \begin{tabular}{rcl r}
        $((\bigwedge_{ j\neq i}\alpha_j)\to R^{(i)}[\overline{h}_i])(w)$ & $=$ & $(\bigwedge_{ j\neq i}\alpha_j)\to \bigwedge_{\overline{a}^i\in \Pi_{j\neq i} W_j}((\bigwedge_{j\neq i}h_j(a_j))\rightarrow R(\overline{a}_w^i))$ &  \\
         & $=$ & $\bigwedge_{\overline{a}^i\in \Pi_{j\neq i} W_j}((\bigwedge_{j\neq i}(\alpha_j\land h_j(a_j)))\rightarrow R(\overline{a}_w^i))$ & Currying \\
         & $=$ & $R^{(i)}[\overline{(\alpha\land h)}_i](w).$ & Def.\ of $R^{(i)}[\cdot]$
    \end{tabular}
\par}}
\end{proof}

A {\em formal}  $\mathbf{A}$-{\em context} or $\mathbf{A}$-{\em polarity} (cf.~\cite{belohlavek}) is a structure $\mathbb{P} = (A, X, I)$ such that $A$ and $X$ are sets and $I: A\times X\to \mathbf{A}$. Any $\mathbf{A}$-polarity induces  maps $(\cdot)^{\uparrow}: \mathbf{A}^A\to \mathbf{A}^X$ and $(\cdot)^{\downarrow}: \mathbf{A}^X\to \mathbf{A}^A$ given by $(\cdot)^{\uparrow} = I^{(1)}[\cdot]$ and $(\cdot)^{\downarrow} = I^{(0)}[\cdot]$. 
%
By Lemma \ref{lemma: basic mv} (2), the pair of maps $(\cdot)^{\uparrow}$ and $(\cdot)^{\downarrow}$ form a  Galois connection of the partial orders $(\mathbf{A}^A, \leq)$ and $(\mathbf{A}^X, \leq)$, where $\leq$ is the pointwise order\footnote{Notice that, since $\mathbf{A}$ is a complete lattice, so are $(\mathbf{A}^A,\leq)$ and $(\mathbf{A}^X,\leq)$.}. 
A {\em formal}  $\mathbf{A}$-{\em concept} of $\mathbb{P}$ is a pair $(h, u)\in \mathbf{A}^A\times \mathbf{A}^X$ which is stable w.r.t.\ this Galois connection, i.e.\ $(h,u)$ is such that $h^{\uparrow} = u$ and $u^{\downarrow} = h$. It follows immediately that, if $(h, u)$ is a formal $\mathbf{A}$-concept, then $h^{\uparrow \downarrow} = h$ and $u^{\downarrow\uparrow} = u$, that is, $h$ and $u$ are {\em stable}. The set of formal $\mathbf{A}$-concepts is partially ordered as follows:
\[(h, u)\leq (k, v)\quad \mbox{ iff }\quad h\subseteq k \quad \mbox{ iff }\quad v\subseteq u. \]
Ordered in this way, the set $\mathbb{P}^+$ of the formal  $\mathbf{A}$-concepts of $\mathbb{P}$ is a complete lattice, and is isomorphic to a complete sub-meet-semilattice of $(\mathbf{A}^A,\leq)$  and $(\mathbf{A}^X,\leq)$. This observation justifies the following lemma.
\begin{lemma}
\label{lemma:galois stable is closed under intersection mv}
For every $\mathbf{A}$-polarity $(A, X, I)$ and any collections $\mathcal{H}$ (resp.\ $\mathcal{U}$) of $\mathbf{A}$-valued subsets of $A$ (resp.\ $X$), the $\mathbf{A}$-valued subsets $\bigwedge \mathcal{H}$ and $\bigwedge \mathcal{U}$ are Galois-stable.
\end{lemma}

The following definition extends those given in  \cite[Section 7.2]{conradie2021rough} and \cite[Definition 2.2]{conradie2019logic}.
\begin{definition}\label{def:polarity:based:frm:mv}
	An {\em $\mathbf{A}$-polarity-based $\mathcal{L}_{\text{LE}}$-frame}  is a tuple $\mathbb{F} = (\mathbb{P}, \mathcal{R}_{\mathcal{F}}, \mathcal{R}_{\mathcal{G}})$, where  $\mathbb{P} = (A, X,  I)$ is an $\mathbf{A}$-polarity, $\mathcal{R}_{\mathcal{F}} = \{R_f\mid f\in \mathcal{F}\}$, and $\mathcal{R}_{\mathcal{G}} = \{R_g\mid g\in \mathcal{G}\}$, such that  for each $f\in \mathcal{F}$ and $g\in \mathcal{G}$, the symbols $R_f$ and  $R_g$ respectively denote $(n_f+1)$-ary and $(n_g+1)$-ary $\mathbf{A}$-relations 
	\begin{equation*}
	R_f : X \times A^{\epsilon_{f}}\to \mathbf{A}    \ \mbox{ and }\ R_g:  A \times X^{\epsilon_{g}}\to \mathbf{A},
	\end{equation*}
	where $A^{\epsilon_{f}}$ 
	and $X^{\epsilon_g}$ are  as in Definition \ref{def:polarity:based:frm}, 
	and all relations $R_f$ and $R_g$ are required to be {\em $I$-compatible}, i.e.\ the following sets are assumed to be  Galois-stable for all $\overline{\alpha}\in \mathbf{A}^{n_f}$, $\overline{\beta}\in \mathbf{A}^{n_g}$, $\alpha_0,\beta_0\in\mathbf{A}$, $\overline{a} \in A^{\epsilon_f}$, and $\overline{x} \in X^{\epsilon_g}$, $a_0\in A$, $x_0\in X$ and every $1\leq i\leq n_f$ (resp.~$1\leq i\leq n_g$): 
	\[
R_f^{(0)}[\overline{\{\alpha\slash{a}\}}] \quad\quad R_f^{(i)}[\{\alpha_0\slash x_0\},\overline{\{\alpha\slash{a}\}}^i] \quad\quad R_g^{(0)}[\overline{\{\beta\slash x\}}] \quad\quad R_g^{(i)}[\{\beta_0\slash a_0\},\overline{\{\beta\slash{x}\}}^i],
	\]
where $\overline{\{\alpha\slash a\}}=(\{\alpha_i\slash a_i\})_{1\leq i\leq n_f}$ and $\alpha_i$ and $a_i$ are the $i$th coordinates of $\overline{\alpha}$ and $\overline{a}$ respectively; similarly for $\overline{\{\beta\slash x\}}$.
	Thus, for all $\Diamond, {\lhd}\in \mathcal{F}$ and all $\Box, {\rhd}\in \mathcal{G}$,
	\begin{equation*}
	R_\Diamond :X \times A \to \mathbf{A}     \quad R_\lhd : X \times X \to \mathbf{A}    \quad R_\Box : A \times X \to \mathbf{A}   \quad R_\rhd : A \times A \to \mathbf{A},
	\end{equation*}
	such that  for all $\alpha \in \mathbf{A}$, $a \in A$ and $x \in X $,
	\[
\begin{array}{rcl}
	(R_\Diamond^{(0)}[\{\alpha \slash a\}])^{\downarrow\uparrow}\subseteq R_\Diamond^{(0)}[\{\alpha \slash a\}] & \mbox{ and } & (R_\Diamond^{(1)}[\{\alpha \slash x\}])^{\uparrow\downarrow}\subseteq R_\Diamond^{(1)}[\{\alpha \slash x\}]\\ 
	(R_\Box^{(0)}[\{\alpha \slash x\}])^{\uparrow\downarrow}\subseteq R_\Box^{(0)}[\{\alpha \slash x\}] & \mbox{ and } &
	(R_\Box^{(1)}[\{\alpha \slash a\}])^{\downarrow \uparrow}\subseteq R_\Box^{(1)}[\{\alpha \slash a\}]\\
	(R_\lhd^{(0)}[\{\alpha \slash x\}])^{\downarrow\uparrow}\subseteq R_\lhd^{(0)}[\{\alpha \slash x\}] & \mbox{ and } & (R_\lhd^{(1)}[\{\alpha \slash x\}])^{\downarrow\uparrow}\subseteq R_\lhd^{(1)}[\{\alpha \slash x\}]\\ 
	(R_\rhd^{(0)}[\{\alpha \slash a\}])^{\uparrow\downarrow}\subseteq R_\rhd^{(0)}[\{\alpha \slash a\}] &  \mbox{ and }  &(R_\rhd^{(1)}[\{\alpha \slash a\}])^{\uparrow\downarrow }\subseteq R_\rhd^{(1)}[\{\alpha \slash a\}].\\
	\end{array}
	\]
\end{definition}

	The {\em complex algebra} of an $\mathbf{A}$-polarity-based $\mathcal{L}_{\mathrm{LE}}$-frame $\mathbb{F}$ as above is the $\mathcal{L}_{\mathrm{LE}}$-algebra 
$$\mathbb{F}^+ = (\mathbb{L}, \{f_{R_f}\mid f\in \mathcal{F}\}, \{g_{R_g}\mid g\in \mathcal{G}\}),$$
where $\mathbb{L}: = \mathbb{P}^+$, and for all $f \in \mathcal{F}$ and $g \in \mathcal{G}$,

\begin{enumerate}
	\setlength{\itemsep}{0.2pt}
        \setlength{\parskip}{0pt}
        \setlength{\parsep}{0pt}
	\item $ f_{R_f}: \mathbb{L}^{n_f}\to \mathbb{L}$ is defined by the assignment $f_{R_f}(\overline{c}) = \left(\left(R_f^{(0)}[\overline{\val{c}}^{\epsilon_f}]\right)^{\downarrow}, R_f^{(0)}[\overline{\val{c}}^{\epsilon_f}]\right)$;
	
	\item $g_{R_g}: \mathbb{L}^{n_g}\to \mathbb{L}$ is defined by the assignment  $g_{R_g}(\overline{c}) = \left(R_g^{(0)}[\overline{\descr{c}}^{\epsilon_g}], \left(R_g^{(0)}[\overline{\descr{c}}^{\epsilon_g}]\right)^{\uparrow}\right)$. 
	
\end{enumerate}
\noindent
Here, for every $\overline{c}\in \mathbb{L}^{n_f}$, the tuple $\overline{\val{c}}^{\epsilon_f}$ is such that for
each $1 \leq i \leq n_f$,  the $i$-th coordinate of $\overline{\val{c}}^{\epsilon_f}$ is $\val{c_i}$ if $\epsilon_f(i) = 1$ and is $\descr{c_i}$ if $\epsilon_f(i) = \partial$, and $\overline{\descr{c}}^{\epsilon_g}$ is such that for
each $1 \leq i \leq n_g$  the $i$-th coordinate of $\overline{\descr{c}}^{\epsilon_g}$ is $\descr{c_i}$ if $\epsilon_g(i) = 1$, and is  $\val{c_i}$ if $\epsilon_g(i) = \partial$. 
Thus, for all $\Diamond, {\lhd}\in \mathcal{F}$ and all $\Box, {\rhd}\in \mathcal{G}$,
\begin{enumerate}
	\setlength{\itemsep}{0.2pt}
        \setlength{\parskip}{0pt}
        \setlength{\parsep}{0pt}
	\item $ \langle R_\Diamond\rangle, \langle R_\lhd]: \mathbb{L}\to \mathbb{L}$ are respectively defined by the assignments 
	\[
	c\mapsto \left(\left(R_\Diamond^{(0)}[\val{c}]\right)^{\downarrow}, R_\Diamond^{(0)}[\val{c}]\right) \quad\text{ and }\quad c\mapsto \left(\left(R_\lhd^{(0)}[\descr{c}]\right)^{\downarrow}, R_\lhd^{(0)}[\descr{c}]\right);\]
	
	\item $[R_\Box], [R_\rhd\rangle : \mathbb{L}\to \mathbb{L}$ are respectively defined by the assignments 
  \[ c \mapsto \left(R_\Box^{(0)}[\descr{c}], \left(R_\Box^{(0)}[\descr{c}]\right)^{\uparrow}\right) \quad\text{ and }\quad c\mapsto \left(R_\rhd^{(0)}[\val{c}], \left(R_\lhd^{(0)}[\val{c}]\right)^{\uparrow} \right).\]

\end{enumerate}
 The next proposition is the many-valued counterpart of \cite[Proposition 21]{greco2018algebraic} (reported in Proposition \ref{prop:F plus is complete LE}).
	
\begin{proposition}\label{prop:F plus is complete LE-mv}
	If $\mathbb{F} = (\mathbb{P}, \mathcal{R}_\mathcal{F}, \mathcal{R}_\mathcal{G})$ is an $\mathbf{A}$-polarity-based $\mathcal{L}_{\text{LE}}$-frame, then $\mathbb{F}^+ = (\mathbb{L}, \{f_{R_f}\mid f\in \mathcal{F}\}, \{g_{R_g}\mid g\in \mathcal{G}\})$ is a complete normal $\mathcal{L}_{\text{LE}}$-algebra.
\end{proposition}	
\begin{proof}
We only discuss $f_{R_f}$, since the case for $g_{R_g}$ is similar. What we need to show is that $f_{R_f}(\overline{c})$ is well defined, i.e. $R^{(0)}_f[\overline{\val{c}}^{\epsilon_f}]$ is a Galois-stable set, and that $f_{R_f}$ is residuated in each coordinate. Notice preliminarily that $\val{c_i}^{\epsilon_f(i)}=\bigvee_{a\in A^{\epsilon_f(i)}}\{\val{c_i}^{\epsilon_f(i)}(a)\slash a\}$. By Lemma \ref{lemma: almost basic mv}, the set of maps $\{R^{(i)}_f[-]\mid 0\leq i \leq n_f\}$ are residuated to each other in the algebras $\mathbb{A}^A$ and $\mathbb{A}^X$, so $$R^{(0)}_f[\overline{\val{c}}^{\epsilon_f}]=R^{(0)}_f[\overline{\bigvee_{a}\{\val{c}(a)\slash a\}}^{\epsilon_f}]=\bigwedge_{\overline{a}\in \overline{A}^{\epsilon_f}}R_f^{(0)}[\overline{\{\val{c}(a)\slash{a}\}}].$$ The conjuncts on the right hand are Galois-stable by the assumption of $I$-compatibility, hence the conjunction is Galois-stable as the conjunction of Galois stable sets. Analogously, for each $i$, the image of $R^{(i)}_f[-]$ contains only Galois-stable sets. Hence, the $i$-th residual of $f_{R_f}$ is $R^{(i)}_f$ as witnessed again by Lemma \ref{lemma: almost basic mv}. This concludes the proof.   
\end{proof}

\paragraph{Many-valued polarity-based models.}
For any LE-language $\mathcal{L}$ over a  set $\mathsf{Prop}$ of atomic propositions,  an  $\mathbf{A}$-{\em polarity-based} $\mathcal{L}$-{\em model}  is a tuple $\mathbb{M} = (\mathbb{F}, V)$ such that $\mathbb{F} $ is an $\mathbf{A}$-polarity-based $\mathcal{L}$-frame as above, and $V: \mathsf{Prop}\to \mathbb{F}^+$. For every $p\in \mathsf{Prop}$, let $V(p): = (\val{p}, \descr{p})$, where $\val{p}: A\to \mathbf{A}$ and $\descr{p}: X\to\mathbf{A}$, and $\val{p}^\uparrow = \descr{p}$ and $\descr{p}^\downarrow = \val{p}$.
	Every $V$ as above has a unique homomorphic extension, also denoted $V: \mathcal{L} \to \mathbb{F}^+$, defined as follows:
	\smallskip
	
	{{\centering
		\begin{tabular}{rcl c rcl}
			$V(p)$ & = & $(\val{p}, \descr{p})$\\
			$V(\top)$ & = & $(\top^{\mathbf{A}^A}, (\top^{\mathbf{A}^A})^\uparrow)$ &\quad&
			$V(\bot)$ & = & $((\top^{\mathbf{A}^X})^\downarrow, \top^{\mathbf{A}^X})$\\
			$V(\phi\wedge \psi)$ & = & $(\val{\phi}\wedge\val{\psi}, (\val{\phi}\wedge\val{\psi})^\uparrow)$ &&
			$V(\phi\vee \psi)$ & = & $((\descr{\phi}\wedge\descr{\psi})^\downarrow, \descr{\phi}\wedge\descr{\psi})$\\
			$V(f(\overline{\phi}))$ & $ = $ & $\left(\left(R_f^{(0)}[\overline{\val{\phi}}^{\epsilon_f}]\right)^{\downarrow}, R_f^{(0)}[\overline{\val{\phi}}^{\epsilon_f}]\right)$ &\quad&
$V(g(\overline{\phi}))$ & $ = $ & $\left(R_g^{(0)}[\overline{\descr{\phi}}^{\epsilon_g}], \left(R_g^{(0)}[\overline{\descr{\phi}}^{\epsilon_g}]\right)^{\uparrow}\right)$
		\end{tabular}
	\par}}
	\smallskip
	
\noindent Here, for every $\overline{\phi}\in \mathcal{L}^{n_f}$ (resp.~$\overline{\phi}\in \mathcal{L}^{n_g}$), the tuple $\overline{\val{\phi}}^{\epsilon_f}$ is such that for
each $1 \leq i \leq n_f$,  the $i$-th coordinate of $\overline{\val{\phi}}^{\epsilon_f}$ is $\val{\phi_i}$ if $\epsilon_f(i) = 1$, and is $\descr{\phi_i}$ if $\epsilon_f(i) = \partial$, and $\overline{\descr{\phi}}^{\epsilon_g}$ is such that for
each $1 \leq i \leq n_g$  the $i$-th coordinate of $\overline{\descr{\phi}}^{\epsilon_g}$ is $\descr{\phi_i}$ if $\epsilon_g(i) = 1$, and is  $\val{\phi_i}$ if $\epsilon_g(i) = \partial$. Thus, for all $\Diamond, {\lhd}\in \mathcal{F}$ and all $\Box, {\rhd}\in \mathcal{G}$,
\smallskip

{{\centering
\begin{tabular}{rcl c rcl}
$V(\Box\phi)$ & $ = $ & $(R_{\Box}^{(0)}[\descr{\phi}], (R_{\Box}^{(0)}[\descr{\phi}])^{\uparrow})$ & \quad\quad &
$V(\Diamond\phi)$ & $ = $ & $((R_{\Diamond}^{(0)}[\val{\phi}])^\downarrow, R_{\Diamond}^{(0)}[\val{\phi}])$\\
$V({\rhd}\phi)$ & $ = $ & ($R_\rhd^{(0)}[\val{\phi}], (R_\rhd^{(0)}[\val{\phi}])^\uparrow)$ &&
$V({\lhd}\phi)$ & $ = $ & $((R_\lhd^{(0)}[\descr{\phi}])^\downarrow, R_\lhd^{(0)}[\descr{\phi}])$.\\
\end{tabular}
\par}}

 In its turn, the interpretation $V$ induces  $\alpha$-{\em membership relations} for each $\alpha\in \mathbf{A}$ (in symbols: $\mathbb{M}, a\Vdash^\alpha \phi$), and $\alpha$-{\em description relations} for each $\alpha\in \mathbf{A}$ (in symbols: $\mathbb{M}, x\succ^\alpha \phi$) such that for every $\phi\in \mathcal{L}$,
	\[\mathbb{M}, a\Vdash^\alpha \phi \quad \mbox{ iff }\quad \alpha\leq \val{\phi}(a), \quad\quad\quad \mathbb{M}, x\succ^\alpha \phi \quad \mbox{ iff }\quad \alpha\leq \descr{\phi}(x).\]
	This can be equivalently expressed by means of the following  recursive definition:
	\smallskip 
	
	{{\centering
		\begin{tabular}{r c l}
			$\mathbb{M}, a\Vdash^\alpha p$ & iff & $\alpha\leq \val{p}(a)$;\\
			$\mathbb{M}, a\Vdash^\alpha \top$ & iff & $\alpha\leq (\top^{\mathbf{A}^A})(a)$ i.e.~always;\\
			$\mathbb{M}, a\Vdash^\alpha \bot$ & iff & $\alpha\leq (\top^{\mathbf{A}^X})^\downarrow (a) = \bigwedge_{x\in X}(\top^{\mathbf{A}^X}(x)\to I(a, x)) =  \bigwedge_{x\in X} I(a, x)$;\\
			$\mathbb{M}, a\Vdash^\alpha \phi\wedge \psi$ & iff & $\mathbb{M}, a\Vdash^\alpha \phi$ and $\mathbb{M}, a\Vdash^\alpha \psi$;\\
			$\mathbb{M}, a\Vdash^\alpha \phi\vee \psi$ & iff & $\alpha\leq (\descr{\phi}\wedge\descr{\psi})^\downarrow(a) = \bigwedge_{x\in X}(\descr{\phi}(x)\wedge\descr{\psi}(x)\to I(a, x))$;\\
			$\mathbb{M}, a\Vdash^\alpha g(\overline{\phi})$ & iff & $\alpha\leq (R^{(0)}_g[\overline{\descr{\phi}}^{\epsilon_g}])(a) = \bigwedge_{\overline{x}\in X^{\epsilon_g}}(\overline{\descr{\phi}}^{\epsilon_g}(\overline{x})\to R_g(a, \overline{x}))$;\\
			$\mathbb{M}, a\Vdash^\alpha f(\overline{ \phi})$ & iff & $\alpha\leq ((R^{(0)}_f[\overline{\val{\phi}}^{\epsilon_f}])^\downarrow)(a) = \bigwedge_{x\in X}((R^{(0)}_f[\overline{\val{\phi}}^{\epsilon_f}])(x)\to I(a, x))$\\[2.5mm]
			$\mathbb{M}, x\succ^\alpha p$ & iff & $\alpha\leq \descr{p}(x)$;\\
			$\mathbb{M}, x\succ^\alpha \bot$ & iff & $\alpha\leq (\top^{\mathbf{A}^X})(x)$ i.e.~always;\\
			$\mathbb{M}, x\succ^\alpha \top$ & iff & $\alpha\leq (\top^{\mathbf{A}^A})^\uparrow (x) = \bigwedge_{a\in A}(\top^{\mathbf{A}^A}(a)\to I(a, x)) =  \bigwedge_{a\in A} I(a, x)$;\\
			$\mathbb{M}, x\succ^\alpha \phi\vee \psi$ & iff & $\mathbb{M}, x\succ^\alpha \phi$ and $\mathbb{M}, x\succ^\alpha \psi$;\\
			$\mathbb{M}, x\succ^\alpha \phi\wedge \psi$ & iff & $\alpha\leq (\val{\phi}\wedge\val{\psi})^\uparrow(x) = \bigwedge_{a\in A}(\val{\phi}(a)\wedge\val{\psi}(a)\to I(a, x))$;\\
			$\mathbb{M}, x\succ^\alpha f(\overline{ \phi})$ & iff & $\alpha\leq (R^{(0)}_f[\overline{\val{\phi}}^{\epsilon_f}])(x) = \bigwedge_{\overline{a}\in A^{\epsilon_f}}(\overline{\val{\phi}}^{\epsilon_f}(\overline{a})\to R_f(x, \overline{a}))$;\\
			$\mathbb{M}, x\succ^\alpha g(\overline{ \phi})$ & iff & $\alpha\leq ((R^{(0)}_g[\overline{\descr{\phi}}^{\epsilon_g}])^\uparrow)(x) = \bigwedge_{a\in A}((R^{(0)}_g[\overline{\descr{\phi}}^{\epsilon_g}])(a)\to I(a, x))$.\\
		\end{tabular}
	\par}} 
	\smallskip
	
	\noindent Specializing the last four clauses above to the interpretation of $f$-formulas and $g$-formulas when $n_f = n_g = 1$:
	
{{\centering
		\begin{tabular}{r c l}
$\mathbb{M}, a\Vdash^\alpha \Box \phi$ & iff & $\alpha\leq (R^{(0)}_\Box[\descr{\phi}])(a) = \bigwedge_{x\in X}(\descr{\phi}(x)\to R_\Box(a, x))$;\\
$\mathbb{M}, a\Vdash^\alpha {\rhd} \phi$ & iff & $\alpha\leq (R^{(0)}_\rhd[\val{\phi}])(a) = \bigwedge_{b\in A}(\val{\phi}(b)\to R_\rhd(a, b))$;\\
			$\mathbb{M}, a\Vdash^\alpha \Diamond \phi$ & iff & $\alpha\leq ((R^{(0)}_\Diamond[\val{\phi}])^\downarrow)(a) = \bigwedge_{x\in X}((R^{(0)}_\Diamond[\val{\phi}])(x)\to I(a, x))$\\
			$\mathbb{M}, a\Vdash^\alpha {\lhd} \phi$ & iff & $\alpha\leq ((R^{(0)}_\lhd[\descr{\phi}])^\downarrow)(a) = \bigwedge_{x\in X}((R^{(0)}_\lhd[\descr{\phi}])(x)\to I(a, x))$;\\
			$\mathbb{M}, x\succ^\alpha \Diamond \phi$ & iff & $\alpha\leq (R^{(0)}_\Diamond[\val{\phi}])(x) = \bigwedge_{a\in A}(\val{\phi}(a)\to R_\Diamond(x, a))$;\\
			$\mathbb{M}, x\succ^\alpha {\lhd} \phi$ & iff & $\alpha\leq (R^{(0)}_\lhd[\descr{\phi}])(x) = \bigwedge_{y\in X}(\descr{\phi}(y)\to R_\lhd(x, y))$;\\

			$\mathbb{M}, x\succ^\alpha \Box \phi$ & iff & $\alpha\leq ((R^{(0)}_\Box[\descr{\phi}])^\uparrow)(x) = \bigwedge_{a\in A}((R^{(0)}_\Box[\descr{\phi}])(a)\to I(a, x))$;\\
			$\mathbb{M}, x\succ^\alpha {\rhd} \phi$ & iff & $\alpha\leq ((R^{(0)}_\rhd[\val{\phi}])^\uparrow)(x) = \bigwedge_{a\in A}((R^{(0)}_\rhd[\val{\phi}])(a)\to I(a, x))$.\\

		\end{tabular}
	\par}}
	\smallskip
	
	\noindent A sequent $\phi \vdash \psi$ is {\em satisfied} in $\mathbb{M} = (\mathbb{F}, V)$  (notation:  $\mathbb{M} \models \phi \vdash \psi$), if $\val{\phi} \subseteq \val{\psi}$, or equivalently, $\descr{\psi} \subseteq \descr{\phi}$.  
	A sequent $\phi \vdash \psi$ is {\em valid} in an  $\mathbf{A}$-polarity-based frame $\mathbb{F}$ (notation:  $\mathbb{F} \models \phi \vdash \psi$) if   $\phi \vdash \psi$ is true in every model $\mathbb{M}$ based on $\mathbb{F}$.

\section{Composing crisp relations}
\label{sec: composing crisp}

In this section, we lay the ground for an environment in which we can meaningfully compare the first-order correspondents of inductive MRPs across different relational semantics. This will entail representing these first-order correspondents as inclusions of relations associated with complex terms. Since, algebraically, complex terms are interpreted as compositions of operations, this algebraic interpretation should be matched on the relational side by suitable compositions of the relations dually corresponding to those operations. In Section \ref{ssec:lifting properties}, we partly recall, partly introduce several notions of relational compositions, in the setting of polarity-based frames, with different properties, see Remark \ref{remark: i-composition vs pseudo composition}.  
In Section \ref{sec:pseudo:Kripke}, we introduce a non-standard notion of relational composition on Kripke frames which will match the notions of relational composition on polarity-based frames referred to as non $I$-mediated compositions (cf.\ Definition \ref{def:relational composition}.3--8). In Section \ref{ssec:prel and krel}, we introduce algebraic languages of relations on polarity-based frames and Kripke frames, which are interpreted by means of the compositions defined in the previous two sections.

\subsection{Composing relations on polarities}
\label{ssec:lifting properties}

Let us recall that the usual {\em composition} of $R\subseteq S_1\times S_2$ and $T\subseteq S_2\times S_3$ is  $R\circ T: = \{(u, w)\in S_1\times S_3\mid  (u, v)\in R \mbox{ and } (v, w)\in T\mbox{ for some } v\in S_2\}$. Also, for any $R\subseteq S_1\times S_2$, the {\em converse} relation $R^{-1}\subseteq S_2\times S_1$ is defined as $R^{-1}: = \{(w, u)\in S_2\times S_1\mid  (u, w)\in R\}$.  Well known properties of relational composition and converse are $\Delta$ being the left- and right-unit of composition, composition being associative, converse being idempotent, and the following distribution law of converse over composition: $(R\circ T)^{-1} = T^{-1}\circ R^{-1}$.

The following definition expands those given in \cite[Section 3.4]{conradie2021rough}. As will be clear from Lemma \ref{lemma:lifting composition}, the $I$-composition can be understood as the `typed counterpart' of the usual composition of relations in the setting of formal contexts. 

\begin{notation}
\label{notation:crisp_type_relation}
In what follows, for any formal context $\mathbb{P}= (A,X,I)$, relations of type $A \times X$ will be denoted by the capital letter $B$, possibly with sub- or superscripts; 
 and similarly relations of type $X \times A$ will be denoted by $D$, of type $X \times X$ by $L$, and of type $A \times A$ by $M$. Summing up,
 \smallskip
 
 {{\centering $B\subseteq A \times X\quad$ $\quad D \subseteq X \times A\quad$ $\quad L \subseteq X \times X\quad$ $\quad M \subseteq A \times A$,
\par}}
\end{notation}

\begin{definition}
\label{def:relational composition}
For any formal context $\mathbb{P}= (A,X,I)$,
we define the following notions of relational composition:  for all $a,b \in A$ and $x,y \in X$, 
    \[
    \begin{array}{r c c c c cl}
		B_1\, ;_I B_2 \subseteq A \times X && (B_1\, ;_I B_2)^{(0)} [x] = B_1^{(0)}[I^{(0)}[B_2^{(0)} [x]]], & \text{ i.e. }  & a (B_1\, ;_IB_2) x &  \text{ iff }  & a \in B_1^{(0)}[I^{(0)}[B_2^{(0)} [x]]] \\
		D_1\, ;_I D_2 \subseteq X \times A && (D_1\, ;_I D_2)^{(0)} [a] = D_1^{(0)}[I^{(1)}[D_2^{(0)} [a]]], &  \text{ i.e. }  & x (D_1\, ;_I D_2) a &  \text{ iff }  & x \in D_1^{(0)}[I^{(1)}[D_2^{(0)} [a]]] \\
		B\, ;D \subseteq A \times A && (B\, ;D)^{(0)} [b] = B^{(0)}[D^{(0)} [b]], &  \text{ i.e. }  & a (B\, ;D) b &  \text{ iff }  & a \in B^{(0)}[D^{(0)} [b]] \\
		D\, ;B \subseteq X \times X && (D\, ;B)^{(0)} [y] = D^{(0)}[B^{(0)} [y]], &  \text{ i.e. }  & x (D\, ;B) y &  \text{ iff }  & x \in D^{(0)}[B^{(0)} [y]] \\
		M\, ;B \subseteq A \times X && (M\, ;B)^{(0)} [x] = M^{(0)}[B^{(0)} [x]], &  \text{ i.e. }  & a (M\, ; B) x &  \text{ iff }  & a \in M^{(0)}[B^{(0)} [x]] \\
		D\, ;M \subseteq X \times A && (D\, ;M)^{(0)} [a] = D^{(0)}[M^{(0)} [a]], &  \text{ i.e. }  & x (D\, ;M) a &  \text{ iff }  & x \in D^{(0)}[M^{(0)} [a]] \\
		L\, ;D \subseteq X \times A && (L\, ;D)^{(0)} [a] = L^{(0)}[D^{(0)} [a]], &  \text{ i.e. }  & x (L\, ;D) a &  \text{ iff }  & x \in L^{(0)}[D^{(0)} [a]] \\
		B\, ;L \subseteq A \times X && (B\, ;L)^{(0)} [x] = B^{(0)}[L^{(0)} [x]], &  \text{ i.e. }  & a (B\, ;L) x &  \text{ iff }  & a \in B^{(0)}[L^{(0)} [x]]. 
	\end{array}
	\]
\end{definition}

\begin{remark}\label{remark: i-composition vs pseudo composition}
The notions of relational compositions collected in Definition \ref{def:relational composition} do not exhaust the combinatorial space of possible compositions of relations on polarities, but are those relevant to the result about the modal reduction principles that we aim at establishing. While the notions defined in items 1 and 2 of Definition \ref{def:relational composition} are {\em mediated} by $I$ (and this is the reason why we refer to these as $I$-{\em compositions}), the remaining ones are  {\em not mediated} by $I$.  
More notions of $I$-composition can be defined e.g.~for composing $R\subseteq X\times X$ and $T\subseteq A\times A$, by declaring $x (R\, ;_I T)a$ iff $x\in R^{(0)}[I^{(1)}[T^{(0)}[a]]]$, or for composing $R\subseteq X\times A$ and $T\subseteq X\times X$, by declaring $x (R\, ;_I T)y$ iff $x\in R^{(0)}[I^{(0)}[T^{(0)}[y]]]$, and so on. Notice that  $I$-compositions can be defined exactly between relations  of a type that interprets modal operators the composition of which gives rise to a {\em normal} modal operator. In items 1 and 2 of the Definition above, $I$-compositions are defined between relations interpreting normal box operators, or interpreting normal diamond operators, and the operation defined as the composition of normal box (resp.~diamond) operators is a normal box (resp.~diamond) operator. Likewise,  $I$-compositions can be defined between a relation of type $A\times A$ (supporting the interpretation of a unary negative connective ${\rhd}\in \mathcal{G}$) and a relation of type $X\times X$ (supporting the interpretation of a unary negative connective ${\lhd}\in \mathcal{F}$), and the operation defined as the composition of a normal left triangle and a normal  right triangle  is either a normal box or a normal diamond operator. In contrast, compositions not mediated by $I$ can be defined between a relation of type $A\times X$ and one of type $X\times A$, and composing a box and a diamond operator gives rise to a non normal map. 
Since these different notions of composition define binary (fusion-type) operations spanning over relation algebras of different types, they can be framed algebraically in the context of {\em heterogeneous algebras} \cite{birkhoff1970heterogeneous}. 
In what follows, for any polarity $\mathbb{P}= (A,X,I)$,   we will refer generically to relations  $R\subseteq X\times A$, $R\subseteq A\times X$, $R\subseteq X\times X$, $R\subseteq A\times A$ as ``relations on $\mathbb{P}$'' or ``polarity-based relations''. While, as just discussed, each type of polarity-based relation can be composed (in an $I$-mediated or non $I$-mediated way) with  polarity-based relations of any  type, in the remainder of the paper we will only consider the compositions listed in Definition \ref{def:relational composition},  and hence we will define the heterogeneous relation algebras associated with polarity-based $\mathcal{L}$-frames (cf.~Definition \ref{def:rel algebras crisp}) accordingly.  
\end{remark}
Under the assumption of $I$-compatibility, equivalent, alternative formulations of the $I$-composition of relations are available, which are stated in the following lemma. 

\begin{lemma}[cf.~\cite{conradie2021rough}, Lemma 3.11]
\label{lem:equiv:I:coposition}
	If $\mathbb{P}= (A,X,I)$ is a polarity, then for all $I$-compatible relations $B_1, B_2, D_1, D_2$, and all $a \in A$ and $x\in X$,
	
	{{\centering
        \begin{tabular}{cc}
        $x (D_1\, ;_I D_2) a \quad \text{ iff }  \quad a \in D_2^{(1)}[I^{(1)}[D_1^{(1)} [x]]]$ \quad\quad\quad & 
        $a (B_1\, ;_I B_2) x  \quad \text{ iff }  \quad x \in B_2^{(1)}[I^{(0)}[B_1^{(1)} [a]]].$
        \end{tabular} \par}}
\end{lemma}
\begin{proof}
Let us show the second equivalence.

{{\centering
    \begin{tabular}{r c l l}
     $a(R;_IT)x$    & iff  
         & $\{a\}\subseteq R^{(0)}[I^{(1)}[T^{(0)}[x]]]$ & Definition \ref{def:rel algebras crisp} (2) \\
         & iff  
         & $I^{(1)}[T^{(0)}[x]]\subseteq R^{(1)}[a]$ & $R^{(0)}[-]$ and $R^{(1)}[-]$ Galois-adjoint (cf.~\cite[Lemma 2.2.2]{conradie2021rough}) \\
         & iff  
         & $I^{(0)}[R^{(1)}[a]]\subseteq I^{(0)}[I^{(1)}[T^{(0)}[x]]]$ & antitonicity of $I^{(0)}[-]$ (cf.~\cite[Lemma 2.2.1]{conradie2021rough}) \\
         & iff  
         & $I^{(0)}[R^{(1)}[a]]\subseteq T^{(0)}[x]$ & $T$ is $I$-compatible  \\
         & iff  
         & $\{x\}\subseteq  T^{(1)}[I^{(0)}[R^{(1)}[a]]]$ & $T^{(0)}[-]$ and $T^{(1)}[-]$ Galois-adjoint (cf.~\cite[Lemma 2.2.2]{conradie2021rough}) \\
    \end{tabular} \par}}
\end{proof}
\begin{lemma}
\label{lem: unit of I composition}
If $\mathbb{P}= (A,X,I)$ is a polarity, then, letting $J\subseteq X\times A$ be defined by $xJa$ iff $aIx$, for every $I$-compatible relations $B$ and $D$,

{{\centering\begin{tabular}{rl c rl}
1. & $D\,;_I I = D = I\,;_I D$ & \quad\quad & 2. & $B\,;_I J = B = J\,;_I B$
\end{tabular}\par}}

\end{lemma}
\begin{proof} We only show item 1, the proof of the second item being similar.
By definition, $a (R\,;_I I ) x$ iff $a\in R^{(0)}[I^{(1)}[I^{(0)}[x]]] = R^{(0)}[x^{\downarrow\uparrow}] = R^{(0)}[x]$, the last identity holding by  \cite[Lemma 2.6]{conradie2021rough}, because $R$ is $I$-compatible. This shows that $R\,;_I I = R$. The second identity is shown similarly, using the Galois-stability of $R^{(0)}[x]$.
\end{proof}

\begin{proposition}\label{lem:composition:1}
If $\mathbb{P}= (A,X,I)$ is a polarity and $R,T\subseteq A\times X$ (resp. $R,T\subseteq X\times A$),  are  $I$-compatible, then so is $R\, ;_IT$.
\end{proposition}
\begin{proof}
Let $x\in X$ and $a\in A$. By definition, $(R\,;_IT)^{(0)}[x]=R^{(0)}[I^{(1)}[T^{(0)}[x]]]$, which  is Galois-stable by \cite[Lemma 2.6.1]{conradie2021rough} and the $I$-compatibility of $R$. By Lemma \ref{lem:equiv:I:coposition}, $(R\,;_IT)^{(1)}[a]=T^{(1)}[I^{(0)}[R^{1}[a]]]$, which is Galois-stable by \cite[Lemma 2.6.2]{conradie2021rough} and the $I$-compatibility of $T$. This concludes the proof.
\end{proof}

\begin{remark}
Even if $R$ and $T$ are $I$-compatible, $R\, ;T$ (when definable) is not in general $I$-compatible. For instance, if $R\subseteq A\times X$ and $T\subseteq X\times A$, while by definition
$(R;T)^{(0)}[b] = R^{(0)}[T^{(0)}[b]]$ is Galois-stable for every $b\in A$,  the following set does not need to be Galois-stable for all $a\in A$.
\smallskip

{{\centering
\begin{tabular}{r c l}
$(R;T)^{(1)}[a] =\{d\in A\mid a(R;T)d\}$& $ =$ & $\{d\in A\mid a\in  R^{(0)}[T^{(0)}[d]]\}$\\
& $ =$ & $\{d\in A\mid \forall x(x\in T^{(0)}[d]\Rightarrow a\in  R^{(0)}[x])\}$\\ & $ =$ & $ \{d\in A\mid \forall x(x\in T^{(0)}[d]\Rightarrow x\in  R^{(1)}[a])\}$\\ & $ =$ & $ \{d\in A\mid T^{(0)}[d]\subseteq  R^{(1)}[a]\}.$\\
\end{tabular}
\par}}  
\smallskip

For instance, let $\mathbb{P} = (A, X, I)$ such that  $A: = \{a, b, c\}$, $X = \{x, y, z\}$  $I = \{ (a, x), (b, y), (c, z)\}$, and let $R = \{(b, y)\}$ and $T = \{(z, c)\}$. It is easy to see that both $R$ and $T$ are $I$-compatible; however, $(R;T)^{(1)}[a] =\{d\in A\mid T^{(0)}[d]\subseteq R^{(1)}[a]\} = \{a, b\}$ is not Galois-stable.
\end{remark}

For any polarity $\mathbb{P}= (A,X,I)$ and any $R_{\Box}\subseteq A\times X$ and $R_{\Diamond}\subseteq X\times A$, let $R_{\Diamondblack}\subseteq X\times A$ and $R_{\blacksquare}\subseteq A\times X$ be respectively defined as $x R_{\Diamondblack} a$ iff $a R_{\Box} x$ and  $a R_{\blacksquare} x$ iff $x R_{\Diamond}a$. The following lemma shows  distribution laws of converse and $I$-compositions which are analogous to those holding for standard relational composition.  
\begin{lemma}
\label{lem: distribution law composition inverse}
For any polarity $\mathbb{P}= (A,X,I)$ and all $I$-compatible relations $R_{\Box_1}, R_{\Box_2}\subseteq A\times X$ and  $R_{\Diamond_1}, R_{\Diamond_2}\subseteq X\times A$, for all $a\in A$ and $x\in X$
\smallskip

{{\centering
\begin{tabular}{rcl c rcl}
$ (R_{\Box_1}\, ;_I R_{\Box_2})x$ & iff & $x(R_{\Diamondblack_2}\, ;_I R_{\Diamondblack_1})a$
& \quad\quad\quad &
$x(R_{\Diamond_1}\, ;_I R_{\Diamond_2})a$ & iff & $a( R_{\blacksquare_2}\, ;_I R_{\blacksquare_1}) x$.
\end{tabular}
\par}}
\end{lemma}
\begin{proof} We only show item 1.
It immediately follows from the definition of $R_{\Diamondblack}$ and $R_{\blacksquare}$ that $R_{\Box}^{(1)}[B] = R_{\Diamondblack}^{(0)}[B]$ and $R_{\Diamond}^{(1)}[Y] = R_{\blacksquare}^{(0)}[Y]$ for any $B\subseteq A$ and any $Y\subseteq X$. 
Hence, by applying Lemma \ref{lem:equiv:I:coposition}, 
\smallskip

{{\centering 
\begin{tabular}{cc cc cc c}
$a(R_{\Box_1}\, ;_I R_{\Box_2})x$ & 
iff & 
$x\in R^{(1)}_{\Box_2}[I^{(0)}[R^{(1)}_{\Box_1}[a]]$ &
iff & 
$x\in R^{(0)}_{\Diamondblack_2}[I^{(0)}[R^{(0)}_{\Diamondblack_1}[a]]$ &
iff &
$x (R_{\Diamondblack_2}\, ;_I R_{\Diamondblack_1}) a$.
\end{tabular}
\par}}
\end{proof}
However, similar distribution laws  do not hold for composition of relations of different types, such as those defined in items 3 and 4 of Definition \ref{def:relational composition}. If this property were true, we would have, for all $a, b\in A$,
\[ a(R_{\Box}\, ; R_{\Diamond})b  \; \text{ iff }\; b (R_{\blacksquare}\, ; R_{\Diamondblack}) a.\]
That is, we would have, for all $a, b\in A$,
\[ a\in R_{\Box}^{(0)}[R_{\Diamond}^{(0)}[b]] \; \text{ iff }\;  b \in R_{\blacksquare}^{(0)}[R_{\Diamondblack}^{(0)}[ a]] \; \text{ iff }\;  b \in R_{\Diamond}^{(1)}[R_{\Box}^{(1)}[ a]].\]
However, by the basic properties of Galois connections, $a\in R_{\Box}^{(0)}[R_{\Diamond}^{(0)}[b]] $ iff $R_{\Diamond}^{(0)}[b]\subseteq R_{\Box}^{(1)}[a]$, and $b \in R_{\Diamond}^{(1)}[R_{\Box}^{(1)}[ a]]$ iff $R_{\Box}^{(1)}[ a]\subseteq R_{\Diamond}^{(0)}[b]$. Clearly, none of these two inclusions implies the other.

\begin{lemma}[cf.~\cite{conradie2021rough}, Lemma 3.13] 
\label{lemma:comp4}
For any polarity $\mathbb{P} = (A, X, I)$, any $I$-compatible relations $B_1, B_2, D_1, D_2$, any $C \subseteq A$, and any $Y \subseteq X$,
\smallskip

{{\centering 
\begin{tabular}{rcl}
$(B_1;_IB_2)^{(0)}[Y]=B_1^{(0)}[I^{(1)}[B_2^{(0)}[Y]]]$ &  and & $(B_1;_IB_2)^{(1)}[C]=B_1^{(1)}[I^{(0)}[B_2^{(1)}[C]]]$. \\
$(D_1;_ID_2)^{(1)}[Y]=D_1^{(1)}[I^{(1)}[D_2^{(1)}[Y]]]$ & and & $(D_1;_I D_2)^{(0)}[C]=D_1^{(0)}[I^{(0)}[D_2^{(0)}[C]]]$
\end{tabular}
\par}}
\end{lemma}

\begin{lemma}[cf.~\cite{conradie2021rough}, Lemma 3.14] 
\label{lem:composition:3}
For any polarity $\mathbb{P} = (A, X, I)$, and any $I$-compatible relations $B_1,B_2,B_3$, $D_1,D_2,D_3$,
\smallskip

{{\centering 
$(D_1\, ;_ID_2)\, ;_ID_3 = D_1\, ;_I(D_2\, ;_I D_3)$ \quad\quad and \quad\quad  $(B_1\, ;_IB_2)\, ;_IB_3 = B_1\, ;_I(B_2\, ;_I B_3)$
\par}}
\end{lemma}

\begin{remark}
\label{rem: crisp polarity-based heterogeneous associativity does not hold}
Associativity among the various compositions does not hold in general. For instance, 
%
%
let $\mathbb{P}=(A, X, I)$ such that $A = \{a_1, a_2\}$, $X = \{x_1, x_2\}$ and $I = \{(a_1, x_2), (a_2, x_1) \}$.  Let $R, U \subseteq A \times X$ and $T \subseteq X \times A$ be such that $R = \varnothing$, $U = A \times X$, and $T = \{(x_1, a_2), (x_2, a_1)\}$, respectively. All three relations are $I$-compatible (cf.~discussion before Definition \ref{def:complex algebra of LE-frame}). For all $a\in A$ and $x\in X$,
\smallskip

{{\centering
    \begin{tabular}{r c ll}
$ a (R;(T;U)) x $& iff & $a \in R^{(0)}[(T;U)^{(0)}[x]]$ & Definition \ref{def:relational composition} (8)\\
& iff &  $a\in R^{(0)}[T^{(0)}[U^{(0)}[x]]]$ & Definition \ref{def:relational composition} (4) \\
& iff &  $a\in R^{(0)}[T^{(0)}[A]] $ & $U = A \times X$\\
& iff &  $a\in R^{(0)}[\varnothing] = \{b\in A\mid \forall x(x\in\varnothing  \Rightarrow bR x)\} $ & $T = \{(x_1, a_2), (x_2, a_1)\}$  \\
& iff &  $a\in  A$,\\
\end{tabular}
\par}} 
\smallskip

\noindent which is always the case, while  for all $a\in A$ and $x\in X$,
\smallskip

{{\centering
    \begin{tabular}{r c ll}
$ a ((R;T);U) x$ & iff & $ a \in (R;T)^{(0)}[U^{(0)}[x]]$ & Definition \ref{def:relational composition} (5) \\
& iff & $a \in (R;T)^{(0)}[A]$ & $U = A \times X$\\ 
& iff & $ \forall b(b\in A\Rightarrow a (R;T) b)$\\
& iff & $\forall b(b\in A\Rightarrow a \in R^{(0)}[T^{(0)}[b]]),$\\
\end{tabular}
\par}}
\smallskip

\noindent which is never the case for any $a\in A$, given that if $b:=a_1$, then  $R^{(0)}[T^{(0)}[b]] = R^{(0)}[T^{(0)}[a_1]] = R^{(0)}[x_2] = \varnothing$.
\end{remark}
Even if, as shown above, generalized associativity among non $I$-mediated compositions  does not hold, in order to lighten notation,  in what follows we sometimes write e.g.~$R;\, T;\, U $ for $R;\, (T;\, U) $.

\begin{lemma}[cf.~\cite{conradie2021rough}, Lemma 3.15]
\label{lemma:lifting composition}
	For all $R, T\subseteq S\times S$,
	\[
	I_{(R\circ T)^c} = I_{R^c}\, ;_I I_{T^c}
	\quad\quad\quad J_{(R\circ T)^c} = J_{R^c}\, ;_I J_{T^c}.\]
\end{lemma}
\begin{proof}
The first equality is proved in \cite[Lemma 3.15]{conradie2021rough}; the proof of the second is similar, but we report it here for the sake of self-containedness. The notational conventions we use in the proof have been introduced at the beginning of Section \ref{ssec: lifting}. For any $a\in S$, 
\smallskip

{{\centering
\begin{tabular}{ccc}
\begin{tabular}{llll}
    $J_{(R \circ T)^c}^{(0)}[a]$ & = & $( \{x \in S \mid \forall b (x R b \Rightarrow b T^c a )\})_X$\\
	&=&$ (\{x \in S \mid \forall b( b T a \Rightarrow x R^c b ]\})_X$\\
	&=&$ (\{x \in S \mid \forall b(  b\in  T^{-1}[a] \Rightarrow x R^c b )\})_X$\\
	&=&$ \{x \in S_X \mid \forall b(  b\in  (T^{-1}[a])_A \Rightarrow x J_{R^c} b )\}$\\
	&=& $J_{R^c}^{(0)}[(T^{-1}[a])_A]$
\end{tabular}
& \quad &
\begin{tabular}{llll}
     & = & $J_{R^c}^{(0)}[(T^{-1}[a])_A]$\\
			&=& $J_{R^c}^{(0)}[((T^{-1}[a])^{cc})_A]$\\
			&=& $J_{R^c}^{(0)}[(J_{T^c}[a])^{\downarrow}]$\\
			&=& $J_{R^c}^{(0)}[ I_{\Delta^c}^{(0)}[J_{T^c}^{(0)}[a]]]$\\
			&=& $(J_{R^c}\, ;_I J_{T^c})^{(0)}[a]$.
\end{tabular}
\end{tabular}
\par}}
\end{proof}
\subsection{Pseudo-composition of relations on Kripke frames}\label{sec:pseudo:Kripke}
In addition to the usual relational composition, in what follows we will find it useful to introduce the notion of {\em pseudo-composition}, which is understood as the (untyped) counterpart of non $I$-mediated compositions of binary relations on polarities.
For any relation $T\subseteq U\times V$, and any $U'\subseteq U$  and $V'\subseteq V$, let
\begin{equation}\label{eq:def:square brackets}T^{[1]}[U']:=\{v\mid \forall u(u\in U'\Rightarrow uT^cv) \}  \quad\quad T^{[0]}[V']:=\{u\mid \forall v(v\in V'\Rightarrow uT^cv) \}.\end{equation}
\begin{lemma}
\label{lemma:properties of square bracket superscript}
For any relation $T \subseteq U \times W$, any $U'\subseteq U$  and $V\subseteq W$, and for every $u\in U$ and $v \in W$,
\smallskip

{{\centering
\begin{tabular}{rl c rl}
    1. & $T^{[1]}[U'] = (T^c)^{(1)}[U']$ and $T^{[0]}[V'] = (T^c)^{(0)}[V']$. & \quad\quad & 
    4. & $T^{[0]}[V] = (T^{-1}[V])^c = [T\rangle V$. \\
    2. & $T^{[0]}[v]= (T^{(0)}[v])^c$ and $T^{[1]}[u]= (T^{(1)}[u])^c$. &&
    5. & $(T^{[0]}[V^c])^c = T^{-1}[V^c] = \langle T]V$. \\
    3. & $T^{[0]}[V^c] = (T^{-1}[V^c])^c = [T]V$. &&
    6. & $(T^{[0]}[V])^c = T^{-1}[V] = \langle T\rangle V$.
\end{tabular}
\par}}
\end{lemma}
\begin{proof}
Item 1 immediately follows from the definition. As to item 2,
\begin{equation}
\label{eqn:square notation singletons}
\begin{array}{l}
T^{[0]}[v] = \{ u \in U \mid u T^c v \} = (\{ u \in U \mid u T v \})^c = (T^{(0)}[v])^c \\
T^{[1]}[u] = \{ v \in V \mid u T^c v \} = (\{ v \in V \mid u T v \})^c = (T^{(1)}[u])^c \\
\end{array}
\end{equation}
As to item 3,
\[T^{[0]}[V^c]  = (T^c)^{(0)}[V^c]  
     = \{ u \in U \mid \forall y(y \in V^c \Rightarrow u T^c y) \}  
      = \{ u \in U \mid \forall y( u T y \Rightarrow y \in V) \} 
     = (T^{-1}[V^c])^c = [T](V).\]
The remaining items are proved similarly to item 3.
\end{proof}

\begin{definition}
\label{def: pseudo comp on kripke}
For any $R,S \subseteq W \times W$,  the {\em pseudo composition} $(R \star S) \subseteq W \times W$ is defined as follows:
\[
y (R \star S) x \quad \text{iff} \quad y \in (R^{[0]}[S^{[0]}[x]])^c \quad\quad\quad\quad  \text{for all } x,y \in W.
\]
\end{definition}
\begin{lemma}
\label{lemma:characterization of pseudo}
For any $R,S\subseteq W \times W$ and every $x \in W$,
$(R\star S)^{[0]}[x]= R^{[0]}[S^{[0]}[x]]$.
\end{lemma}
\begin{proof}
    $(R \star S)^{[0]}[x]  = \{ y \in W \mid y (R \star S)^c x \}
     = \{ y \in W \mid y \in (R^{[0]}[S^{[0]}[x]])^c \} 
     = R^{[0]}[S^{[0]}[x]].$
\end{proof}

\begin{lemma}
\label{lemma: lifting}
For any Kripke $\mathcal{L}$-frame $\mathbb{F}$ based on  $W$, all $\Box , {\rhd} \in \mathcal{G}^\ast$ and $\Diamond, {\lhd} \in \mathcal{F}^\ast$, and all $B, Y\subseteq W$,
\smallskip

{{\centering
\begin{tabular}{rlcrl}
1. & $J_{R_{\Diamond}^c}^{(0)}[B]  = ((R_{\Diamond}^{-1}[B])^c)_X$. &\quad\quad &3. & $K_{R_{\lhd}^c}^{(0)}[Y]  = ((R_{\lhd}^{-1}[Y])^c)_X$. \\
2. & $I_{R_\Box^c}^{(0)}[Y] = ((R_{\Box}^{-1}[Y])^c)_A$. &&

4. & $H_{R_\rhd^c}^{(0)}[B] = ((R_{\rhd}^{-1}[B])^c)_A$.
\end{tabular}
\par}}
\end{lemma}
\begin{proof} We only prove item 2, the remaining items being proved similarly. For any $Y\subseteq W$,
\smallskip

{{\centering
\begin{tabular}{r c l}
$I_{R_\Box^c}^{(0)}[Y]$& $ = $& $\{a\in W_A\mid \forall y(y\in Y\Rightarrow a I_{R_\Box^c} y)\}$\\ 
& $ = $& $ (\{a\in W\mid \forall y(y\in Y\Rightarrow a R_\Box^c y)\})_A $\\
& $ = $& $ (\{a\in W\mid \forall y(a R_\Box y \Rightarrow y\notin Y)\})_A$\\
 & $ = $& $ ((R_{\Box}^{-1}[Y])^c)_A.$\\
 \end{tabular}
 \par}}
\end{proof}
The following corollary is an immediate consequence of the lemma above.
 \begin{corollary}
\label{cor: lifting of delta}
For any Kripke $\mathcal{L}$-frame $\mathbb{F}$ based on  $W$,  and all $B, Y\subseteq W$,
\smallskip

{{\centering
\begin{tabular}{rlcrl}
1. & $J_{\Delta^c}^{(0)}[B]  = ((\Delta^{-1}[B])^c)_X = (B^c)_X$. &\quad\quad& 3. & $K_{\Delta^c}^{(0)}[Y]  = ((\Delta^{-1}[Y])^c)_X = (Y^c)_X$. \\
2. & $I_{\Delta^c}^{(0)}[Y] = ((\Delta^{-1}[Y])^c)_A = (Y^c)_A$. && 4. & $H_{\Delta^c}^{(0)}[B] = ((\Delta^{-1}[B])^c)_A = (B^c)_A$.\\
\end{tabular}
\par}}
 \end{corollary}
 
\begin{lemma}
\label{Lemma: lifting heterogeneous relations commute}
        For any relations $R,S \subseteq W \times W$,
        \smallskip
        
    {{\centering
         \begin{tabular}{rl c rl c rl}
         1. & $I_{R^c} \, ; J_{S^c} = H_{(R \, \star S)^c}$ & \quad\quad & 3. & $K_{R^c} \, ; J_{S^c} = J_{(R \, \star S)^c}$ & \quad\quad & 5. & $J_{R^c} \, ; H_{S^c} = J_{(R \, \star S)^c}$ \\
         2. & $J_{R^c} \, ; I_{S^c} = K_{(R \, \star S)^c}$ &&
         4. & $H_{R^c} \, ; I_{S^c} = I_{(R \, \star S)^c}$ &&
         6. & $I_{R^c} \, ; K_{S^c} = I_{(R \, \star S)^c}$
    \end{tabular}
    \par}}
\end{lemma}
\begin{proof}
We only prove items 1 and 4, the proofs of the other items being similar.
\smallskip

{{\centering
\begin{tabular}{cc}
\begin{tabular}{rclr}
     && $(I_{R^c} \, ; J_{S^c})^{(0)}[a]$ & \\ & $=$ & $I_{R^c}^{(0)}[ J_{S^c}^{(0)}[a]]$ 
     &\hfill Definition \ref{def:relational composition} (3) \\ 
     & $=$ & $((R^c)^{(0)}[(S^c)^{(0)}[a]])_A$ 
     &\hfill Definition \ref{def:liftings relations} (1,2)\\
     & $=$ & $(R^{[0]}[S^{[0]}[a]])_A$ 
     &\hfill Lemma \ref{lemma:properties of square bracket superscript} (1)\\
     & $=$ & $((R \star S)^{[0]}[a])_A$ 
     &\hfill Lemma \ref{lemma:characterization of pseudo} \\
     & $=$ & $(((R \star S)^c)^{(0)}[a])_A$ 
     &\hfill Lemma \ref{lemma:properties of square bracket superscript} (1)\\
     & $=$ & $H_{(R \, \star S)^c}^{(0)}[a]$
     &\hfill Definition \ref{def:liftings relations} (3)
\end{tabular}
&

\begin{tabular}{rclr}
     && $(H_{R^c} \, ; I_{S^c})^{(0)}[x]$ & \\ & $=$ & $H_{R^c}^{(0)}[ I_{S^c}^{(0)}[x]]$ 
     & \hfill Definition \ref{def:relational composition} (5) \\ 
     & $=$ & $((R^c)^{(0)}[(S^c)^{(0)}[x]])_A$ 
     & \hfill Definition \ref{def:liftings relations} (1,3)\\
     & $=$ & $(R^{[0]}[S^{[0]}[x]])_A$ 
     & \hfill Lemma \ref{lemma:properties of square bracket superscript} (1)\\
     & $=$ & $((R \star S)^{[0]}[x])_A$ 
     & \hfill Lemma \ref{lemma:characterization of pseudo} \\
     & $=$ & $(((R \star S)^c)^{(0)}[x])_A$ 
     & \hfill Lemma \ref{lemma:properties of square bracket superscript} (1)\\
     & $=$ & $I_{(R \, \star S)^c}^{(0)}[x]$
     & \hfill Definition \ref{def:liftings relations} (1)
\end{tabular}
\end{tabular}
\par}}

\end{proof}

\begin{remark} 
\label{rem: pseudo-compositions not associative} As mentioned in the beginning of the present section, the pseudo-composition of binary relations on Kripke frames is intended as the (untyped) counterpart of (typed) non $I$-mediated compositions of relations on polarities. Since generalized associativity among non $I$-mediated compositions does not hold (cf.~Remark \ref{rem: crisp polarity-based heterogeneous associativity does not hold}), it is  expected that  pseudo-composition will not be associative either. We will exhibit a counterexample to 
$ (R_\Box \star R_\Diamond) \star R_\Box = R_\Box \star (R_\Diamond \star R_\Box)$ by showing that it `lifts' to the example discussed in Remark \ref{rem: crisp polarity-based heterogeneous associativity does not hold} for showing the failure of generalized associativity on polarities. Consider the Kripke frame $\mathbb{X} = (S, R, T, U)$ such that $S: = \{s_1, s_2\}$, and $R, T, U\subseteq S\times S$ are defined as follows: $R: = S\times S$, $U: = \varnothing$ and $T: = \Delta_S$. Then, clearly, $\mathbb{F}_{\mathbb{X}}: = (S_A, S_X, I_{\Delta^c}, I_{R^c}, J_{T^c}, I_{U^c})$ is such that $S_A$ and $S_X$ are set-isomorphic to $A$ and $X$ in the polarity discussed in Remark \ref{rem: crisp polarity-based heterogeneous associativity does not hold}, and likewise, each lifted relation is set-isomorphic to its corresponding relation in that example. Clearly, \[ (R \star T) \star U = R \star (T \star U) \quad \text{ iff } \quad  I_{((R \star T) \star U))^c} = I_{(R \star (T \star U))^c}.\] Hence, it is enough to show that the latter equality fails.
\smallskip

{{\centering
\begin{tabular}{ccc}
\begin{tabular}{rclr}
     && $I_{((R\star T)\star U)^c}$ &\\ & $=$ & $H_{(R\star T)^c} ; I_{U^c}$ & Lemma \ref{Lemma: lifting heterogeneous relations commute} (4) \\
     & $=$ & $(I_{R^c} ; J_{T^c}) ; I_{U^c}$ & Lemma \ref{Lemma: lifting heterogeneous relations commute} (1) \\
\end{tabular}
& \quad\quad & 
\begin{tabular}{rclr}
     && $I_{(R \star (T \star U))^c}$ & \\ & $=$ & $I_{R^c} ; K_{(T \star U)^c}$ & Lemma \ref{Lemma: lifting heterogeneous relations commute} (6) \\
     & $=$ & $I_{R^c} ; (J_{T^c} ; I_{U^c})$, & Lemma \ref{Lemma: lifting heterogeneous relations commute} (2) 
\end{tabular}
\end{tabular}
\par}}
\smallskip

\noindent and, as discussed in Remark \ref{rem: crisp polarity-based heterogeneous associativity does not hold}, the identity $I_{R^c} ; (J_{T^c} ; I_{U^c}) = (I_{R^c} ; J_{T^c}) ; I_{U^c}$ fails.
\end{remark}

\subsection{(Heterogeneous) relation algebras and their propositional languages}
\label{ssec:prel and krel}
We conclude the present section by introducing mathematical structures in the context of which the notions of composition introduced above can be studied systematically from a universal-algebraic or category-theoretic perspective. Besides being of potential independent interest, these structures can naturally be associated with {\em (multi-type) propositional languages} which will be key tools to build a common ground where the first-order correspondents of modal axioms in various semantic contexts can be compared and systematically related.
\begin{definition}
\label{def:rel algebras crisp}
For any LE-language $\mathcal{L}_{\mathrm{LE}} = \mathcal{L}_{\mathrm{LE}}(\mathcal{F}, \mathcal{G})$, 
\begin{enumerate}
\setlength{\itemsep}{0.2pt}
\setlength{\parskip}{0pt}
\setlength{\parsep}{0pt}
\item any
polarity-based $\mathcal{L}_{\mathrm{LE}}$-frame   $\mathbb{F} = (\mathbb{P}, \mathcal{R}_{\mathcal{F}}, \mathcal{R}_{\mathcal{G}})$ based on  $\mathbb{P} = (A, X,  I)$ induces the heterogeneous relation algebra 
\[
\mathbb{F}^\ast: = (\mathcal{P}(A\times X), \mathcal{P}(X\times A), \mathcal{P}(A\times A), \mathcal{P}(X\times X),  \mathcal{R}_{\mathcal{F}^\ast}, \mathcal{R}_{\mathcal{G}^\ast},  ; _I^{\Diamond}, ;_I^{\Box}, ;^{\Box\Diamond},  ;^{\Diamond\Box}, ;^{{\rhd}\Box}, ;^{\Diamond{\rhd}}, ;^{{\lhd}\Diamond}, ;^{\Box{\lhd}}, I, J), 
\]
where, abusing notation, we understand $ \mathcal{R}_{\mathcal{F}^\ast}: = \{R_{\Diamond} \mid \Diamond\in \mathcal{F}\}\cup \{R_{\Diamondblack}\mid  \Box\in \mathcal{G}\}\subseteq \mathcal{P}(X\times A)$ and $\mathcal{R}_{\mathcal{G}^\ast}: = \{R_{\Box} \mid \Box\in \mathcal{G}\}\cup\{ R_{\blacksquare}\mid  \Diamond\in \mathcal{F}\}\subseteq \mathcal{P}(A\times X)$, 
where $R_{\Diamondblack}$ and $R_{\blacksquare}$ are defined as indicated in the discussion above Lemma \ref{lem: distribution law composition inverse},
and the operations are all binary and defined as in Definition \ref{def:relational composition}. Superscripts  in the operations indicate the types of the two inputs, which completely determine the output type. These superscripts will be dropped whenever this does not cause ambiguities.
\item Any
Kripke $\mathcal{L}_{\mathrm{LE}}$-frame   $\mathbb{X} = (W, \mathcal{R}_{\mathcal{F}}, \mathcal{R}_{\mathcal{G}})$ induces the relation algebra 
\[\mathbb{X}^\ast: = (\mathcal{P}(W\times W), \mathcal{R}_{\mathcal{F}^{\ast}}, \mathcal{R}_{\mathcal{G}^{\ast}}, \circ, \star, \Delta), \]
where, abusing notation, we understand $ \mathcal{R}_{\mathcal{F}^\ast}: = \{R_{\Diamond} \mid \Diamond\in \mathcal{F}\}\cup \{R_{\Diamondblack}\mid  \Box\in \mathcal{G}\}\subseteq \mathcal{P}(W\times W)$ and $\mathcal{R}_{\mathcal{G}^\ast}: = \{R_{\Box} \mid \Box\in \mathcal{G}\}\cup\{ R_{\blacksquare}\mid  \Diamond\in \mathcal{F}\}\subseteq \mathcal{P}(W\times W)$, 
where $R_{\Diamondblack}$ and $R_{\blacksquare}$ are defined as indicated in the discussion above Lemma \ref{lem: distribution law composition inverse},
moreover, $\circ$ denotes the standard composition of  relations, and  $\star$ is defined as in Definition \ref{def: pseudo comp on kripke}.
\end{enumerate}
\end{definition}



For any LE-language $\mathcal{L} = \mathcal{L}(\mathcal{F}, \mathcal{G})$, consider the  following sets of constant symbols:  
\[
\mathsf{DRel}_{\mathcal{L}}: = \{R_\Diamond  \mid \Diamond \in \mathcal{F}^\ast\}\quad\quad  \mathsf{BRel}_{\mathcal{L}}: = \{R_{\Box}  \mid \Box  \in \mathcal{G}^\ast\},
\]
where by e.g.\ $\Diamond \in \mathcal{F}^\ast$ we include also the adjoints $\Diamondblack$ of any $\Box \in \mathcal{G}$, and likewise for $\Box  \in \mathcal{G}^\ast$.

The {\em multi-type propositional language}  $\mathsf{PRel}_{\mathcal{L}}$ of the heterogeneous relation algebras associated with polarity-based $\mathcal{L}_{\mathrm{LE}}$-frames contains terms of types $\mathsf{T}_{A\times X}$, $\mathsf{T}_{X\times A}$, $\mathsf{T}_{A\times A}$ and $\mathsf{T}_{X\times X}$, and is defined by the following simultaneous recursions:

{{\centering
\begin{tabular}{rcl c lcl}
$\mathsf{T}_{A\times X} \ni \beta$ & $:: =$ &\, $I \mid R_{\Box}\mid \beta\, ;_I \beta \mid \rho\, ; \beta \mid \beta ; \lambda$ &\quad\quad& 
$\mathsf{T}_{A\times A}\ni \rho$ & $:: =$ &\,  $\delta \, ;\beta$ \\
$\mathsf{T}_{X\times A}\ni \delta$ & $:: = $ &\, $J \mid R_{\Diamond} \mid \delta\, ;_I \delta\mid  \lambda\, ; \delta \mid \delta ; \rho$ &&
$\mathsf{T}_{X\times X}\ni \lambda$ & $:: =$ &\, $\beta\, ; \delta $,
\end{tabular}
\par}} 

\noindent where $I$ (resp.\ $J$) is the constant symbol denoting the incidence relation (resp.\ the converse of the incidence relation) of the polarity on which its model is based, and $R_{\Box}\in \mathsf{BRel}_{\mathcal{L}}$ and $R_{\Diamond}\in \mathsf{DRel}_{\mathcal{L}}$. Notice that we can omit the superscripts in the compositions because the type of the input terms in each coordinate takes care of the disambiguation. 

The {\em propositional language} $\mathsf{KRel}_{\mathcal{L}}$ of the relation algebras associated with Kripke $\mathcal{L}$-frames  is defined by the following  recursion: 
\begin{align*}
\mathsf{KRel}_{\mathcal{L}} \ni \xi :: = &\, \Delta \mid R \mid  \xi\circ \xi  \mid \xi \star \xi ,
\end{align*}
where $\Delta$ is the constant symbol for the identity relation, and $R\in \mathsf{BRel}_{\mathcal{L}}\cup \mathsf{DRel}_{\mathcal{L}}$. 
Term inequalities of $\mathsf{KRel}_{\mathcal{L}}$ will be written as $\xi_1\subseteq \xi_2$ and will be interpreted on Kripke $\mathcal{L}$-frames in the obvious way, and in particular,
\[
\mathbb{X}\models \xi_1\subseteq \xi_2\quad \text{ iff } \quad\mathbb{X}^\ast \models \xi_1\subseteq \xi_2.
\] 
Likewise, term inequalities of $\mathsf{PRel}_{\mathcal{L}}$ such as
$\beta_1\subseteq \beta_2$ and $\delta_1\subseteq \delta_2$ are necessarily homogeneous, in the sense that the left-hand term and the right-hand term must be of the same type, and  will be interpreted on polarity-based $\mathcal{L}$-frames in the obvious way, and in particular, \[\mathbb{F}\models \zeta_1\subseteq \zeta_2\quad \text{ iff } \quad\mathbb{F}^\ast \models \zeta_1\subseteq \zeta_2.\] 

In the next section we will represent the first-order correspondents of inductive modal reduction principles as term inequalities of $\mathsf{PRel}_{\mathcal{L}}$ and $\mathsf{KRel}_{\mathcal{L}}$.
Intuitively,  the various compositions considered above will be used as ``abbreviations'' to achieve a more compact representation of the correspondents which is also more amenable to computation. Let us illustrate this idea with examples before moving on to the next section.
\begin{example}
\label{example: classical case}
The modal reduction principle $\Diamond\Box p\leq \Box\Diamond p$ is analytic inductive, and corresponds, on Kripke frames such that $R_{\Box} = R = R_{\Diamond}$, to the well known confluence condition
\[\forall x\forall y(\exists z(z Rx \ \&\ z R y)\Rightarrow \exists w(x R w\ \& \ yRw)),\]
which can equivalently be rewritten as 
\[\forall x\forall y(x(R^{-1}\circ R) y\Rightarrow x ( R \circ R^{-1})y),\]
and, as $R^{-1}=R_\Diamondblack$,  it can be equivalently rewritten as the following $\mathsf{KRel}_\mathcal{L}$-inequality:
\[R_\Diamondblack \circ R_\Diamond \subseteq R_\Diamond \circ R_\Diamondblack.\]

The modal reduction principle $\Box p\leq \Diamond p$ is analytic inductive, and corresponds, on Kripke frames such that $R_{\Box} = R = R_{\Diamond}$, to the well known seriality condition
\[\forall x\exists y( x R y),\]
which can equivalently be rewritten as 
\[\forall x(x\Delta x \Rightarrow \exists y(x R y\ \& \ yR^{-1}x)),\]
and, as $R^{-1}=R_\Diamondblack$,  it can be equivalently rewritten as the following $\mathsf{KRel}_\mathcal{L}$-inequality:
\[\Delta \subseteq R_\Diamond \circ R_\Diamondblack.\]

The modal reduction principle $ p\leq \Diamond\Box p$ is  inductive (but not analytic) of shape (a), and, by Proposition \ref{prop:albaoutput}, it is equivalent to
$\forall\nomj(\nomj \leq \Diamond\Box \nomj )$, which can be rewritten as follows: 
\smallskip

{{\centering
\begin{tabular}{rll}
&$\forall x \big(\{x\} \subseteq \langle R_\Diamond \rangle [ R_\Box ] \{ x \} \big)$ & \\
iff & $\forall x\big( \{x\} \subseteq \langle R_\Diamond \rangle R_\Box^{[0]}[\{x\}^c] \big)$ & Lemma \ref{lemma:properties of square bracket superscript} (3) \\
iff & $\forall x\big( \{x\} \subseteq  (R_\Diamond^{[0]}[ R_\Box^{[0]}[\{x\}^c]])^c \big)$ & Lemma \ref{lemma:properties of square bracket superscript} (6) \\
iff & $\forall x\big(  R_\Diamond^{[0]}[ R_\Box^{[0]}[\{x\}^c]] \subseteq \{x\}^c \big)$ & contraposition \\
iff & $\forall x\big(  R_\Diamond^{[0]}[ R_\Box^{[0]}[\Delta^{[0]}[x]]] \subseteq \Delta^{[0]}[x] \big)$ & Lemma \ref{lemma:properties of square bracket superscript} (4) \\
iff & $\forall x\big(  R_\Diamond^{[0]}[ (R_\Box \star \Delta)^{[0]}[x] ] \subseteq \Delta^{[0]}[x] \big)$ & Lemma \ref{lemma:characterization of pseudo} \\
iff & $\forall x\big(  (R_\Diamond \star R_\Box \star \Delta)^{[0]}[x] \subseteq \Delta^{[0]}[x] \big)$ & Lemma \ref{lemma:characterization of pseudo} \\
iff & $\forall x\big(\Delta^{(0)}[x] \subseteq (R_\Diamond \star R_\Box \star \Delta)^{(0)}[x] \big)$ & Lemma \ref{lemma:properties of square bracket superscript} (2) and contraposition \\
iff & $\Delta \subseteq R_\Diamond \star R_\Box \star \Delta$ & Definition of $(\cdot)^{(0)}$
\end{tabular}
\par}}
\smallskip

The modal reduction principle $ p\leq \Diamond\Box \Diamond p$ is  inductive (but not analytic) of shape (a), and an ALBA run yields $\forall \nomj(\nomj \leq \Diamond\Box\Diamond\nomj)$, which can be rewritten as follows:
\smallskip

{{\centering
    \begin{tabular}{rll}
    & $\forall x \big( \{ x \} \subseteq \langle R_\Diamond \rangle [ R_\Box ] \langle R_\Diamond \rangle \{ x \}  \big )$ & \\
    iff & $\forall x \big( \{ x \} \subseteq \langle R_\Diamond \rangle [ R_\Box ] (R_\Diamond^{[0]}[ x ])^c  \big )$ & Lemma \ref{lemma:properties of square bracket superscript} (6) \\
    iff & $\forall x \big( \{ x \} \subseteq \langle R_\Diamond \rangle R_\Box^{[0]}[ R_\Diamond^{[0]}[ x ]]  \big )$ & Lemma \ref{lemma:properties of square bracket superscript} (3) \\
    iff & $\forall x \big( \{ x \} \subseteq (R_\Diamond^{[0]} [ R_\Box^{[0]}[ R_\Diamond^{[0]}[ x ]]])^c  \big )$ & Lemma \ref{lemma:properties of square bracket superscript} (6) \\
    iff & $\forall x \big( R_\Diamond^{[0]} [ R_\Box^{[0]}[ R_\Diamond^{[0]}[ x ]]] \subseteq \{ x \}^c  \big )$ & contraposition \\
    iff & $\forall x \big( R_\Diamond^{[0]} [ R_\Box^{[0]}[ R_\Diamond^{[0]}[ x ]]] \subseteq \Delta^{[0]}[x]  \big )$ & Lemma \ref{lemma:properties of square bracket superscript} (4) \\
    iff & $\forall x \big( (R_\Diamond \star R_\Box \star  R_\Diamond)^{[0]}[ x ] \subseteq \Delta^{[0]}[x]  \big )$
    & Lemma \ref{lemma:characterization of pseudo} twice \\
    iff & $\forall x \big( \Delta^{(0)}[x] \subseteq (R_\Diamond \star R_\Box \star  R_\Diamond)^{(0)}[ x ]  \big )$
    & Lemma \ref{lemma:properties of square bracket superscript} (2) and contraposition \\
    iff & $ \Delta \subseteq R_\Diamond \star R_\Box \star  R_\Diamond  $
    & Definition of $(\cdot)^{(0)}$ \\
    \end{tabular}
\par}}
\smallskip

The modal reduction principle $\Box\Diamond p \leq p$ is inductive (but not analytic) of shape (b), so by Proposition \ref{prop:albaoutput} it is equivalent to $\Box\Diamond\cnomm \leq \cnomm$, which can be rewritten as follows:
\smallskip

{{\centering
\begin{tabular}{rll}
& $\forall x( [R_\Box] \langle R_\Diamond \rangle \{x\}^c \subseteq \{x\}^c )$ & \\  
iff & $\forall x( [R_\Box] (R^{[0]}_\Diamond[\{x\}^c])^c \subseteq \{x\}^c )$ & Lemma \ref{lemma:properties of square bracket superscript} (6)\\  
iff & $\forall x( R_\Box^{[0]}[ R^{[0]}_\Diamond[\{x\}^c]] \subseteq \{x\}^c )$ & Lemma \ref{lemma:properties of square bracket superscript} (3)\\  
iff & $\forall x( R_\Box^{[0]}[ R^{[0]}_\Diamond[\Delta^{[0]}[x]]] \subseteq \Delta^{[0]}[x] )$ & Lemma \ref{lemma:properties of square bracket superscript} (4)\\ 
iff & $\forall x( (R_\Box \star R_\Diamond \star \Delta)^{[0]}[x] \subseteq \Delta^{[0]}[x] )$ &  Lemma \ref{lemma:characterization of pseudo} twice \\
iff & $\forall x(  \Delta^{(0)}[x] \subseteq (R_\Box \star R_\Diamond \star \Delta)^{(0)}[x] )$ & Lemma \ref{lemma:properties of square bracket superscript} (2) and contraposition \\
iff & $\Delta \subseteq R_\Box \star R_\Diamond \star \Delta$ & Definition of $(\cdot)^{(0)}$
\end{tabular}
\par}}
\end{example}
\section{Crisp correspondents of inductive modal reduction principles}
\label{sec:crisp}
With the definitions of the previous section in place, we are now in a position to concretely represent the first-order correspondents of inductive modal reduction principles  on polarity-based $\mathcal{L}$-frames and on  Kripke $\mathcal{L}$-frames. This representation  will  enable us  to show that the first-order correspondents on Kripke frames and on polarity-based frames of the same inductive modal reduction principle are  connected via  the same embedding connecting Kripke frames  to polarity-based frames.  Specifically, in  Sections \ref{ssec: correspondents on polarity-based frames} and \ref{ssec: correspondents on kripke}, we show that 

\begin{proposition}
\label{prop: pure inclusions crisp}
For any LE-language $\mathcal{L}$, the first-order correspondent of any inductive modal reduction principle of $\mathcal{L}$ on  polarity-based $\mathcal{L}$-frames (resp.~Kripke $\mathcal{L}$-frames)  can be represented as a term inequality in $\mathsf{PRel}_{\mathcal{L}}$ (resp.~$\mathsf{KRel}_{\mathcal{L}}$) 
\end{proposition}
In Section \ref{ssec: connection via lifting}, we  show that  a first-order correspondent on polarity-based $\mathcal{L}$-frames of any inductive modal reduction principle is a `lifting' of its first-order correspondent on Kripke $\mathcal{L}$-frames; in Section \ref{ssec:parametric corr mrp crisp}, we take conceptual stock  of this result, in terms of the notion of {\em parametric correspondence}.   

\subsection{Correspondents on polarity-based frames}
\label{ssec: correspondents on polarity-based frames}

\begin{notation}
\label{notation:phipsi}
We will often use the convention that formulas consisting of a finite (possibly empty) sequence of diamonds in $\mathcal{F}^*$ (resp.\ boxes in $\mathcal{G}^*$) are denoted by $\varphi$ (resp.\ $\psi$) possibly with sub- or super-scripts, and formulas consisting of a finite concatenation of connectives in $\mathcal{F}^* \cup \mathcal{G}^*$ starting with a diamond in $\mathcal{F}^*$ (resp.\ box in $\mathcal{G}^*$) are denoted by $\chi$ (resp.\ $\zeta$). Each use of this convention will be reminded of to the reader.
\end{notation}

\begin{definition}\label{def:pol-based rels associated with terms}
Let us associate terms in $\mathsf{PRel}_{\mathcal{L}}$ with ${\mathcal{L^*}}$-terms of certain syntactic shapes as follows. For all $\varphi(!y), \psi(!y), \chi(!x) = \phi_1\psi_1\ldots\phi_{n_\chi}\psi_{n_\chi}(!x), \zeta(!x) = \psi_1\phi_1\ldots\psi_{n_\zeta}\phi_{n_\zeta}(!x)$ as in Notation \ref{notation:phipsi}, let us define $R_\varphi, R_\psi, R_\chi, R_\zeta \in \mathsf{PRel}_\mathcal{L}$ such that:
\smallskip

{{\centering
\begin{tabular}{lcl}
if $\phi: = y$, then $R_{\phi}: = J$ & \quad & if $\psi_{n_\chi}$ is empty,
$
    R_{\chi} \coloneqq R_{\varphi_1} ; R_{\psi_1} ; \cdots ; R_{\varphi_{n_\chi}}
$, while \\
if $\phi: = \Diamond\phi'$, then $R_{\phi}: = R_\Diamond ;_I R_{\phi'}$, with $\Diamond \in \mathcal{F}^*$ &&if $\psi_{n_\chi}$ is nonempty,
$
    R_{\chi} \coloneqq R_{\varphi_1} ; R_{\psi_1} ; \cdots ; R_{\varphi_{n_\chi}} ; R_{\psi_{n_\chi}} ; J
$. \\[1.5mm]
if $\psi: = y$, then $R_{\psi}: = I$; & & if $\phi_{n_\zeta}$ is empty,
$
    R_{\zeta} \coloneqq R_{\psi_1} ; R_{\phi_1} ; \cdots ; R_{\psi_{n_\zeta}}
$, while\\
if $\psi: = \Box\psi'$, then $R_{\psi}: = R_\Box ;_I R_{\psi'}$, with $\Box \in \mathcal{G}^*$ &&if $\phi_{n_\zeta}$ is nonempty,
$
    R_{\zeta} \coloneqq R_{\psi_1} ; R_{\phi_1} ; \cdots ; R_{\psi_{n_\zeta}} ; R_{\phi_{n_\zeta}} ; I
$.
\end{tabular}
\par}}

\end{definition}

\begin{lemma}
\label{lemma:molecular polarity}
For any $\mathcal{L}$-frame $\mathbb{F}$ based on the polarity $\mathbb{P} = (A,X,I)$, any $\varphi(!z)$, $\psi(!z)$, $\chi(!z)$, $\zeta(!z)$ as in Definition \ref{def:pol-based rels associated with terms}, any $B\subseteq A$ and $Y \subseteq X$, and any $a \in A$ and $x \in X$,
\smallskip

{{\centering
\begin{tabular}{rl c rl}
    1. & $\descr{\varphi(!z)}(z \coloneqq B) =  R^{(0)}_{\varphi}[B];$ & \quad\quad & 
    3. & $\descr{\chi[\nomj/!z]}(\nomj \coloneqq a^{\uparrow\downarrow}) = R^{(0)}_{\chi}[a^{\uparrow\downarrow}]=  R^{(0)}_{\chi}[a];$ \\
    2. & $\val{\psi(!z)}(z \coloneqq Y) = R^{(0)}_\psi[Y]$ &&
    4. & $\val{\zeta[\cnomm/!z]}(\cnomm \coloneqq x^{\downarrow\uparrow}) = R^{(0)}_{\zeta}[x^{\downarrow\uparrow}] = R^{(0)}_{\zeta}[x].$
\end{tabular}
\par}} 
\end{lemma}
\begin{proof}
Items 1 and 2 can be shown straightforwardly by induction on $\varphi$ and $\psi$, using Lemma \ref{lemma:comp4}. Items 3 and 4 can be shown  by induction on the number of alternations of concatenations of box and diamond connectives, using  items 1 and 2, and Definition \ref{def:relational composition}. That $R^{(0)}_{\chi}[a^{\uparrow\downarrow}]=  R^{(0)}_{\chi}[a]$ follows from the fact that $R_{\chi}$ is a composition of an odd number of $I$-compatible relations, via induction. Indeed, if $R$ is an $I$-compatible relation, then if $b\in a^{\uparrow\downarrow}$, it follows that $b^{\uparrow\downarrow}\subseteq a^{\uparrow\downarrow}$, and hence $R^{(0)}[a^{\uparrow\downarrow}]\subseteq R^{(0)}[b^{\uparrow\downarrow}]$, so by the $I$-compatibility of $R$, $R^{(0)}[a]\subseteq R^{(0)}[b]$. Now assume that $R$ is a composition of $2n+1$ $I$-compatible relations and $R^{(0)}[a]\subseteq R^{(0)}[b]$. Then $S^{(0)}[R^{(0)}[b]]\subseteq S^{(0)}[R^{(0)}[a]]$ and so $T^{(0)}[S^{(0)}[R^{(0)}[a]]]\subseteq T^{(0)}[S^{(0)}[R^{(0)}[b]]]$. The case for $\zeta$ is shown similarly.
\end{proof}

As discussed in Section \ref{sec:mrp}, the inductive modal reduction principles of any LE-language $\mathcal{L}$ have one of the following shapes,  which we can describe compactly using the terminology and notation of Section \ref{Inductive:Fmls:Section}:
\begin{enumerate}
\setlength{\itemsep}{0.2pt}
\setlength{\parskip}{0pt}
\setlength{\parsep}{0pt}
\item[(a)] $\phi [\alpha (p)/!y]\leq \psi [\chi (p)/!x]$ with $+p\prec +\alpha\prec +s$ and $-p\prec -\chi\prec -t$, and $+\alpha(p)$ PIA-branch, $+\phi(!y)$ and $-\psi(!x)$ Skeleton-branches, and either $\chi(p): = p$ or the main connective of $-\chi(p)$ is non-Skeleton (i.e.~$\chi(p): = \Diamond \chi'(p)$ for some $\Diamond\in \mathcal{F}$);
\item[(b)] $\phi [\zeta (p)/!y]\leq \psi [\delta (p)/!x]$ with $+p\prec +\zeta\prec +s$ and $-p\prec -\delta\prec -t$, and $-\delta(p)$ PIA-branch, $+\phi(!y)$ and $-\psi(!x)$ Skeleton-branches, and either $\zeta(p): = p$ or the main connective of $+\zeta(p)$ is non-Skeleton (i.e.~$\zeta(p): = \Box \zeta'(p)$ for some $\Box\in \mathcal{G}$).
\end{enumerate}
By Proposition \ref{prop:albaoutput}, modal reduction principles of shape (a) are equivalent to the following pure inequality:
\begin{equation}
\label{eq: alba output a}
\forall \nomj\left(\mathsf{LA}(\psi)[\phi[\nomj/!y]/!u] \leq \chi[\mathsf{LA}(\alpha)[\nomj/!u]/p]\right).
\end{equation}
When interpreting the condition above on concept lattices arising from a given polarity $\mathbb{P} = (A, X, I)$, we need to recall  the definition of the order on concept lattices, and that the nominal variable $\nomj$ ranges over the formal concepts of the form $(a^{\uparrow\downarrow}, a^{\uparrow})$ for $a\in A$. Then, the condition above translates as follows:
\begin{equation}
\label{eq: alba output a translated}
\forall a \left(\descr{\chi[\mathsf{LA}(\alpha)[\nomj/!u]/p] }(\nomj: =a^{\uparrow\downarrow})\subseteq \descr{\mathsf{LA}(\psi)[\phi[\nomj/!y]/!u]}(\nomj: =a^{\uparrow\downarrow})\right).
\end{equation}
Since $\mathsf{LA}(\psi)[\phi(!y)/!u]$ is a finite concatenation of diamond connectives in $\mathcal{F}^*$, and $\chi[\mathsf{LA}(\alpha)/p]$ is a finite concatenation of diamond and box connectives in $\mathcal{F}^\ast\cup \mathcal{G}^\ast$ the outermost connective of which is a diamond in $\mathcal{F}^*$,
by items 1 and 3 of Lemma \ref{lemma:molecular polarity},
\smallskip

{{\centering
\begin{tabular}{rcccl}
$\descr{\mathsf{LA}(\psi)[\phi[\nomj/!y]/!u]}(\nomj: = a^{\uparrow\downarrow})$ & $=$ & $R_{\mathsf{LA}(\psi)[\phi/!u]}^{(0)}[a^{\uparrow\downarrow}]$ & $=$ & $(R_{\mathsf{LA}(\psi)} \,;_I R_{\phi})^{(0)}[a]$. \\

$\descr{\chi[\mathsf{LA}(\alpha)[\nomj/!u]/p] }(\nomj:= a^{\uparrow\downarrow})$ & $=$ & $R_{\chi[\mathsf{LA}(\alpha)/p]}^{(0)}[a^{\uparrow\downarrow}]$ & $=$ & $R_{\chi[\mathsf{LA}(\alpha)/p]}^{(0)}[a]$.
\end{tabular}
\par}}
\smallskip

\noindent Hence, \eqref{eq: alba output a translated} can be rewritten as follows:
\begin{equation}
\label{eq: alba output a polarity after lemma}
\forall a \left(R_{\chi[\mathsf{LA}(\alpha)/p]}^{(0)}[a]\subseteq (R_{\mathsf{LA}(\psi)} \,;_I R_{\phi})^{(0)}[a]\right),
\end{equation}
which is equivalent to the following pure inclusion of binary relations of type $X\times A$:
\begin{equation}
\label{eq final a polarity}
R_{\chi[\mathsf{LA}(\alpha)/p]}\subseteq R_{\mathsf{LA}(\psi)} \,;_I R_{\phi}.
\end{equation}
Similarly, by Proposition \ref{prop:albaoutput}, modal reduction principles of shape (b) are equivalent to the following pure inequality:
\begin{equation}
\label{eq: alba output b}
\forall \cnomm\left(\zeta[\mathsf{RA}(\delta)[\cnomm/!u]/p] \leq \mathsf{RA}(\phi)[\psi[\cnomm/!y]/!u]\right),
\end{equation}
and, when interpreting the condition above on concept lattices, recalling that the conominal variable $\cnomm$ ranges over the formal concepts $(x^{\downarrow}, x^{\downarrow\uparrow})$ for $x\in X$, the condition above translates as follows:
\begin{equation}
\label{eq: alba output b translated}
\forall x\left (\val{\zeta[\mathsf{RA}(\delta)[\cnomm/!u]/p]}(\cnomm: = x^{\downarrow\uparrow}) \subseteq \val{\mathsf{RA}(\phi)[\psi[\cnomm/!y]/!u]}(\cnomm: = x^{\downarrow\uparrow})\right ).
\end{equation}
By items 4 and 2 of Lemma \ref{lemma:molecular polarity}, 
\smallskip

{{\centering
\begin{tabular}{rcccl}
     $\val{\mathsf{RA}(\phi)[\psi[\cnomm/!y]/!u]}(\cnomm:= x^{\downarrow\uparrow})$ & $=$ &  $R_{\mathsf{RA}(\phi)[\psi/!u]}^{(0)}[x^{\downarrow\uparrow}]$ & $=$ & $(R_{\mathsf{RA}(\phi)} \,;_I R_{\psi})^{(0)}[x]$. \\
$\val{\zeta[\mathsf{RA}(\delta)[\cnomm/!u]/p] }(\cnomm:= x^{\downarrow\uparrow})$ &  $=$ &  $R_{\zeta[\mathsf{RA}(\delta)/p]}^{(0)}[x^{\downarrow\uparrow}]$ & $=$ & $R_{\zeta[\mathsf{RA}(\delta)/p]}^{(0)}[x]$.
\end{tabular}
\par}} 
\smallskip

\noindent Therefore, we can rewrite \eqref{eq: alba output b translated} as follows:
\[
\forall x \left( R_{\zeta[\mathsf{RA}(\delta)/p]}^{(0)}[x]\subseteq (R_{\mathsf{RA}(\phi)} \,;_I R_{\psi})^{(0)}[x]\right),
\]
 which is equivalent to the following pure inclusion of binary relations of type $A\times X$:
\begin{equation}
\label{eq final b polarity}
R_{\zeta[\mathsf{RA}(\delta)/p]}\subseteq R_{\mathsf{RA}(\phi)} \,;_I R_{\psi}.
\end{equation}
Finally, notice that, if the modal reduction principle is also analytic, the shape of $\chi(p)\neq p$ (resp.\ $\zeta(p)\neq p$)  simplifies, in case (a), to $\chi(p) = \phi_{n_\chi} \psi_{n_\chi}(p)$ with $n_\chi = 1$ and $\psi_{n_\chi}$ empty 
and, in case (b), to $\zeta(p) = \psi_{n_\zeta} \phi_{zeta}(p)$ with $n_\zeta = 1$ and $\phi_{n_\zeta}$ empty. 
Hence, \eqref{eq final a polarity}  and \eqref{eq final b polarity}  simplify to the following inclusions, respectively:
\begin{equation}
\label{eq final analytic}
R_{\phi_{n_\chi}}\, ;_IR_{\mathsf{LA}(\alpha)}\subseteq R_{\mathsf{LA}(\psi)} \,;_I R_{\phi} \quad\quad R_{\psi_{n_\zeta}}\, ;_I R_{\mathsf{RA}(\delta)}\subseteq R_{\mathsf{RA}(\phi)} \,;_I R_{\psi}.
\end{equation}

\begin{example}
\label{ex: p leq diamond box p}
The modal reduction principle $p\leq \Diamond \Box p$  is inductive of shape (a), where $\phi (y): =  y$, and $\alpha (p): = p$, hence $\mathsf{LA}(\alpha)(u): = u$, and $\psi(x): =  x$, hence $\mathsf{LA}(\psi)(v): =  v$, and $\chi(p) := \Diamond \Box p$. Then, by Proposition \ref{prop:albaoutput}, this inequality is equivalent to
\begin{equation*}
\forall \nomj\left([[\nomj/y]/v] \leq \Diamond\Box[[\nomj/u]/p]\right),
\end{equation*}
which can be rewritten as
\begin{equation*}
\label{eq: alba output a example 1}
\forall \nomj\left(\nomj \leq \Diamond \Box\nomj\right).
\end{equation*}
By the definition of order on concept lattices, this condition is equivalent to
\begin{equation*}
\forall \nomj \left(\descr{\Diamond \Box\nomj } \subseteq \descr{\nomj}\right),
\end{equation*}
which translates as 
\begin{equation*}
\begin{array}{rl}
& \forall a \left(R_{\Diamond}^{(0)}[R_{\Box}^{(0)}[J^{(0)}[a]]]\subseteq J^{(0)}[a]\right),\\
\mbox{i.e.} & \forall a \left((R_{\Diamond}\, ;R_{\Box}\, ; J)^{(0)}[a]\subseteq J^{(0)}[a]\right),
\end{array}
\end{equation*}
which is equivalent to the following pure inclusion of binary relations:
\begin{equation*}
R_{\Diamond}\, ;R_{\Box}\, ; J\subseteq J.
\end{equation*}
\end{example}

\begin{example}
\label{ex: diamond p leq box  p}
The modal reduction principle $\Diamond p\leq \Box  p$  is inductive of shape (b), where $\phi (y): =  \Diamond y$, hence $\mathsf{RA}(\phi)(v): =  \blacksquare v$, and $\zeta (p): =  p$,  and $\psi(x): = \Box x$, and $\delta(p) :=   p$, hence $\mathsf{RA}(\delta)(u): =  u$. Then, by Proposition \ref{prop:albaoutput}, this inequality is equivalent to
\begin{equation*}
\forall \cnomm\left( [[\cnomm/u]/p]\leq \blacksquare\Box [[\cnomm/y]/v] \right),
\end{equation*}
which can be rewritten as
\begin{equation*}
\label{eq: alba output a example 2}
\forall \cnomm\left( \cnomm \leq\blacksquare \Box\cnomm \right).
\end{equation*}
By the definition of order on concept lattices, this condition is equivalent to
\begin{equation*}
\forall \cnomm \left(\val{\cnomm } \subseteq \val{\blacksquare \Box\cnomm}\right),
\end{equation*}
which translates as 
\begin{equation*}
\begin{array}{rl}
&\forall x \left( I^{(0)}[x] \subseteq R_{\blacksquare}^{(0)}[I^{(1)}[R_{\Box}^{(0)}[x]]]\right), \\
\mbox{i.e.} & \forall x \left( I^{(0)}[x] \subseteq (R_{\blacksquare};_I R_{\Box})^{(0)}[x]\right),
\end{array}
\end{equation*}
which is equivalent to the following pure inclusion of binary relations:
\begin{equation*}
I\subseteq R_{\blacksquare};_I R_{\Box}.
\end{equation*}
\end{example}
\begin{example}
\label{ex: p leq diamond box diamond p}
The modal reduction principle $p\leq \Diamond \Box \Diamond p$  is inductive of shape (a), where $\phi (y): =  y$, and $\alpha (p): = p$, hence $\mathsf{LA}(\alpha)(u): = u$, and $\psi(x): =  x$, hence $\mathsf{LA}(\psi)(v): =  v$, and $\chi(p) := \Diamond \Box\Diamond p$. Then, by Proposition \ref{prop:albaoutput}, this inequality is equivalent to
\begin{equation*}
\forall \nomj\left([[\nomj/y]/v] \leq \Diamond\Box\Diamond [[\nomj/u]/p]\right),
\end{equation*}
which can be rewritten as
\begin{equation*}
\label{eq: alba output a example 3}
\forall \nomj\left(\nomj \leq \Diamond \Box\Diamond\nomj\right).
\end{equation*}
By the definition of order on concept lattices, this condition is equivalent to
\begin{equation*}
\forall \nomj \left(\descr{\Diamond \Box\Diamond\nomj } \subseteq \descr{\nomj}\right),
\end{equation*}
which translates as 
\begin{equation*}
\begin{array}{rl}
& \forall a \left(R_{\Diamond}^{(0)}[R_{\Box}^{(0)}[R_{\Diamond}^{(0)}[a]]]\subseteq J^{(0)}[a]\right), \\
\mbox{i.e.} & \forall a \left((R_{\Diamond}\, ;R_{\Box}\, ; R_{\Diamond})^{(0)}[a]\subseteq J^{(0)}[a]\right),
\end{array}
\end{equation*}
which is equivalent to the following pure inclusion of binary relations:
\begin{equation*}
R_{\Diamond}\, ;R_{\Box}\, ; R_{\Diamond}\subseteq J.
\end{equation*}
\end{example}

\begin{example}
The modal reduction principle $\Diamond p\leq \Box\Diamond \Box p$  is inductive of shape (a), where $\phi (y): = \Diamond y$, and $\alpha (p): = p$, hence $\mathsf{LA}(\alpha)(u): = u$, and $\psi(x): = \Box x$, hence $\mathsf{LA}(\psi)(v): = \Diamondblack v$, and $\chi(p) := \Diamond \Box p$. Then, by Proposition \ref{prop:albaoutput}, this inequality is equivalent to
\begin{equation*}
\forall \nomj\left(\Diamondblack[\Diamond[\nomj/y]/v] \leq \Diamond\Box[[\nomj/u]/p]\right),
\end{equation*}
which can be rewritten as
\begin{equation*}
\label{eq: alba output a example 4}
\forall \nomj\left(\Diamondblack\Diamond\nomj \leq \Diamond \Box\nomj\right).
\end{equation*}
By the definition of order on concept lattices, this condition is equivalent to
\begin{equation*}
\forall \nomj \left(\descr{\Diamond \Box\nomj } \subseteq \descr{\Diamondblack\Diamond\nomj}\right),
\end{equation*}
which translates as 
\begin{equation*}
\begin{array}{rl}
& \forall a \left(R_{\Diamond}^{(0)}[R_{\Box}^{(0)}[J^{(0)}[a]]]\subseteq R_{\Diamondblack}^{(0)}[I^{(1)}[R_{\Diamond}^{(0)}[a]]]\right),\\
\mbox{i.e.} &\forall a \left((R_{\Diamond}\, ;R_{\Box}\, ; J)^{(0)}[a]\subseteq (R_{\Diamondblack}\,;_I R_{\Diamond})^{(0)}[a]\right),
\end{array}
\end{equation*}
which is equivalent to the following pure inclusion of binary relations:
\begin{equation*}
R_{\Diamond}\, ;R_{\Box}\, ; J\subseteq R_{\Diamondblack}\,;_I R_{\Diamond}.
\end{equation*}
\end{example}

\begin{example}
The modal reduction principle $\Box\Diamond p \leq \Box \Diamond\Diamond p $  is inductive of shape (b) 
with $\phi(y): = y$, hence $\mathsf{RA}(\phi)(v): = v$ and $\psi (x): = \Box x$, and $\delta(p): = \Diamond\Diamond p$, hence $\mathsf{RA}(\delta)(u): = \blacksquare\blacksquare u$, and $\zeta(p): = \Box\Diamond p$.
\smallskip

{{\centering
			\begin{tabular}{r r l l l}
				$\forall p  [\Box\Diamond p \leq \Box \Diamond\Diamond p ]$
				%
				&iff& $\forall \cnomm  [\Box\Diamond \blacksquare\blacksquare \cnomm \le \Box \cnomm]$
				& Proposition \ref{prop:albaoutput}\\
			&	iff& $\forall \cnomm  \left(\val{\Box\Diamond \blacksquare\blacksquare \cnomm} \subseteq \val{\Box \cnomm} \right)$\\
			&	i.e. &$\forall x\left (R^{(0)}_{\Box}[R^{(0)}_{\Diamond}[R_{\blacksquare}^{(0)}[ I^{(1)}[R_{\blacksquare}^{(0)}[x] ]]]\subseteq R_{\Box}^{(0)}[x]\right )$\\
&	iff&$\forall x \left (  R^{(0)}_{\Box}[R^{(0)}_{\Diamond}[(R_{\blacksquare};_I  R_{\blacksquare})^{(0)}[x] ]] \subseteq R^{(0)}_{\Box}[x] \right )$ &  \\
&	iff&$\forall x \left ( (R_{\Box}\, ;R_{\Diamond}\, ; (R_{\blacksquare}\, ;_IR_{\blacksquare}))^{(0)}[x] \subseteq  R^{(0)}_{\Box}[x] \right )$ &  \\
&	iff&$R_{\Box}\, ;R_{\Diamond}\, ; (R_{\blacksquare}\, ;_IR_{\blacksquare}) \subseteq R_{\Box}$. &  \\
				\end{tabular}
		\par}}

\end{example}

\subsection{Correspondents on Kripke frames}
\label{ssec: correspondents on kripke}
In the present subsection, we specialize the discussion of the previous subsection to the environment of Kripke frames, so as to obtain a specific representation of the first-order correspondents of inductive modal reduction principles on Kripke frames as pure inclusions of binary relations on Kripke frames.  

\begin{definition}\label{def:kripke-based rels associated with terms}
Let us associate terms in $\mathsf{KRel}_{\mathcal{L}}$ with ${\mathcal{L^*}}$-terms of certain syntactic shapes as follows. For all $\varphi(!y), \psi(!y), \chi(!x) = \phi_1\psi_1\ldots\phi_{n_\chi}\psi_{n_\chi}(!x), \zeta(!x) = \psi_1\phi_1\ldots\psi_{n_\zeta}\phi_{n_\zeta}(!x)$ as in Notation \ref{notation:phipsi}, let us define $R_\varphi, R_\psi, R_\chi, R_\zeta \in \mathsf{KRel}_\mathcal{L}$ such that:
\smallskip

{{\centering
\begin{tabular}{lcl}
if $\phi: = y$, then $R_{\phi}: = \Delta$ & \quad & if $\psi_{n_\chi}$ is empty,
$
    R_{\chi} \coloneqq R_{\varphi_1} \star R_{\psi_1} \star \cdots \star R_{\varphi_{n_\chi}}
$, while \\
if $\phi: = \Diamond\phi'$, then $R_{\phi}: = R_\Diamond \circ R_{\phi'}$, with $\Diamond \in \mathcal{F}^*$ &&if $\psi_{n_\chi}$ is nonempty,
$
    R_{\chi} \coloneqq R_{\varphi_1} \star R_{\psi_1} \star \cdots \star R_{\varphi_{n_\chi}} \star R_{\psi_{n_\chi}} \star \Delta
$. \\[1.5mm]
if $\psi: = y$, then $R_{\psi}: = \Delta$; & & if $\phi_{n_\zeta}$ is empty,
$
    R_{\zeta} \coloneqq R_{\psi_1} \star R_{\phi_1} \star \cdots \star R_{\psi_{n_\zeta}}
$, while\\
if $\psi: = \Box\psi'$, then $R_{\psi}: = R_\Box \circ R_{\psi'}$, with $\Box \in \mathcal{G}^*$ &&if $\phi_{n_\zeta}$ is nonempty,
$
    R_{\zeta} \coloneqq R_{\psi_1} \star R_{\phi_1} \star \cdots \star R_{\psi_{n_\zeta}} \star R_{\phi_{n_\zeta}} \star \Delta
$.
\end{tabular}
\par}}
\end{definition}

\begin{lemma}
\label{lemma:molecular kripke}
For any Kripke $\mathcal{L}$-frame $\mathbb{X}$, formulas $\varphi, \psi, \chi, \zeta$ such as in Notation \ref{notation:phipsi},
\smallskip

{{\centering
\begin{tabular}{rl c rl}
1. & $\val{\varphi(!x)}(x \coloneqq S) = \langle R_{\varphi} \rangle S$ & \quad\quad & 3. & $\val{\chi[\nomj/!x]}(\nomj \coloneqq a) = \langle R_{\chi} \rangle \{a\}$ \\
2. & $\val{\psi(!x)}(x \coloneqq S) = [R_\psi]S$ && 4. & $\val{\zeta[\cnomm/!x]}(\cnomm \coloneqq z^c) = [ R_{\zeta} ] \{z\}^c.$
\end{tabular}
\par}}
\end{lemma}
\begin{proof}
Item 1 and item 2 are proved by straightforward induction on the complexity of $\varphi$ and $\psi$ respectively, i.e., on the number of diamond (resp.\ box) connectives.

As to item 3, we proceed by induction on the number of alternations of boxes and diamonds. If $\chi = x$, $R_\chi = \Delta$, so the statement follows trivially. For the inductive case, $\chi\neq x$, and then $\chi = \varphi_1 \psi_1 \cdots \varphi_n \psi_n(!x)$ such that $n\geq 1$, and the $\varphi_i$ (resp.\ $\psi_i$) are finite, nonempty, concatenations of diamond (resp.\ box) operators, except for $\psi_n$ which is possibly empty. If $n=1$ and $\psi_1$ empty, then $R_\chi = R_{\varphi_1}$, hence the statement follows from item 1. If $n=1$ and $\psi_1(!z)$ is not empty, then, $R_\chi = R_{\varphi_1} \star R_{\psi_1} \star \Delta$ by item 1 and item 2,
\smallskip

{{\centering
\begin{tabular}{rrlr}
 $\val{\chi[\nomj/x]}(\nomj \coloneqq a)$ & 
$=$ & $\val{\varphi_1(!y)}(y\coloneqq \val{\psi_1[\nomj/z]}(\nomj \coloneqq a))$ & \\
&$=$ & $\langle R_{\varphi_1} \rangle \val{\psi_1[\nomj/z]} (\nomj \coloneqq a)$ & item 1 \\
&$=$ & $\langle R_{\varphi_1} \rangle [R_{\psi_1}]\{a\} $ & item 2\\
&$=$ & $\langle R_{\varphi_1} \rangle R^{[0]}_{\psi_1}[\{a\}^c] $ & Lemma \ref{lemma:properties of square bracket superscript} (3) \\
&$=$ & $ (R^{[0]}_{\varphi_1}[ R^{[0]}_{\psi_1}[\{a\}^c]])^c $ & Lemma \ref{lemma:properties of square bracket superscript} (6) \\
&$=$ & $ (R^{[0]}_{\varphi_1}[ R^{[0]}_{\psi_1}[\Delta^{[0]}[a]]])^c $ & Lemma \ref{lemma:properties of square bracket superscript} (2) \\
&$=$ & $((R_{\varphi_1} \star R_{\psi_1} \star \Delta)^{[0]}[a] )^c$ & Definition \ref{def: pseudo comp on kripke} \\
&$=$ & $(R_{\chi}^{[0]}[a] )^c$ & $\chi=\varphi_1\psi_1$ \\
&$=$ & $\langle R_{\chi} \rangle \{a\}$ 
& Lemma \ref{lemma:properties of square bracket superscript} (6) \\
\end{tabular}
\par}}
\smallskip

\noindent 
When $n>1$, $\chi = \varphi_1 \psi_1 \chi'(!x)$, where both $\varphi_1$ and $\psi_1$ are nonempty, $R_\chi = R_{\varphi_1} \star R_{\psi_1} \star R_{\chi'}$, and we can apply the induction hypothesis on $\chi'$. Hence:
\smallskip

{{\centering
    \begin{tabular}{rrlr}
          $\val{\chi[\nomj/x]}(\nomj \coloneqq a)$  
         &$=$ & $\val{\varphi_1\psi_1\chi'[\nomj/x]}(\nomj \coloneqq a)$ & \\
&$=$ & $\val{\varphi_1(!y)}(y\coloneqq \val{\psi_1\chi'[\nomj/x]}(\nomj \coloneqq a))$ & \\
&$=$ & $\val{\varphi_1(!y)}(y\coloneqq \val{\psi_1(!z)}(z \coloneqq \val{\chi'[\nomj/x]}(\nomj \coloneqq a)))$ & \\
&$=$ & $\langle R_{\varphi_1} \rangle \val{\psi_1(!z)}(z \coloneqq \val{\chi'[\nomj/x]}(\nomj \coloneqq a))$ & item 1 \\
&$=$ & $\langle R_{\varphi_1} \rangle [R_{\psi_1}]\val{\chi'[\nomj/x]}(\nomj \coloneqq a) $ & item 2\\
&$=$ & $\langle R_{\varphi_1} \rangle [R_{\psi_1}]\langle R_{\chi'} \rangle \{a\}$ & inductive hypothesis\\
&$=$ & $\langle R_{\varphi_1} \rangle [R_{\psi_1}]( R^{[0]}_{\chi'}[a])^c$ & Lemma \ref{lemma:properties of square bracket superscript} (6)\\
&$=$ & $\langle R_{\varphi_1} \rangle R^{[0]}_{\psi_1}[R^{[0]}_{\chi'}[a]]$ & Lemma \ref{lemma:properties of square bracket superscript} (3)\\
&$=$ & $(R^{[0]}_{\varphi_1} [R^{[0]}_{\psi_1}[R^{[0]}_{\chi'}[a]]])^c$ & Lemma \ref{lemma:properties of square bracket superscript} (6)\\
&$=$ & $((R_{\varphi_1} \star R_{\psi_1} \star R_{\chi'})^{[0]}[a])^c$ & Definition \ref{def: pseudo comp on kripke}\\
&$=$ & $(R_{\chi}^{[0]}[a])^c$ & $\chi=\varphi_1\psi_1\chi'$\\
&$=$ & $\langle R_{\chi} \rangle \{a\}$ & \ref{lemma:properties of square bracket superscript} (6)\\
    \end{tabular}
\par}}
\smallskip

\noindent Item 4 is proved similarly; we  only show the case when $n=1$ and $\zeta = \psi_1\varphi_1(!x)$ with $\varphi_1(!w)$ nonempty.
\smallskip

{{\centering
\begin{tabular}{rrlr}
 $\val{\zeta[\cnomm/x]}(\cnomm \coloneqq z^c)$ 
&$=$ & $\val{\psi_1(!y)}(y\coloneqq \val{\varphi_1[\cnomm/w]}(\cnomm \coloneqq z^c))$ & \\
&$=$ & $[ R_{\psi_1} ] \val{\varphi_1[\cnomm/w]} (\cnomm \coloneqq z^c)$ & item 2 \\
&$=$ & $ [R_{\psi_1}] \langle R_{\varphi_1}\rangle \{z\}^c $ & item 1\\
&$=$ & $[ R_{\psi_1} ] (R^{[0]}_{\varphi_1}[\{z\}^c])^c $ & Lemma \ref{lemma:properties of square bracket superscript} (6) \\
&$=$ & $ R^{[0]}_{\psi_1}[ R^{[0]}_{\varphi_1}[\{z\}^c]] $ & Lemma \ref{lemma:properties of square bracket superscript} (3) \\
&$=$ & $ R^{[0]}_{\psi_1}[ R^{[0]}_{\varphi_1}[\Delta^{[0]}[z]]] $ & Lemma \ref{lemma:properties of square bracket superscript} (2) \\
&$=$ & $(R_{\psi_1} \star R_{\varphi_1} \star \Delta)^{[0]}[z]$ & Definition \ref{def: pseudo comp on kripke} \\
&$=$ & $R_{\zeta}^{[0]}[z]$ & $\zeta=\psi_1\varphi_1$ \\
&$=$ & $[ R_{\zeta} ] \{z\}^c$ 
& Lemma \ref{lemma:properties of square bracket superscript} (3) \\
\end{tabular}
\par}}
\end{proof}

By Proposition \ref{prop:albaoutput}, the modal reduction principles of type (a) are equivalent to the pure inequalities
\begin{equation}
\label{eq: alba output a 2}
\forall \nomj\left(\mathsf{LA}(\psi)[\phi[\nomj/!y]/!u] \leq \chi[\mathsf{LA}(\alpha)[\nomj/!u]/p]\right).
\end{equation}
When interpreting the condition above on concept lattices arising from a given polarity $\mathbb{P} = (A, X, I)$, the condition above can be  rewritten as follows:
\begin{equation}
\forall a \left(\descr{\chi[\mathsf{LA}(\alpha)[\nomj/!u]/p] }(\nomj:= a^{\uparrow\downarrow})\subseteq \descr{\mathsf{LA}(\psi)[\phi[\nomj/!y]/!u]}(\nomj:= a^{\uparrow\downarrow})\right).
\end{equation}
When the given polarity is the lifting of a Kripke frame (i.e.~$A = X = W$ and $I = I_{\Delta^c}$), recalling that $\descr{\cdot} = \val{\cdot}^c$ and the formal concepts $(a^{\uparrow\downarrow}, a^{\uparrow}) = (a, a^c)$ for $a\in W$, the condition above can be rewritten as 
\begin{equation}
\forall a \left(\val{\chi[\mathsf{LA}(\alpha)[\nomj/!u]/p] }^c(\nomj: = a)\subseteq \val{\mathsf{LA}(\psi)[\phi[\nomj/!y]/!u]}^c(\nomj: = a)\right).
\end{equation}
which is equivalent to
\begin{equation}
\label{eq: alba output a translated boolean}
\forall a \left(\val{\mathsf{LA}(\psi)[\phi[\nomj/!y]/!u]}(\nomj: = a)\subseteq \val{\chi[\mathsf{LA}(\alpha)[\nomj/!u]/p] }(\nomj: = a)\right).
\end{equation}
By items 1 and 3 of Lemma \ref{lemma:molecular kripke}, \eqref{eq: alba output a translated boolean} is equivalent to:
\begin{equation}
    \forall a(\langle R_{\mathsf{LA}(\psi)} \circ R_\varphi \rangle\{a\} \subseteq \langle R_{\chi[\mathsf{LA}(\alpha)/p]} \rangle \{a\}),
\end{equation}
which, by Lemma \ref{lemma:properties of square bracket superscript} (6), can be rewritten as
\[
\forall a \left((R_{\mathsf{LA}(\psi)} \circ R_{\phi})^{-1}[a]\subseteq R_{\chi[\mathsf{LA}(\alpha)/p]}^{-1}[a]\right),
\]
which is equivalent to the following pure inclusion in $\mathsf{KRel}_{\mathcal{L}}$, as required:
\begin{equation}
\label{eq final a boolean}
 R_{\mathsf{LA}(\psi)} \circ R_{\phi} \subseteq R_{\chi[\mathsf{LA}(\alpha)/p]}.
\end{equation}

\noindent By Proposition \ref{prop:albaoutput}, modal reduction principles of type (b) are equivalent to the following pure inequalities:
\begin{equation}
\label{eq: alba output b boolean}
\forall \cnomm\left(\zeta[\mathsf{RA}(\delta)[\cnomm/!u]/p] \leq \mathsf{RA}(\phi)[\psi[\cnomm/!y]/!u]\right),
\end{equation}
and, recalling that  $(x^{\downarrow}, x^{\downarrow\uparrow}) = (x, x^c)$ for $x\in W$, the condition above translates as follows:
\begin{equation}
\label{eq: alba output b translated boolean}
\forall x\left(\val{\zeta[\mathsf{RA}(\delta)[\cnomm/!u]/p]}(\cnomm:= x^c) \subseteq \val{\mathsf{RA}(\phi)[\psi[\cnomm/!y]/!u]}(\cnomm:= x^c)\right).
\end{equation}

By items 4 and 2 of Lemma \ref{lemma:molecular kripke}, \eqref{eq: alba output b translated boolean} can be rewritten as
\[
\forall x(
[R_{\zeta[\mathsf{RA}(\delta)/p]}]\{x\}^c \subseteq [R_{\mathsf{RA}(\varphi)} \circ R_{\psi}]\{x\}^c
),
\]
which, by Lemma \ref{lemma:properties of square bracket superscript} (3), can be rewritten as follows:
\[
\forall x \left( R_{\zeta[\mathsf{RA}(\delta)/p]}^{[0]}[x]\subseteq (R_{\mathsf{RA}(\phi)} \,\circ  R_{\psi})^{[0]}[x]\right),
\]
which by contraposition is equivalent to
\[
\forall x \left( (R_{\mathsf{RA}(\phi)} \,\circ  R_{\psi})^{(0)}[x] \subseteq R_{\zeta[\mathsf{RA}(\delta)/p]}^{(0)}[x] \right),
\]
 which is equivalent to the following pure inclusion of binary relations:
\begin{equation}
\label{eq final b boolean}
     R_{\mathsf{RA}(\phi)} \,\circ R_{\psi} \subseteq R_{\zeta[\mathsf{RA}(\delta)/p]}.
\end{equation}
Finally, notice that, if the modal reduction principle is also analytic, the shape of $\chi(p)\neq p$ (resp.\ $\zeta(p) \neq p$) simplifies, in case (a), to $\chi(p) = \phi_{n_\chi} \psi_{n_\chi}(p)$ with $n_\chi = 1$ and $\psi_{n_\chi}$ empty 
and, in case (b), to $\zeta(p) = \psi_{n_\zeta} \phi_{n_\zeta}(p)$ with $n_\zeta = 1$ and $\phi_{n_\zeta}$ empty. 
Hence, \eqref{eq final a boolean}  and \eqref{eq final b boolean}  simplify to the following inclusions, respectively:
\begin{equation}
\label{eq final analytic boolean}
   R_{\mathsf{LA}(\psi)} \, \circ R_{\phi}  \subseteq R_{\phi_{n_\chi}}\, \circ R_{\mathsf{LA}(\alpha)} \quad\quad    R_{\mathsf{RA}(\phi)} \,\circ R_{\psi}\subseteq R_{\psi_{n_\zeta}}\, \circ R_{\mathsf{RA}(\delta)}. 
\end{equation}

\begin{example}
\label{ex: p leq diamond box p boolean}
The modal reduction principle $p\leq \Diamond \Box p$  is inductive of shape (a), where $\phi (y): =  y$, and $\alpha (p): = p$, hence $\mathsf{LA}(\alpha)(u): = u$, and $\psi(x): =  x$, hence $\mathsf{LA}(\psi)(v): =  v$, and $\chi(p) := \Diamond \Box p$. Then, by Proposition \ref{prop:albaoutput}, it is equivalent to the following pure  inequality
\begin{equation*}
\forall \nomj\left([[\nomj/y]/v] \leq \Diamond\Box[[\nomj/u]/p]\right),
\end{equation*}
which can be rewritten as
\begin{equation*}
\label{eq: alba output a example 7}
\forall \nomj\left(\nomj \leq \Diamond \Box\nomj\right).
\end{equation*}
By Lemma \ref{lemma:molecular kripke}, this condition translates on Kripke frames as 
\[
\forall a  (\langle \Delta\rangle\{a\}  \subseteq \langle R_{\Diamond\Box} \rangle \{a\}), 
\]
i.e.\ (cf.\ Lemma \ref{lemma:properties of square bracket superscript} (6)) as
\[
\forall a  ( (\Delta^{[0]}[a])^c \subseteq (R_{\Diamond\Box }^{[0]}[a] )^c), 
\]
i.e., by contraposition, as 
\[
\forall a  (   R_{\Diamond\Box }^{[0]}[a]\subseteq \Delta^{[0]}[a] ). 
\]
By Definition \ref{def:kripke-based rels associated with terms} (3), this is equivalent to
\[
\forall a  (   (R_{\Diamond}\star R_{\Box }\star \Delta)^{[0]}[a]\subseteq \Delta^{[0]}[a] ), 
\]
which, by Lemma \ref{lemma:properties of square bracket superscript} (2) and contraposition,  is equivalent to the following inclusion of binary relations:
\[
\Delta \subseteq R_{\Diamond}  \star R_{\Box} \star \Delta.
\]
\end{example}
\begin{example}
\label{ex: diamond p leq box  p-kf}
The MRP $\Diamond p\leq \Box  p$  is inductive of shape (b), where $\phi (y): =  \Diamond y$, hence $\mathsf{RA}(\phi)(v): =  \blacksquare v$, and $\zeta (p): =  p$,  and $\psi(x): = \Box x$, and $\delta(p) :=   p$, hence $\mathsf{RA}(\delta)(u): =  u$. Then,
\smallskip

{{\centering
\begin{tabular}{rrlr}
 $\forall(\Diamond p\leq \Box  p)$ & 
iff & $\forall \cnomm\left( [[\cnomm/u]/p]\leq \blacksquare\Box [[\cnomm/y]/v] \right)$ & Proposition \ref{prop:albaoutput} \\
&iff & $\forall x \left( [\Delta](\{x\}^c) \subseteq [R_{\blacksquare} \circ R_{\Box}](\{x\}^c)\right),$ & Lemma \ref{lemma:molecular kripke} \\
&iff & $\forall x \left( \Delta^{[0]}[x] \subseteq (R_{\blacksquare}\circ  R_{\Box})^{[0]}[x]\right)$ & Lemma \ref{lemma:properties of square bracket superscript} (3)\\
&iff & $\forall x \left( (\Delta^{(0)}[x])^c \subseteq ((R_{\blacksquare}\circ  R_{\Box})^{(0)}[x])^c\right)$ & Lemma \ref{lemma:properties of square bracket superscript} (2) \\
&iff & $\forall x \left(  (R_{\blacksquare}\circ  R_{\Box})^{(0)}[x]\subseteq \Delta^{(0)}[x] \right)$ & \\
&iff & $R_{\blacksquare}\circ R_{\Box}\subseteq\Delta.$ &
\end{tabular}
\par}}
\end{example}
\begin{example}
\label{ex: p leq diamond box diamond p 2}
The modal reduction principle $p\leq \Diamond \Box \Diamond p$  is inductive of shape (a), where $\phi (y): =  y$, and $\alpha (p): = p$, hence $\mathsf{LA}(\alpha)(u): = u$, and $\psi(x): =  x$, hence $\mathsf{LA}(\psi)(v): =  v$, and $\chi(p) := \Diamond \Box\Diamond p$. 
\smallskip

{{\centering
	\begin{tabular}{rc ll}
		 $\forall p(p\leq \Diamond \Box \Diamond p)$ & 
		iff & $\forall \nomj\left([[\nomj/y]/v] \leq \Diamond\Box\Diamond [[\nomj/u]/p]\right)$ & Proposition \ref{prop:albaoutput} \\
		&iff &  $\forall \nomj\left(\nomj \leq \Diamond \Box\Diamond\nomj\right)$  &  \\
		&iff &  $\forall a \left( \langle \Delta \rangle \{a \} \subseteq \langle R_{\Diamond \Box\Diamond} \rangle \{a \}\right)$  & Lemma \ref{lemma:molecular kripke}  \\
		&iff &  $\forall a \left( (\Delta^{[0]}[a])^c  \subseteq (R^{[0]}_{\Diamond \Box\Diamond} [a])^c \right)$  &Lemma \ref{lemma:properties of square bracket superscript} (6)  \\
		&iff &  $\forall a \left(  R^{[0]}_{\Diamond \Box\Diamond} [a]  \subseteq  \Delta^{[0]}[a] \right)$ & contraposition \\
		&iff &  $\forall a \left(  (R_{\Diamond} \star R_{\Box} \star R_{\Diamond})^{[0]} [a]  \subseteq  \Delta^{[0]}[a] \right)$ &Definition \ref{def:kripke-based rels associated with terms} (3)  \\
		&iff &  $\Delta \subseteq R_{\Diamond} \star R_{\Box} \star R_{\Diamond}$ & Lemma \ref{lemma:properties of square bracket superscript} (2) and contraposition \\
	\end{tabular}
\par}}
\end{example}

\begin{example}
The modal reduction principle $\Diamond p\leq \Box\Diamond \Box p$  is inductive of shape (a), where $\phi (y): = \Diamond y$, and $\alpha (p): = p$, hence $\mathsf{LA}(\alpha)(u): = u$, and $\psi(x): = \Box x$, hence $\mathsf{LA}(\psi)(v): = \Diamondblack v$, and $\chi(p) := \Diamond \Box p$.
\smallskip

{{\centering
    \begin{tabular}{rc ll}
          $\forall p(\Diamond p\leq \Box\Diamond \Box p)$ & 
         iff & $\forall \nomj\left(\Diamondblack[\Diamond[\nomj/y]/v] \leq \Diamond\Box[[\nomj/u]/p]\right)$ & Proposition \ref{prop:albaoutput} \\
    & iff &  $\forall \nomj\left(\Diamondblack\Diamond\nomj \leq \Diamond \Box\nomj\right)$  & \\
     &i.e. & $\forall a\left(\langle R_{\Diamondblack\Diamond}\rangle\{a\}\subseteq \langle R_{\Diamond \Box}\rangle\{a\}\right)$ &  Lemma \ref{lemma:molecular kripke} \\
     &iff  & $\forall a\left(( R_{\Diamondblack\Diamond}^{[0]}[a])^c\subseteq  (R_{\Diamond \Box}^{[0]}[a])^c\right)$& Lemma \ref{lemma:properties of square bracket superscript} (6)  \\
     &iff  & $\forall a\left(R_{\Diamond \Box}^{[0]}[a]\subseteq R_{\Diamondblack\Diamond}^{[0]}[a]\right)$& contraposition \\
     &iff  & $\forall a\left((R_{\Diamond \Box}^{(0)}[a])^c\subseteq (R_{\Diamondblack\Diamond}^{(0)}[a])^c\right)$& Lemma \ref{lemma:properties of square bracket superscript} (2)  \\
     &iff  & $\forall a\left(R_{\Diamondblack\Diamond}^{(0)}[a]\subseteq R_{\Diamond \Box}^{(0)}[a]\right)$& contraposition  \\
     &iff  & $R_{\Diamondblack\Diamond}\subseteq R_{\Diamond \Box}$& definition of $(\cdot)^{(0)}$  \\
     &iff  & $R_{\Diamondblack}\circ R_{\Diamond}\subseteq R_{\Diamond}\star R_{ \Box}\star \Delta$& Definition \ref{def:kripke-based rels associated with terms}  \\
    \end{tabular}
\par}}

\end{example}

\begin{example}
The modal reduction principle $\Box\Diamond p \leq \Box \Diamond\Diamond p $  is inductive of shape (b) 
with $\phi(y): = y$, hence $\mathsf{RA}(\phi)(v): = v$ and $\psi (x): = \Box x$, and $\delta(p): = \Diamond\Diamond p$, hence $\mathsf{RA}(\delta)(u): = \blacksquare\blacksquare u$, and $\zeta(p): = \Box\Diamond p$.
\smallskip

{{\centering
			\begin{tabular}{r l l l}
				&$\forall p  \left(\Box\Diamond p \leq \Box \Diamond\Diamond p \right)$\\
            iff& $\forall \cnomm  \left(\Box\Diamond \blacksquare\blacksquare \cnomm \le \Box \cnomm\right)$
				
				& Proposition \ref{prop:albaoutput}\\
				i.e. &$\forall x\left ( [R_{\Box\Diamond \blacksquare\blacksquare}](\{x\}^c)\subseteq [R_{\Box}](\{x\}^c)\right)$ & Lemma \ref{lemma:molecular kripke} \\
iff	&$\forall x\left ( R_{\Box\Diamond \blacksquare\blacksquare}^{[0]}[x]\subseteq R_{\Box}^{[0]}[x]\right)$ & Lemma \ref{lemma:properties of square bracket superscript} (3) \\			iff	&$ R_{\Box} \subseteq R_{\Box\Diamond \blacksquare\blacksquare}$ & Lemma \ref{lemma:properties of square bracket superscript} (2), contraposition, and def. of $(\cdot)^{(0)}$ \\	
	iff&$ R_{\Box} \subseteq R_{\Box}\, \star R_{\Diamond}\, \star (R_{\blacksquare}\, \circ R_{\blacksquare}) $. & Definition \ref{def:kripke-based rels associated with terms}  \\
				\end{tabular}
		\par}}

\end{example}

\subsection{Connection via lifting}
\label{ssec: connection via lifting}
In the two previous subsections, we have shown that, both in the setting of polarity-based frames and in that of Kripke frames, the first-order correspondents of inductive modal reduction principles can be expressed as pure inclusions of (compositions and pseudo compositions of) binary relations associated with certain terms. Besides providing a more compact way to represent first-order sentences, 
this relational representation has also specific {\em correspondence-theoretic} consequences:   
it allows us to formulate and establish a systematic  connection between the first-order correspondents, in the two semantic settings, of any inductive modal reduction principle. Establishing this connection is the focus of the present subsection. Specifically, we will show that, for any inductive modal reduction principle $s(p)\leq t(p)$,  its first-order correspondent on polarity-based frames is  the ``lifted version" (in a sense that will be made precise below) of its first-order correspondent on Kripke frames. That is, thanks to the relational representation, the interesting phenomenon observed in Proposition 5 of  \cite{conradie2021rough}  can be generalized  (and made much more explicit in the process) to all inductive modal reduction principles. Finally, besides its technical benefits, this connection has also {\em conceptual} ramifications (which we will discuss in Section \ref{ssec:parametric corr mrp crisp}): for instance, it allows us to recognize when first-order sentences in the frame-correspondence languages of different structures encode the same ``modal content", and  to meaningfully transfer and represent well known relational properties such as reflexivity, transitivity, seriality, confluence, density, across different semantic contexts.

\medskip

For any Kripke $\mathcal{L}$-frame $\mathbb{X}$, any polarity-based $\mathcal{L}$-frame $\mathbb{F}$, every $\xi \in \mathsf{KRel}$ and any $\beta,\delta \in \mathsf{PRel}$ of types $\mathsf{T}_{A\times X}$ and $\mathsf{T}_{X \times A}$ respectively, we let $\xi^\mathbb{X}$ denote the interpretation of $\xi$ in $\mathbb{X}$, and $\beta^{\mathbb{F}}$ and $\delta^{\mathbb{F}}$ denote the interpretations of $\beta$ and $\delta$ in $\mathbb{F}$ respectively.

\begin{definition}
\label{def:ij lifting}
For any LE-language $\mathcal{L}$, let $\xi_1\subseteq \xi_2$   be a $\mathsf{KRel}_{\mathcal{L}}$-inequality, and $\beta_1\subseteq \beta_2$ (resp.~$\delta_1\subseteq \delta_2$)  be a $\mathsf{PRel}_{\mathcal{L}}$-inequality of type $\mathsf{T}_{A\times X}$ (resp.~$\mathsf{T}_{X\times A}$).  
\begin{enumerate}
\setlength{\itemsep}{0.2pt}
\setlength{\parskip}{0pt}
\setlength{\parsep}{0pt}
\item The inequality $\beta_1\subseteq \beta_2$ is the $I$-{\em lifting} of $\xi_1\subseteq \xi_2$ if for any Kripke $\mathcal{L}$-frame $\mathbb{X}$,
\[
\beta_1^\mathbb{F_X} = I_{(\xi_2^\mathbb{X})^c} 
\ \ \ \ \ \ \ \ \ 
\text{and}
\ \ \ \ \ \ \ \ \
\beta_2^\mathbb{F_X} = I_{(\xi_1^\mathbb{X})^c} 
\]
\item The inequality $\delta_1\subseteq \delta_2$ is the $J$-{\em lifting} of $\xi_1\subseteq \xi_2$ if for any Kripke $\mathcal{L}$-frame $\mathbb{X}$,
\[
\delta_1^\mathbb{F_X} = J_{(\xi_2^\mathbb{X})^c} 
\ \ \ \ \ \ \ \ \ 
\text{and}
\ \ \ \ \ \ \ \ \
\delta_2^\mathbb{F_X} = J_{(\xi_1^\mathbb{X})^c} 
\]
\end{enumerate}
\end{definition}

\begin{example}
The $\mathsf{PRel}_{\mathcal{L}}$-inequality $R_{\Box}\subseteq I$ is the $I$-lifting of the $\mathsf{KRel}_{\mathcal{L}}$-inequality $\Delta\subseteq R_{\Box}$ (representing reflexivity) on Kripke frames. Indeed, $\Delta\subseteq R_{\Box}$ is equivalent to $R_{\Box}^c\subseteq \Delta^c$ which is equivalent to $I_{R_{\Box}^c}\subseteq I_{\Delta^c}$, which instantiates the term inequality $R_\Box \subseteq I$ on all polarity-based frames which are lifted Kripke frames.

The $\mathsf{PRel}_{\mathcal{L}}$-inequality $R_{\Box}\subseteq R_{\Box}\, ;_I R_{\Box}$  is the $I$-lifting of the $\mathsf{KRel}_{\mathcal{L}}$-inequality $R_{\Box}\circ R_{\Box}\subseteq R_{\Box}$ (representing transitivity) on Kripke frames. Indeed, $R_{\Box}\circ R_{\Box}\subseteq R_{\Box}$ is equivalent to $R_{\Box}^c\subseteq (R_{\Box}\circ R_{\Box})^c$, which is equivalent to $I_{R_{\Box}^c}\subseteq I_{(R_{\Box}\circ R_{\Box})^c}$, which, by Lemma \ref{lemma:lifting composition}, is equivalent to $I_{R_{\Box}^c}\subseteq I_{R_{\Box}^c}\, ;_I I_{ R_{\Box}^c}$, which instantiates the term inequality $R_{\Box}\subseteq R_{\Box}\, ;_I R_{\Box}$ on all polarity-based frames which are lifted Kripke frames.
\end{example}
Table \ref{fig:MPS:Famous} lists some  well-known Sahlqvist modal reduction principles and their corresponding relational inequalities in $\mathsf{KRel}_{\mathcal{L}}$, and $\mathsf{PRel}_{\mathcal{L}}$. Since these principles are all analytic inductive, in Table \ref{fig:MPS:Famous}, we report both the relational inequalities resulting from treating them as inductive of shape (a), displayed in the shaded rows, and as inductive of shape (b), displayed in the unshaded rows. The $\mathsf{PRel}_{\mathcal{L}}$-inequality in each shaded (resp.~unshaded) row is the $J$-lifting (resp.~$I$-lifting) of the $\mathsf{KRel}_{\mathcal{L}}$-inequality in the same row.
\begin{table}[h]
\begin{center}
\begin{tabular}{|c|c|c|c|}
\hline
\textbf{Property} & \textbf{Modal reduction principle}&  \textbf{$\mathsf{KRel}_{\mathcal{L}}$} correspondent & \textbf{$\mathsf{PRel}_{\mathcal{L}}$} correspondent \\
\hline
\multirow{4}{*}{\textbf{Reflexivity}}  & \multirow{2}{*}{$p\leq \Diamond p$} &\cellcolor[HTML]{EFEFEF} $\Delta \subseteq
R_{\Diamond}$  &\cellcolor[HTML]{EFEFEF}  $R_{\Diamond} \subseteq J$  \\ \cline{3-4}
& & $\Delta\subseteq R_{\blacksquare}$ & $R_{\blacksquare}\subseteq I$\\ \cline{2-4}
& \multirow{2}{*}{$\Box p\leq p$ } &\cellcolor[HTML]{EFEFEF} $\Delta\subseteq R_{\Diamondblack}$ &\cellcolor[HTML]{EFEFEF} $R_{\Diamondblack}\subseteq J$\\ \cline{3-4}
& &$\Delta \subseteq R_{\Box}$  & $R_{\Box} \subseteq I$ \\ \hline
\multirow{4}{*}{\textbf{Transitivity}}  & \multirow{2}{*}{$\Diamond\Diamond p\leq \Diamond p$} &\cellcolor[HTML]{EFEFEF} $R_{\Diamond}\circ R_{\Diamond}\subseteq R_{\Diamond}$  &\cellcolor[HTML]{EFEFEF} $ R_{\Diamond}\subseteq R_{\Diamond}\, ;_I \, R_{\Diamond}$ \\ \cline{3-4} 
&&$R_{\blacksquare}\circ R_{\blacksquare}\subseteq R_{\blacksquare}$& $R_{\blacksquare}\subseteq R_{\blacksquare}\, ;_I \, R_{\blacksquare}$\\
\cline{2-4}
& \multirow{2}{*}{$ \Box p\leq \Box\Box p$} &\cellcolor[HTML]{EFEFEF} $R_{\Diamondblack}\circ R_{\Diamondblack}\subseteq R_{\Diamondblack}$&\cellcolor[HTML]{EFEFEF} $ R_{\Diamondblack}\subseteq R_{\Diamondblack}\, ;_I \, R_{\Diamondblack}$ \\ \cline{3-4} 
&& $R_{\Box}\circ R_{\Box}\subseteq R_{\Box}$  & $ R_{\Box}\subseteq R_{\Box}\, ;_I \, R_{\Box}$\\
\hline
\multirow{4}{*}{\textbf{Symmetry}} & \multirow{2}{*}{$p \leq \Box\Diamond p$} &\cellcolor[HTML]{EFEFEF} $R_\Diamondblack \subseteq R_\Diamond$&\cellcolor[HTML]{EFEFEF} $R_\Diamond \subseteq R_\Diamondblack$\\ \cline{3-4}  
&&$R_{\Box}\subseteq R_{\blacksquare}$ & $R_{\blacksquare}\subseteq R_{\Box}$\\
\cline{2-4}
&\multirow{2}{*}{$\Diamond \Box p \leq p$ }&\cellcolor[HTML]{EFEFEF} $R_{\Diamond}\subseteq R_{\Diamondblack}$ &\cellcolor[HTML]{EFEFEF} $R_{\Diamondblack}\subseteq R_{\Diamond}$ \\
\cline{3-4}  
&&$R_{\blacksquare}\subseteq R_{\Box}$ & $R_{\Box}\subseteq R_{\blacksquare}$\\
\hline
\multirow{2}{*}{\textbf{Seriality }} & \multirow{2}{*}{$\Box p \leq \Diamond p$} &\cellcolor[HTML]{EFEFEF} $\Delta \subseteq R_\Diamond \circ R_\Diamondblack$ &\cellcolor[HTML]{EFEFEF} $R_\Diamond ;_I R_\Diamondblack \subseteq J$ \\ \cline{3-4}
& & $\Delta\subseteq R_{\blacksquare}\circ R_{\Box}$ &  $ R_{\blacksquare}\, ;_I\, R_{\Box}\subseteq I$ 
 \\ 
\hline
\multirow{2}{*}{\textbf{Partial functionality}} &\multirow{2}{*}{$\Diamond p \leq \Box p$} &\cellcolor[HTML]{EFEFEF} $R_{\Diamondblack}\circ R_{\Diamond}\subseteq \Delta$ &\cellcolor[HTML]{EFEFEF} $J\subseteq R_{\Diamondblack}\, ;_I\, R_{\Diamond}$ \\ \cline{3-4}
& & $R_{\blacksquare}\circ R_{\Box}\subseteq \Delta$ &  $I\subseteq R_{\blacksquare}\, ;_I\, R_{\Box}$ \\ \hline
\multirow{4}{*}{\textbf{Euclideanness}} & \multirow{2}{*}{$\Diamond p \leq \Box\Diamond p$}
&\cellcolor[HTML]{EFEFEF}$R_\Diamondblack \circ R_\Diamond \subseteq R_\Diamond$ &\cellcolor[HTML]{EFEFEF}  $R_\Diamond \subseteq R_\Diamondblack\, ;_I\,    R_\Diamond$\\ 
\cline{3-4}  
&&$R_{\blacksquare}\circ R_{\Box}\subseteq R_{\blacksquare}$ & $ R_{\blacksquare} \subseteq R_{\blacksquare}\, ;_I\, R_{\Box}$\\
\cline{2-4}
&\multirow{2}{*}{$\Diamond\Box p\leq \Box p$} &\cellcolor[HTML]{EFEFEF}$R_\Diamondblack \circ R_\Diamond \subseteq R_\Diamondblack$ &\cellcolor[HTML]{EFEFEF}  $R_\Diamondblack \subseteq R_\Diamondblack\, ;_I\,    R_\Diamond$\\ 
\cline{3-4}  
&&$R_\blacksquare \circ R_\Box \subseteq R_\Box$ & $R_\Box \subseteq R_\blacksquare \circ R_\Box $\\
 \hline
\multirow{2}{*}{\textbf{Confluence}} & \multirow{2}{*}{$\Diamond\Box p\leq \Box\Diamond p$} &\cellcolor[HTML]{EFEFEF} $R_\Diamondblack \circ R_\Diamond \subseteq R_\Diamond \circ R_\Diamondblack$ &\cellcolor[HTML]{EFEFEF} $R_\Diamond \, ;_I\, R_\Diamondblack \subseteq R_\Diamondblack \, ;_I\, R_\Diamond $ \\ \cline{3-4}
& & $R_\blacksquare \circ R_\Box \subseteq R_\Box \circ R_\blacksquare$ & $R_\Box \, ;_I\, R_\blacksquare\subseteq R_\blacksquare \, ;_I\, R_\Box $ \\ \hline
\multirow{4}{*}{\textbf{Denseness}}  & \multirow{2}{*}{$\Diamond p\leq \Diamond\Diamond p$} &\cellcolor[HTML]{EFEFEF} $R_{\Diamond}\subseteq R_{\Diamond}\circ R_{\Diamond}$  &\cellcolor[HTML]{EFEFEF} $ R_{\Diamond}\, ;_I\, R_{\Diamond}\subseteq R_{\Diamond}$ \\ \cline{3-4} 
&&$R_{\blacksquare}\subseteq R_{\blacksquare}\circ R_{\blacksquare}$& $R_{\blacksquare}\, ;_I \, R_{\blacksquare}\subseteq R_{\blacksquare}$\\
\cline{2-4}
& \multirow{2}{*}{$ \Box \Box p\leq \Box p$} &\cellcolor[HTML]{EFEFEF} $R_{\Diamondblack}\subseteq R_{\Diamondblack}\circ R_{\Diamondblack}$  &\cellcolor[HTML]{EFEFEF} $  R_{\Diamondblack}\, ;_I \, R_{\Diamondblack}\subseteq R_{\Diamondblack}$ \\ \cline{3-4} 
&&$R_{\Box}\subseteq R_{\Box}\circ R_{\Box}$& $R_{\Box}\, ;_I \, R_{\Box}\subseteq R_{\Box}$\\
\hline
\end{tabular}
\end{center}
\caption{Well-known modal reduction principles and their correspondents as relational inequalities}
    \label{fig:MPS:Famous}
\end{table}

\begin{lemma}
\label{lemma:connection via lifting (finally)}
For any Kripke $\mathcal{L}$-frame $\mathbb{X}$, and formulas $\varphi, \psi, \chi, \zeta$ such as in Notation \ref{notation:phipsi},
\begin{center}
\begin{tabular}{rl c rl c rl c rl}
     1. & $R^{\mathbb{F_X}}_\varphi = J_{(R^{\mathbb{X}}_\varphi)^c}$ & \quad & 2. & $R^{\mathbb{F_X}}_\psi = J_{(R^{\mathbb{X}}_\psi)^c}$  & \quad &
     3. & $R^{\mathbb{F_X}}_\chi = I_{(R^{\mathbb{X}}_\chi)^c}$ & \quad & 4. & $R^{\mathbb{F_X}}_\zeta = I_{(R^{\mathbb{X}}_\zeta)^c}$
\end{tabular}
\end{center}
\end{lemma}
\begin{proof}
Items 1 and 2 are proved straightforwardly by induction on $\varphi$ and $\psi$ using Lemma \ref{lemma:lifting composition}.

Item 3 and item 4 are proved similarly, thus we only show the proof of item 3. We proceed by induction on the number of alternations of boxes and diamonds. If $\chi = x$, then $R^\mathbb{X}_\chi = \Delta^\mathbb{X}$, and $R^\mathbb{F_X}_\chi = I_{(\Delta^\mathbb{X})^c}$, which proves the statement. For the inductive case, $\chi\neq x$, and then $\chi = \varphi_1 \psi_1 \cdots \varphi_n \psi_n(!x)$ such that $n\geq 1$, and the $\varphi_i$ (resp.\ $\psi_i$) are finite, nonempty, concatenations of diamond (resp.\ box) operators, except for $\psi_n$ which is possibly empty. If $n=1$ and $\psi_1$ empty, then $R^\mathbb{X}_\chi = R^\mathbb{X}_{\varphi_1}$, hence the statement follows from item 1. If $n=1$ and $\psi_1(!z)$ is not empty, then $R^\mathbb{X}_\chi = R^\mathbb{X}_{\varphi_1} \star R^\mathbb{X}_{\psi_1} \star \Delta^\mathbb{X}$.
\smallskip

{{\centering
\begin{tabular}{ccc}
\begin{tabular}{rclr}
&& $J_{(R^\mathbb{X}_\chi)^c}$ & \\ & $=$ & $J_{(R^\mathbb{X}_{\varphi_1} \star R^\mathbb{X}_{\psi_1} \star \Delta^\mathbb{X})^c} $ & \\
& $=$ & $J_{(R^\mathbb{X}_{\varphi_1})^c} ; H_{(R^\mathbb{X}_{\psi_1} \star \Delta^\mathbb{X})^c} $ & Lemma \ref{Lemma: lifting heterogeneous relations commute} (5) \\
& $=$ & $J_{(R^\mathbb{X}_{\varphi_1})^c} ; I_{(R^\mathbb{X}_{\psi_1})^c} ; J_{(\Delta^\mathbb{X})^c} $ & Lemma \ref{Lemma: lifting heterogeneous relations commute} (1) 
\end{tabular}
     & \quad &
\begin{tabular}{rclr}
& $=$ & $J_{(R^\mathbb{X}_{\varphi_1})^c} ; I_{(R^\mathbb{X}_{\psi_1})^c} ; J_{(\Delta^\mathbb{X})^c} $ & Lemma \ref{Lemma: lifting heterogeneous relations commute} (1) \\
& $=$ & $R^\mathbb{F_X}_{\varphi_1} ; R^\mathbb{F_X}_{\psi_1} ; J^\mathbb{F_X} $ & items 1 and 2
\\
& $=$ & $(R_{\varphi_1} ; R_{\psi_1} ; J)^\mathbb{F_X} $ & 
\\
& $=$ & $R^\mathbb{F_X}_{\chi}$ & Definition \ref{def:pol-based rels associated with terms} (3)
\end{tabular}
\end{tabular}
\par}}

\noindent When $n>1$, $\chi = \varphi_1 \psi_1 \chi'(!x)$, where both $\varphi_1$ and $\psi_1$ are nonempty, $R_\chi = R_{\varphi_1} \star R_{\psi_1} \star R_{\chi'}$, and we can apply the induction hypothesis on $\chi'$. Hence:
\smallskip

{{\centering
\begin{tabular}{ccc}
\begin{tabular}{rclr}
$J_{(R^\mathbb{X}_\chi)^c}$ & $=$ & $J_{(R^\mathbb{X}_{\varphi_1} \star R^\mathbb{X}_{\psi_1} \star R_{\chi'}^\mathbb{X})^c} $ & \\
& $=$ & $J_{(R^\mathbb{X}_{\varphi_1})^c} ; H_{(R^\mathbb{X}_{\psi_1} \star R_{\chi'}^\mathbb{X})^c} $ & Lemma \ref{Lemma: lifting heterogeneous relations commute} (5) \\
& $=$ & $J_{(R^\mathbb{X}_{\varphi_1})^c} ; I_{(R^\mathbb{X}_{\psi_1})^c} ; J_{(R_{\chi'}^\mathbb{X})^c} $ & Lemma \ref{Lemma: lifting heterogeneous relations commute} (1) \\
& $=$ & $R^\mathbb{F_X}_{\varphi_1} ; R^\mathbb{F_X}_{\psi_1} ; J_{(R_{\chi'}^\mathbb{X})^c} $ & items 1 and 2
\end{tabular}
& &
\begin{tabular}{rclr}
& $=$ & $R^\mathbb{F_X}_{\varphi_1} ; R^\mathbb{F_X}_{\psi_1} ; J_{(R_{\chi'}^\mathbb{X})^c} $ & items 1 and 2
\\
& $=$ & $R^\mathbb{F_X}_{\varphi_1} ; R^\mathbb{F_X}_{\psi_1} ; R_{\chi'}^\mathbb{F_X} $ & inductive hyp.
\\
& $=$ & $(R_{\varphi_1} ; R_{\psi_1} ; R_{\chi'}^\mathbb{F_X})^\mathbb{F_X} $ & 
\\
& $=$ & $R^\mathbb{F_X}_{\chi}$ & Def. \ref{def:pol-based rels associated with terms} (3)
\end{tabular}
\end{tabular}
\par}}
\end{proof}

\begin{theorem}
\label{prop: sahlqvist lifting}
For any LE-language $\mathcal{L}$ and  every inductive modal reduction principle  $s(p)\leq t(p)$ of $\mathcal{L}$,
\begin{enumerate}
\setlength{\itemsep}{0.2pt}
\setlength{\parskip}{0pt}
\setlength{\parsep}{0pt}
\item if $s(p)\leq t(p)$ is of type (a), then  the $\mathsf{PRel}_{\mathcal{L}}$-inequality of type $\mathsf{T}_{X\times A}$ encoding its first-order correspondent  on polarity-based frames  is the $J$-lifting of the $\mathsf{KRel}_{\mathcal{L}}$-inequality encoding its first-order correspondent  on Kripke frames;
\item if $s(p)\leq t(p)$ is of type (b), then the $\mathsf{PRel}_{\mathcal{L}}$-inequality of type $\mathsf{T}_{A\times X}$ encoding its first-order correspondent  on polarity-based frames  is the $I$-lifting of the $\mathsf{KRel}_{\mathcal{L}}$-inequality encoding its first-order correspondent  on Kripke frames.
\end{enumerate}
\end{theorem}
\begin{proof}
We only show item 1, the proof of item 2 being similar. If   $s(p)\leq t(p)$ is of type (a), then it is of the form $\phi [\alpha (p)/!y]\leq \psi [\chi (p)/!x]$, and the interpretations of its first-order correspondents on a Kripke frame $\mathbb{X}$  and its corresponding polarity-based frames $\mathbb{F_X}$  are, respectively:

\begin{equation}
R^\mathbb{X}_{\mathsf{LA}(\psi)} \circ R^\mathbb{X}_{\phi} \subseteq R^\mathbb{X}_{\chi[\mathsf{LA}(\alpha)/p]}
\quad\quad\quad  
R^\mathbb{F_X}_{\chi[\mathsf{LA}(\alpha)/p]}\subseteq R^\mathbb{F_X}_{\mathsf{LA}(\psi)} \,;_I R^\mathbb{F_X}_{\phi}
\end{equation}
The first inclusion can be equivalently rewritten as $ (R^\mathbb{X}_{\chi[\mathsf{LA}(\alpha)/p]})^c\subseteq (R^\mathbb{X}_{\mathsf{LA}(\psi)} \circ R^\mathbb{X}_{\phi})^c$, which equivalently lifts as $J_{(R^\mathbb{X}_{\chi[\mathsf{LA}(\alpha)/p]})^c}\subseteq J_{(R^\mathbb{X}_{\mathsf{LA}(\psi)} \circ R^\mathbb{X}_{\phi})^c}$, which, by items 1 and 3 of Lemma \ref{lemma:connection via lifting (finally)}, is the required inequality
\[
R^\mathbb{F_X}_{\chi[\mathsf{LA}(\alpha)/p]}
\subseteq R^\mathbb{F_X}_{\mathsf{LA}(\psi)} ;_I R^\mathbb{F_X}_{\phi}.
\]
\end{proof}

\begin{example}
The MRP $p\leq \Diamond \Box p$  is inductive of shape (a), where $\phi (y): =  y$, and $\alpha (p): = p$, hence $\mathsf{LA}(\alpha)(u): = u$, and $\psi(x): =  x$, hence $\mathsf{LA}(\psi)(v): =  v$, and $\chi(p) := \Diamond \Box p$, and, as discussed in Example \ref{example: classical case}, its first-order correspondent on classical Kripke frames can be represented as the following $\mathsf{KRel}$-inequality:
\[\Delta  \subseteq R_{\Diamond}\star R_{\Box}\star\Delta, \]
while as discussed in Example \ref{ex: p leq diamond box p}, its first-order correspondent on polarity-based frames can be represented as the following term inequality of $\mathsf{PRel}$:
\[R_{\Diamond}\, ;R_{\Box}\, ; J\subseteq J.\]
Let us show that $R_{\Diamond}\, ;R_{\Box}\, ; J\subseteq J$ is the $J$-lifting of $\Delta  \subseteq R_{\Diamond}\star R_{\Box}\star\Delta$.
\smallskip

{{\centering
\begin{tabular}{r c l l}
   $\Delta  \subseteq R_{\Diamond}\star R_{\Box}\star\Delta$  
& iff &$\forall  a (R_{\Diamond}\star R_{\Box}\star\Delta)^{[0]}[a] \subseteq \Delta^{[0]}[a]   $\\
& iff &$\forall a (R_{\Diamond}^{[0]}[ R_{\Box}^{[0]}[\Delta^{[0]}[a]]]  \subseteq \Delta^{[0]}[a])$& Definition \ref{def: pseudo comp on kripke}\\
& iff &$\forall a (J_{R_\Diamond^c}^{(0)}[I_{R_\Box^c}^{(0)}[J_{\Delta^c}^{(0)}[a]]] \subseteq J_{\Delta^c}^{(0)}[a] )$  & Lemma \ref{lemma: lifting}\\
& iff &$\forall a( (J_{R_\Diamond^c};I_{R_\Box^c};J_{\Delta^c})^{(0)}[a] \subseteq  J_{\Delta^c}^{(0)}[a]  )$& Definition \ref{def:relational composition}\\
& iff &$J_{R_{\Diamond}^{c}}\, ; I_{R_{\Box}^{c}}\, ; J_{\Delta^{c}} \subseteq  J_{\Delta^{c}}$. & \\
\end{tabular}
\par}}
\end{example}
\begin{example}
\label{ex: diamond p leq box  p-lifting}
The modal reduction principle $\Diamond p\leq \Box  p$  is inductive of shape (b), where $\phi (y): =  \Diamond y$, hence $\mathsf{RA}(\phi)(v): =  \blacksquare v$, and $\zeta (p): =  p$,  and $\psi(x): = \Box x$, and $\delta(p) :=   p$, hence $\mathsf{RA}(\delta)(u): =  u$. As discussed in Example \ref{ex: diamond p leq box  p-kf}, its first-order correspondent on classical Kripke frames can be represented as the following term inequality of $\mathsf{KRel}$:
\[
 R_{\blacksquare}\circ R_{\Box}\subseteq\Delta, \]
while as discussed in Example \ref{ex: diamond p leq box  p}, its first-order correspondent on polarity-based frames can be represented as the following term inequality of $\mathsf{PRel}$:
\[
I\subseteq R_{\blacksquare}\, ;_I\, R_{\Box}.
\]
Let us show that $I\subseteq R_{\blacksquare}\, ;_I\, R_{\Box}$ is the $I$-lifting of $R_{\blacksquare}\circ R_{\Box}\subseteq\Delta$.
\begin{center}
\begin{tabular}{r c l l}
 $R_{\blacksquare}\circ R_{\Box}\subseteq\Delta$ 
& iff &$ \Delta^{c}\subseteq (R_{\blacksquare}\circ R_{\Box})^{c}$& \\
& iff &$ I_{\Delta^{c}}\subseteq I_{(R_{\blacksquare}\circ R_{\Box})^{c}}$& \\
& iff &$ I_{\Delta^{c}}\subseteq I_{R_{\blacksquare}^c}\, ;_I\, I_{ R_{\Box}^{c}}$& Lemma \ref{lemma:lifting composition}\\
\end{tabular}
\end{center}
\end{example}
\begin{example}
The modal reduction principle $p\leq \Diamond \Box\Diamond  p$  is inductive of shape (a), where $\phi (y): =  y$, and $\alpha (p): = p$, hence $\mathsf{LA}(\alpha)(u): = u$, and $\psi(x): =  x$, hence $\mathsf{LA}(\psi)(v): =  v$, and $\chi(p) := \Diamond \Box\Diamond  p$, and, as discussed in Example \ref{example: classical case}, its first-order correspondent on classical Kripke frames can be represented as the following term inequality of $\mathsf{KRel}$:
\[\Delta  \subseteq R_{\Diamond}\star R_{\Box}\star R_{\Diamond}, \]
while as discussed in Example \ref{ex: p leq diamond box diamond p}, its first-order correspondent on polarity-based frames can be represented as the following term inequality of $\mathsf{PRel}$:
\[R_{\Diamond}\, ;R_{\Box}\, ; R_{\Diamond}\subseteq J.\]
Let us show that $R_{\Diamond}\, ;R_{\Box}\, ; R_\Diamond\subseteq J$ is the $J$-lifting of $\Delta  \subseteq R_{\Diamond}\star R_{\Box}\star R_{\Diamond}$.
\smallskip

{{\centering
\begin{tabular}{r c l l}
  $\Delta  \subseteq R_{\Diamond}\star R_{\Box}\star R_{\Diamond}$  
& iff &$\forall  a (R_{\Diamond}\star R_{\Box}\star R_{\Diamond})^{[0]}[a] \subseteq \Delta^{[0]}[a]   $\\
& iff &$\forall a  (R_{\Diamond}^{[0]}[ R_{\Box}^{[0]}[R_{\Diamond}^{[0]}[a]]]  \subseteq \Delta^{[0]}[a])$& Definition \ref{def: pseudo comp on kripke}\\
& iff &$\forall a (J_{R_\Diamond^c}^{(0)}[I_{R_\Box^c}^{(0)}[J_{R_{\Diamond}^c}^{(0)}[a]]] \subseteq J_{\Delta^c}^{(0)}[a] )$  & Lemma \ref{lemma: lifting}\\
& iff &$\forall a( (J_{R_\Diamond^c};I_{R_\Box^c};J_{R_{\Diamond}^c})^{(0)}[a] \subseteq  J_{\Delta^c}^{(0)}[a]  )$& Definition \ref{def:relational composition}\\
& iff &$J_{R_{\Diamond}^{c}}\, ; I_{R_\Box^{c}}\, ; J_{R_{\Diamond}^{c}} \subseteq  J_{\Delta^{c}}$. & \\
\end{tabular}
\par}}
\end{example}

\subsection{Parametric correspondence for modal reduction principles}
\label{ssec:parametric corr mrp crisp}

We are now in a position to explain systematically some phenomena, observed for instance in \cite{conradie2016categories, TarkPaper2017, conradie2021rough}, regarding a number of modal reduction principles, especially when the modal operators were given a specific (e.g.\ epistemic) interpretations.
Firstly, thanks to the introduction of the relational languages $\mathsf{KRel}$ and $\mathsf{PRel}$, and the notion of $I$-lifting and $J$-lifting (cf.\ Definition \ref{def:ij lifting}), certain well known first-order conditions on Kripke frames (e.g.\ reflexivity, symmetry, transitivity, seriality, denseness, confluence) can be systematically associated with their {\em lifted counterparts} on polarity-based frames. Hence, we can now identify and speak of certain first-order conditions on polarity-based frames as e.g.\ $I$-reflexivity and $J$-transitivity, i.e.~as the {\em parametric versions} of 
the  conditions on Kripke frames of which they are the liftings.

Secondly, Theorem \ref{prop: sahlqvist lifting} establishes a systematic link between the {\em first-order  correspondents} of inductive modal reduction principles on Kripke frames and on polarity-based frames, precisely in terms of these liftings.
This result encompasses and provides a more systematic framework for a series of specific instances that were found and collected in \cite[Proposition 4.3]{conradie2021rough} in the context of the definition of  enriched formal contexts, and paves the way for a theory of {\em parametric correspondence}, in which the shift from one semantic context to another can be systematically tracked via the shift from one parameter to another. For example, as we saw above, the shift from Kripke frames to polarity-based frames is achieved by a change in parameter from the identity relation $\Delta$ to the relations $I$ and $J$. Yet other parametric shifts are seen in moving from the crisp semantics we have been discussing until now to the many-valued semantics which we will study in Section \ref{sec:MV}, where the parameter is provided by the choice of truth value algebra, and also in the move from Kripke frames to graph-based frames as we discuss in the conclusions.      

Thirdly,  Theorem \ref{prop: sahlqvist lifting} also paves the way to a better {\em conceptual} understanding of how the  interpretations of modal reduction principles in different contexts relate to one another. These  connections already cropped up in  \cite{conradie2021rough, conradie2020non}, in the context of the epistemic reading    of specific modal reduction principles, e.g.\ $\Box p \leq p$, and $\Box p \leq \Box\Box p$, but  Theorem \ref{prop: sahlqvist lifting} provides them with a more systematic underpinning. Let us illustrate with examples how these interpretations can be  connected. As is well known, in classical epistemic logic where $\Box\varphi$ is read as `the agent knows  that $\varphi$ is the case', the modal reduction principles above encode the factivity and positive introspection of knowledge, respectively. As discussed in  \cite{conradie2021rough, conradie2020non}, besides being respectively recognized as the $I$-liftings of reflexivity and transitivity on Kripke frames, the first-order sentences corresponding to these modal reduction principles on polarity-based frames express conditions which arguably reflect factivity and positive introspection when projected to the setting of formal concepts.  Indeed, if we interpret the polarity-based relation $R_{\Box}$ so that $a R_{\Box} x$ means `the agent attributes feature $x$ to object $a$', or in other words, `the agent recognizes $a$ as an $x$-object', then  $R_{\Box}\subseteq I$, which is the $I$-lifting of the reflexivity condition $\Delta\subseteq R_{\Box}$ on Kripke frames,  informally says that all the agent's feature-attributions also hold objectively, and are hence grounded in {\em facts}. Similarly, $R_{\Box} \subseteq R_{\Box};_I R_{\Box}$, which is the $I$-lifting of the transitivity condition $R_{\Box}\circ R_{\Box}\subseteq R_{\Box}$ on Kripke frames,  says that for any object $a$ and any feature $x$, if the agent recognizes $a$ as a member of the `actual' category of $x$-objects, then she  attributes to $a$ also all the features that, according to her, are shared by $x$-objects. So $R_{\Box} \subseteq R_{\Box};_I R_{\Box}$ arguably captures a `cognitive economy' property referred to as {\em default} in the categorization literature \cite{hsu2011typecasting}, which can be also understood as a property of an agent capable of computing the consequences (and hence being aware) of her own attributions.

Finally, under the classical epistemic logic interpretation, the modal reduction principle $p\leq \Box\Diamond p$ expresses the  condition that if a proposition holds, then the agent knows that it is conceivably true. As is well known, this axiom corresponds to the condition  $R_{\Diamondblack}\subseteq R_{\Diamond}$ on Kripke frames, 
expressing that the epistemic indiscernibility relation is symmetric. In other words, every epistemic state is indistinguishable from every epistemic state which is indistinguishable from it.   
Reading $p\leq \Box\Diamond p$ as an axiom of the  epistemic logic of categories, and $x R_{\Diamond} a$ as `the agent recognizes $x$ as an $a$-property', we can argue that, just like $\Box p$ and $\Diamond p$ provide two different viewpoints on an agent's  epistemic stance towards facts, (i.e.~knowledge and conceivability),  $\Box p$ and $\Diamond p$  provide two different viewpoints on an agent's  epistemic stance towards categories:    {\em extensionally oriented}, or object-based,  and  {\em intensionally oriented}, feature-based, respectively. We propose the intuitive reading of $\Box \varphi$ (resp.~$\Diamond\varphi$) as `category $\varphi$ as concretely (resp.~abstractly) understood by the agent'. Under this reading, $p\leq \Box\Diamond p$
 expresses the  condition that every category is a subcategory of the concretization of its abstraction  according to the agent.
The first-order correspondent of  
$p\leq \Box\Diamond p$ on polarity-based frames is $R_{\Diamond}\subseteq R_{\Diamondblack}$,  
which says that for any object $a$ and feature $x$, if the agent recognizes $x$ as an $a$-property, then she recognizes $a$ as an $x$-object.



\section{Many-valued correspondents of inductive MRPs}
\label{sec:MV}
In this section, we show that, for any arbitrary but fixed 
complete frame-distributive and dually frame-distributive Heyting algebra $\mathbf{A} = (D, 1, 0, \vee, \wedge, \to)$, the first-order correspondent on polarity-based  $\mathbf{A}$-frames of any inductive modal reduction principles is ``verbatim the same'' as its first-order correspondent on crisp polarity-based  frames. As we have seen it was the case in the crisp setting discussed in the previous section, key to being able to formulate and establish this result is the representation of the first-order correspondents on polarity-based  $\mathbf{A}$-frames as  pure inclusions of binary $\mathbf{A}$-relations (cf.~Theorem \ref{thm:verbatim crisp mv}).

\subsection{$I$-compatible $\mathbf{A}$-relations}
\label{ssec:icompatible arel}

\begin{lemma}\label{lemma:equivalents of I-compatible mv}
For any $\mathbf{A}$-polarity $(A, X, I)$, any relations $R: A\times X \to \mathbf{A}$, $T: X \times A \to \mathbf{A}$, any $\alpha \in \mathbf{A}$,

\hspace{-1cm}
\begin{tabular}{c}
	\begin{tabular}{l}
	   1.  The following are equivalent: \\
		 \phantom{1. } a. $R^{(0)}[\{\alpha \slash x\}]:A \to \mathbf{A}$ is Galois-stable for all $x\in X$; \\
		 \phantom{1. } b. $R^{(0)} [u]$ is Galois-stable for all $u:X\to \mathbf{A}$; \\
		 \phantom{1. } c. $R^{(1)}[h]=R^{(1)}[h^{\uparrow\downarrow}]$ for all  $h: A\to \mathbf{A}$.
	\end{tabular}
	\begin{tabular}{l}
	    2.  The following are equivalent: \\
		 \phantom{2. } a. $T^{(1)}[\{\alpha \slash a\}]: X \to \mathbf{A}$ is Galois-stable for all $a\in A$; \\
		 \phantom{2. } b. $T^{(1)} [h]$ is Galois-stable for all $h:A \to \mathbf{A}$; \\
		 \phantom{2. } c. $T^{(0)}[u]=T^{(0)}[u^{\downarrow\uparrow}]$ for all $u: X \to \mathbf{A}$.
	\end{tabular}
\end{tabular}
\end{lemma}

\begin{proof} We only prove item 1, the proof of item 2 being similar. For $(a)\Rightarrow (b)$, since $u = \bigvee_{x\in X}\{u(x) \slash x\}$, then $R^{(0)}[u] = R^{(0)}[\bigvee_{x\in X}\{u(x) \slash x\}]$, hence for every $a\in A$,
\begin{center}
\begin{tabular}{rrlr}
 $R^{(0)}[\bigvee_{x\in X}\{u(x) \slash x\}](a)$ &  
$=$ & $\bigwedge_{y \in X}\left( (\bigvee_{x\in X}\{u(x) \slash x\})(y) \rightarrow R(a,y) \right)$ & definition of $R^{(0)}[\cdot]$ \\
&$=$ & $\bigwedge_{y \in X}\left( \bigvee_{x\in X}(\{u(x) \slash x\}(y)) \rightarrow R(a,y) \right)$ & definition of $\{\cdot \slash \cdot \}$ \\
&$=$ & $\bigwedge_{y \in X}\bigwedge_{x\in X}\left( \{u(x) \slash x\}(y) \rightarrow R(a,y) \right)$ & basic property of Heyting $\rightarrow$\\
&$=$ & $\bigwedge_{x \in X}\bigwedge_{y\in X}\left( \{u(x) \slash x\}(y) \rightarrow R(a,y) \right)$ & \\
&$=$ & $\bigwedge_{x \in X} R^{(0)}[\{u(x) \slash x \}](a)$. & definition of $R^{(0)}[\cdot]$\\
\end{tabular}
\end{center}
Then the statement follows from Lemma \ref{lemma:galois stable is closed under intersection mv}.
The converse direction is immediate.
	
	$(a)\Rightarrow (c)$. Since $(\cdot)^{\uparrow\downarrow}$ is a closure operator, $u\leq u^{\uparrow\downarrow}$. Hence, Lemma \ref{lemma: basic mv} (1) implies that $R^{(1)}[u^{\uparrow\downarrow}]\leq R^{(1)}[u]$. For the converse inclusion, since $\mathbf{A}^X$ is join generated by the elements $\{\alpha \slash x \}$ for any $\alpha \in \mathbf{A}$ and $x \in X$, it is enough to show that, for any $\alpha \in \mathbf{A}$ and $x \in X$, if $\{\alpha \slash x \} \leq R^{(1)}[h]$, then $\{\alpha \slash x \} \leq R^{(1)}[h^{\uparrow\downarrow}]$.
	\begin{center}
	\begin{tabular}{rrlr}
	 $\{\alpha \slash x \} \leq R^{(1)}[h]$ & 
	iff & $h \leq R^{(0)}[\{\alpha \slash x \}]$ & Galois connection\\
&	iff & $h^{\uparrow\downarrow} \leq R^{(0)}[\{\alpha \slash x \}]^{\uparrow\downarrow}$ & $(\cdot)^{\uparrow\downarrow}$ closure operator\\
&	iff & $h^{\uparrow\downarrow} \leq R^{(0)}[\{\alpha \slash x \}]$ & assumption {\em (i)} \\
&	iff & $\{\alpha \slash x \} \leq R^{(1)}[h^{\uparrow\downarrow}]$. & Galois connection \\
	\end{tabular}
	\end{center}
	
	$(c)\Rightarrow (a)$. It is enough to show that $(R^{(0)}[\{\alpha \slash x\}])^{\uparrow\downarrow}\leq R^{(0)}[\{\alpha \slash x\}]$ for any $x\in X$ and $\alpha \in \mathbf{A}$.  By Lemma \ref{lemma: basic mv} (2), $R^{(0)}[\{\alpha \slash x\}]\leq R^{(0)}[\{\alpha \slash x\}]$ is equivalent to $\{\alpha \slash x\}\leq R^{(1)}[R^{(0)}[\{\alpha \slash x\}]]$. By assumption, $R^{(1)}[R^{(0)}[\{\alpha \slash x\}]]=R^{(1)}[(R^{(0)}[\{\alpha \slash x\}])^{\uparrow\downarrow}]$, hence $\{\alpha \slash x\}\leq R^{(1)}[(R^{(0)}[\{\alpha \slash x\}])^{\uparrow\downarrow}]$. Again by Lemma \ref{lemma: basic mv} (2), this is equivalent to $(R^{(0)}[\{\alpha \slash x\}])^{\uparrow\downarrow}\leq R^{(0)}[\{\alpha \slash x\}]$, as required.	
\end{proof}
Clearly, an analogous characterization of $I$-compatible relations of type $X \times A$ can be given. However, we omit the statement of this lemma and refer to the lemma above for all types of polarity-based relations.

\begin{lemma} \label{lemma: many valued icompatibility meaning}
For any $\mathbf{A}$-polarity $(A,X,I)$, any $I$-compatible $\mathbf{A}$-relations $R: X \times A \to \mathbf{A}$ and $S: A \times X \to \mathbf{A}$, and any $a \in A$ and $x \in X$,
\smallskip

\begin{tabular}{rl}
1. & $x R a  =\bigwedge_{b \in A}( \bigwedge_{y \in X}( y R a \rightarrow b I y)\rightarrow b I x) = \bigwedge_{b \in A}( \bigwedge_{y \in X}( a I y \rightarrow b I y)\rightarrow b R x)$ \\ 
2. & $x R a  =\bigwedge_{y \in X}( \bigwedge_{b \in A}( y R a \rightarrow b I y)\rightarrow b I x) = \bigwedge_{y \in X}( \bigwedge_{b \in A}( a I y \rightarrow b I y)\rightarrow b R x)$\\
3. & $a S x  =\bigwedge_{y \in X}( \bigwedge_{b \in A}( b S x \rightarrow b I y)\rightarrow a I y)= \bigwedge_{y \in X}( \bigwedge_{b \in A}( b I x \rightarrow b I y)\rightarrow a S y)$ \\
4. & $a S x  =\bigwedge_{b \in A}( \bigwedge_{y \in X}( b S x \rightarrow b I y)\rightarrow a I y) = \bigwedge_{b \in A}( \bigwedge_{y \in X}( b I x \rightarrow b I y)\rightarrow a S y)$
\end{tabular}
\end{lemma}
  \begin{proof}
 We only prove 1, the proof of the remaining items being similar.
 \smallskip
 
 {{\centering
 \begin{tabular}{rclr}
      $xRa$ & $=$ & $1 \rightarrow xRa$ & Heyting algebra tautology \\
      & $=$ & $\bigwedge_{b\in A}(\{1/a\}(b)\rightarrow bRx)$ \\
      & $=$ & $R^{(0)}[\{1 \slash a \}](x)$ & basic definitions (cf.\ Section \ref{ssec:MV-polarities}) \\
      & $=$ & $I^{(1)}[I^{(0)}[R^{(0)}[\{1 \slash a \}]]](x)$ & $I$-compatibility of $R$ \\
      & $=$ & $\bigwedge_{b \in A}( (\bigwedge_{y \in X}( y R a \rightarrow b I y))\rightarrow b I x)$ & basic definitions (cf.\ Section \ref{ssec:MV-polarities})
 \end{tabular}    
 \par}}
 \smallskip
 
 \noindent As to the second identity,
 \smallskip
 
 {{\centering
 \begin{tabular}{rclr}
      $xRa$ & $=$ & $1 \rightarrow xRa$ & Heyting algebra tautology \\
      & $=$ & $R^{(0)}[\{1 \slash a \}](x)$ & basic definitions (cf.\ Section \ref{ssec:MV-polarities}) \\
      & $=$ & $R^{(0)}[I^{(0)}[I^{(1)}[\{1 \slash a \}]]](x)$ & Lemma \ref{lemma:equivalents of I-compatible mv} (2)({\em  iii})\\
      & $=$ & $\bigwedge_{b \in A}( \bigwedge_{y \in X}( a I y \rightarrow b I y)\rightarrow x R b)$ & basic definitions (cf.\ Section \ref{ssec:MV-polarities})
 \end{tabular}    
 \par}}
\end{proof}

\subsection{Composing many-valued relations on many-valued polarities}
\label{sec: composing mv}
Let us introduce  the many-valued versions of the various relational compositions presented in Section \ref{sec: composing crisp}
in the setting of polarity-based frames. The definitions and notation we introduce here will be key to establish a meaningful comparison of the first-order correspondents of inductive modal reduction principles on crisp and many-valued polarity-based frames.

\begin{notation}
\label{notation:mv_type_relation}
In what follows, we introduce  similar conventions as those introduced in Notation \ref{notation:crisp_type_relation}. Namely, for any formal $\mathbf{A}$-context $\mathbb{P}= (A,X,I)$, we will use the capital letter $B$, possibly with sub- or superscripts, for $\mathbf{A}$-relations of type $A \times X \rightarrow \mathbf{A}$; likewise, relations of type $X \times A \rightarrow \mathbf{A}$ will be denoted by $D$, of type $X \times X \rightarrow \mathbf{A}$ by $L$, and of type $A \times A \rightarrow \mathbf{A}$ by $M$. Summing up,
\smallskip

 {{\centering $B: A \times X \rightarrow \mathbf{A} \quad$ $\quad D : X \times A\rightarrow \mathbf{A}\quad$ $\quad L : X \times X\rightarrow \mathbf{A}\quad$ $\quad M : A \times A\rightarrow \mathbf{A}$, 
 \par}}

\end{notation}

\begin{definition}
\label{def:relational composition mv}
	For any formal $\mathbf{A}$-context $\mathbb{P}= (A,X,I)$, we define the following notions of relational composition:  for all $a,b \in A$ and $x,y \in X$,
	\[
	\begin{array}{rclr}
    	B_1 \,;_I B_2 : A\times X \rightarrow \mathbf{A} &\quad& a (B_1\, ;_IB_2) x  = \bigwedge_{y\in X}\left (\bigwedge_{b\in A}(bB_2x\rightarrow bIy)\rightarrow aB_1y \right) & (1) \\ 
        D_1 \,;_I D_2 : X \times A \rightarrow \mathbf{A} && x (D_1\, ;_ID_2) a  = \bigwedge_{b\in A}\left (\bigwedge_{y\in X}(yD_2a\rightarrow bIy)\rightarrow xD_1b \right) & (2)    \\
        B\,;D : A \times A \rightarrow \mathbf{A} &&
        a (B\, ;D) b  = \bigwedge_{x\in X}\left (xDb\rightarrow aBx \right) & (3) \\
        D\,;B : X \times X \rightarrow \mathbf{A} && x (D\, ;B) y  = \bigwedge_{a\in A}\left (aBy\rightarrow xDa \right) & (4)\\
        M\,;B: A \times X \rightarrow \mathbf{A} && a (M\, ;B) x  = \bigwedge_{b\in A}\left (bBx\rightarrow aMb \right) & (5) \\
        D\, ;M: X \times A \rightarrow \mathbf{A} && x (D\, ;M) a  = \bigwedge_{b\in A}\left (bMa\rightarrow xDb \right) & (6) \\ 
        L\, ;D: X \times A \rightarrow \mathbf{A} && x (L\, ;D) a  = \bigwedge_{y\in X}\left (yDa\rightarrow xLy \right) & (7) \\
        B\, ;L : A \times X \rightarrow \mathbf{A} && a (B\, ;L) x  = \bigwedge_{y\in X}\left (yLx\rightarrow aBy \right). & \quad\quad\quad(8)
	\end{array}
	\]
\end{definition}
\begin{remark}
The clauses above translate the defining first-order clauses of the various relational compositions in the crisp setting into the algebraic language of $\mathbf{A}$. For instance, in the crisp setting, for all  $R, T\subseteq X\times A$ and all $a\in A$ and $x\in X$, 
\begin{center}
\begin{tabular}{r c l l}
$x(R\, ;_I T)a$ & iff & $x\in R^{(0)}[I^{(0)}[T^{(0)}[a]]]$\\
& iff & $I^{(0)}[T^{(0)}[a]]\subseteq R^{(1)}[x]$ & adjunction\\
& iff & $\forall b(b\in I^{(0)}[T^{(0)}[a]] \Rightarrow b\in R^{(1)}[x])$\\
& iff & $\forall b(T^{(0)}[a]\subseteq I^{(1)}[b] \Rightarrow b\in R^{(1)}[x])$ & adjunction\\
& iff & $\forall b(\forall y(y\in T^{(0)}[a]\Rightarrow y\in I^{(1)}[b] ) \Rightarrow b\in R^{(1)}[x])$\\
& iff & $\forall b(\forall y(yTa\Rightarrow bI y )\Rightarrow xRb).$\\
\end{tabular}
\end{center}
\end{remark}

\begin{lemma}
\label{lem:fuz:equiv:I:coposition}
	For any $\mathbf{A}$-polarity $(A,X,I)$, all relations $R,T : A \times X \to \mathbb{A}$, $S,U: X \times A \to \mathbb{A}$, any $a \in A$, $x \in X$, \\[1mm]
	\begin{tabular}{rl}
	    1. & $a (R\, ;_I T) x  = R^{(0)}[I^{(1)}[T^{(0)} [\{1\slash x\}]](a)= T^{(1)}[I^{(0)}[R^{(1)} [\{1\slash a\}]]](x);$ \\
	    2. & $(S\, ;_I U) a = S^{(0)}[I^{(0)}[U^{(0)} [\{1\slash a\}]]](x)= U^{(1)}[I^{(1)}[S^{(1)} [\{1\slash x\}]]](a).$
	\end{tabular}
\end{lemma}
\begin{proof}
We only show item 1, the proof of 2 being similar. As to the first identity from the left, 
\begin{center}
    \begin{tabular}{r cl}
     $R^{(0)}[I^{(1)}[T^{(0)} [\{1\slash x\}]]](a)$    & $=$ & $\bigwedge_{y\in X}(I^{(1)}[T^{(0)} [\{1\slash x\}]](y)\rightarrow aRy)$\\
     & $=$ & $\bigwedge_{y\in X}(\bigwedge_{b\in A}(T^{(0)} [\{1\slash x\}](b)\rightarrow bIy)\rightarrow aRy)$\\
     & $=$ & $\bigwedge_{b\in A}(\bigwedge_{y\in X}(\bigwedge_{z\in X}(\{1/x\}(z)\rightarrow bTz)\rightarrow bIy)\rightarrow aRy)$\\
         & $=$ & $\bigwedge_{y\in X}(\bigwedge_{b\in A}((1\rightarrow bTx)\rightarrow bIy)\rightarrow aRy)$\\
         & $=$ & $\bigwedge_{y\in X}(\bigwedge_{b\in A}(bTx\rightarrow bIy)\rightarrow aRy)$\\
         & $=$ & $a (R\, ;_IT) x$.\\
    \end{tabular}
\end{center}
Spelling out the definitions, the second identity can be rewritten as follows:
$$\bigwedge_{y\in X}((\bigwedge_{b\in A}(bTx \to bIy))\to aRy)= \bigwedge_{b\in A}((\bigwedge_{y\in X}(aRy \to bIy))\to bTx).$$
As to showing that
$$\bigwedge_{y\in X}((\bigwedge_{b\in A}(bTx \to bIy))\to aRy)\leq \bigwedge_{b\in A}((\bigwedge_{y\in X}(aRy \to bIy))\to bTx),$$  equivalently, let us show that for all $c\in A$,
$$\bigwedge_{y\in X}((\bigwedge_{b\in A}(bTx \to bIy))\to aRy)\leq (\bigwedge_{y\in X}(aRy \to cIy))\to cTx.$$

\begin{center}
 \begin{tabular}{clr}
  & $\bigwedge_{y\in X}((\bigwedge_{b\in A}(bT x\to bIy)\to aRy) ) \leq \bigwedge_{y\in Y}(aRy\to cIy)\to cTx$  & \\
  iff & $\bigwedge_{y\in X}((\bigwedge_{b\in A}(bT x\to bIy)\to aRy) )\land\bigwedge_{y\in Y}(aRy\to cIy) \leq cTx$  & residuation\\
  iff & $\bigwedge_{y\in X}((\bigwedge_{b\in A}(bT x\to bIy)\to aRy)\land (aRy\to cIy)) \leq cTx$ &   \\
 
    if  & $\bigwedge_{y\in X}((\bigwedge_{b\in A}(bT x\to bIy)\to cIy))\leq cTx$,  & $(\alpha\to \beta)\land(\beta\to\gamma)\leq \alpha\to \gamma$ \\
    \end{tabular}    
 \end{center}
 and the last inequality holds  by Lemma \ref{lemma: many valued icompatibility meaning}. To show the converse inequality
$$\bigwedge_{b\in A}((\bigwedge_{y\in X}(aRy \to bIy))\to bTx) \leq \bigwedge_{y\in X}((\bigwedge_{b\in A}(bTx \to bIy))\to aRy)$$ we work similarly and show that, for all $z\in X$,
$$\bigwedge_{b\in A}((\bigwedge_{y\in X}(aRy \to bIy))\to bTx) \leq (\bigwedge_{b\in A}(bTx \to bIz))\to aRz.$$

\begin{center}
 \begin{tabular}{clr}
  & $\bigwedge_{b\in A}((\bigwedge_{y\in X}(aR y\to bIy)\to bTx) ) \leq \bigwedge_{b\in A}(bTx\to bIz)\to aRz$  & \\
  iff &$\bigwedge_{b\in A}((\bigwedge_{y\in X}(aR y\to bIy)\to bTx) )\land\bigwedge_{b\in A}(bTx\to bIz) \leq aRz$ & residuation \\
  iff & $\bigwedge_{b\in A}((\bigwedge_{y\in X}(aR y\to bIy)\to bTx)\land (bTx\to bIz)) \leq aRz$ & \\
  
    if  & $\bigwedge_{b\in A}((\bigwedge_{y\in X}(aR y\to bIy)\to bIz))\leq aRz$, & $(\alpha\to \beta)\land(\beta\to\gamma)\leq \alpha\to \gamma$  \\
 \end{tabular}    
 \end{center}
 and the last inequality holds by Lemma \ref{lemma: many valued icompatibility meaning}.
This concludes the proof.
\end{proof}

The following lemma provides a simpler definition of $I$-compatibility for $\mathbf{A}$-relations.

\begin{lemma}\label{lem:easier definition of mv icomp}
For any $\mathbf{A}$-polarity $(A, X, I)$ and any order type $\epsilon$ on $n\in\mathbb{N}^{>0}$,
\begin{enumerate}
        \setlength{\itemsep}{0.2pt}
        \setlength{\parskip}{0pt}
        \setlength{\parsep}{0pt}
		\item If $R: X\times A^\epsilon\to \mathbf{A}$, then $R^{(0)}[\overline{\{1\slash a\}}]$ (resp.\ $R^{(i)}[\{1\slash x\},\overline{\{1\slash a\}}_i]$) is Galois-stable if and only if   $R^{(0)}[\overline{\{\alpha\slash a\}}]$ (resp.\ $R^{(i)}[\{\alpha_0\slash x\},\overline{\{\alpha\slash a\}}_i]$) is Galois-stable for every $\alpha_0,\overline{\alpha}$.
		\item If $R:A\times X^\epsilon \to \mathbf{A}$, then $R^{(0)}[\overline{\{1\slash x\}}]$ (resp.\ $R^{(i)}[\{1\slash x\},\overline{\{1\slash a\}}_i]$) is Galois-stable if and only if   $R^{(0)}[\overline{\{\alpha\slash a\}}]$ (resp.\ $R^{(i)}[\{\alpha_0\slash x\},\overline{\{\alpha\slash a\}}_i]$) is Galois-stable for every $\alpha_0,\overline{\alpha}$.
	\end{enumerate}
\end{lemma}
\begin{proof}
We only show item 1, item 2 is shown similarly. Let us show the non-trivial direction:
\begin{center}
    \begin{tabular}{r c l r}
    $R^{(0)}[\overline{\{\alpha\slash a\}}](x)$ & $=$ & $(\bigwedge_{1\leq i\leq n}\alpha_i)\to xR\overline{a}$ & \\
    & $=$ & $((\bigwedge_{1\leq i\leq n}\alpha_i)\to R^{(0)}[\overline{\{1\slash a\}}])(x)$ & \\
    & $=$ & $((\bigwedge_{1\leq i\leq n}\alpha_i)\to I^{(1)}[I^{(0)}[R^{(0)}[\overline{\{1\slash a\}}]]])(x)$ & by assumption\\
    & $=$ & $ I^{(1)}[(\bigwedge_{1\leq i\leq n}\alpha_i)\land I^{(0)}[R^{(0)}[\overline{\{1\slash a\}}]]](x)$ & by Lemma \ref{lem:keytouniversemysteries}\\
    & $=$ & $ I^{(1)}[I^{(0)}[I^{(1)}[(\bigwedge_{1\leq i\leq n}\alpha_i)\land I^{(0)}[R^{(0)}[\overline{\{1\slash a\}}]]]]](x)$ & $I^{(1)}[h]$ is Galois-stable\\
    & $=$ & $ I^{(1)}[I^{(0)}[(\bigwedge_{1\leq i\leq n}\alpha_i)\to I^{(1)}[ I^{(0)}[R^{(0)}[\overline{\{1\slash a\}}]]]]](x)$ & by Lemma \ref{lem:keytouniversemysteries}\\
    & $=$ & $ I^{(1)}[I^{(0)}[(\bigwedge_{1\leq i\leq n}\alpha_i)\to R^{(0)}[\overline{\{1\slash a\}}]]](x)$ & by assumption\\
    & $=$ & $ I^{(1)}[I^{(0)}[R^{(0)}[\overline{\alpha\land\{1\slash a\}}]]](x)$ & by Lemma \ref{lem:keytouniversemysteries}\\
    & $=$ & $ I^{(1)}[I^{(0)}[R^{(0)}[\overline{\{\alpha\slash a\}}]]](x)$. &
    \end{tabular}
\end{center}
\end{proof}
\begin{proposition} \label{Composition is compatible}
For any  $(A, X, I)$  $\mathbf{A}$-polarity, $B_1\,;_I B_2$ and $D_1\,;_I D_2$ are  $I$-compatible  if $B_1, B_2, D_1, D_2$ are.
\end{proposition}
\begin{proof}
We only prove item 1, the proof of item 2 being similar.

By Lemma \ref{lem:easier definition of mv icomp}, it is enough to show that $(R\,;_I T)^{(0)}[\{1\slash a\}]$ and $(R\,;_IT)^{(1)}[\{1\slash x\}]$ are Galois-stable. By Lemma \ref{lem:fuz:equiv:I:coposition}, $(R\,;_I T)^{(0)}[\{1\slash a\}]=R^{(0)}[I^{(0)}[T^{(0)}[\{1\slash a\}]]]$ and $(R\,;_IT)^{(1)}[\{1\slash x\}]=T^{1}[I^{(1)}[R^{(1)}[\{1\slash x\}]]]$, which are both Galois-stable by the $I$-compatibility of $R$ and $T$ respectively.
\end{proof}

For any $\mathbf{A}$-polarity $\mathbb{P}= (A,X,I)$ and any $R_{\Box}: A\times X\to \mathbf{A}$ and $R_{\Diamond}: X\times A\to \mathbf{A}$, let $R_{\Diamondblack}: X\times A\to \mathbf{A}$ and $R_{\blacksquare}: A\times X\to \mathbf{A}$ be respectively defined as $x R_{\Diamondblack} a = a R_{\Box} x$ and  $a R_{\blacksquare} x = x R_{\Diamond}a$ for any $a\in A$ and  $x\in X$.
The following lemma is an immediate consequence of the definitions above.
\begin{lemma}
\label{lemma:elementary property adjoints mv}
For any $\mathbf{A}$-polarity $\mathbb{P}= (A,X,I)$ and any $R_{\Box}: A\times X\to \mathbf{A}$ and $R_{\Diamond}: X\times A\to \mathbf{A}$, any $a\in A$ and  $x\in X$, and any $f: A \to \mathbf{A}$ and $u: X \to \mathbf{A}$,
   \smallskip
   
   {{\centering 
   $R_{\Box}^{(1)}[f] = R_{\Diamondblack}^{(0)}[f]\quad \text{ and }\quad R_{\Diamond}^{(1)}[u] = R_{\blacksquare}^{(0)}[u].$
   \par}}
\end{lemma}
\begin{proof}
%
We only show the first identity, the proof of the second one being similar.
\smallskip

{{\centering
$R_{\Box}^{(1)}[f](x) = \bigwedge_{b\in A}(f(b)\to bR_{\Box}x) = \bigwedge_{b\in A}(f(b)\to xR_{\Diamondblack}b) =  R_{\Diamondblack}^{(0)}[f](x).$
\par}}
\end{proof}

\begin{lemma}
\label{lem: distribution law composition inverse mv}
For any $\mathbf{A}$-polarity $\mathbb{P}= (A,X,I)$ and all $I$-compatible relations $R_{\Box_1}, R_{\Box_2}: A\times X\to \mathbf{A}$ and  $R_{\Diamond_1}, R_{\Diamond_2}: X\times A\to \mathbf{A}$, for all $a\in A$ and $x\in X$,  
\smallskip

{{\centering
$a (R_{\Box_1}\, ;_I R_{\Box_2})x = x(R_{\Diamondblack_2}\, ;_I R_{\Diamondblack_1})a 
\quad\quad\quad
x(R_{\Diamond_1}\, ;_I R_{\Diamond_2})a = a( R_{\blacksquare_2}\, ;_I R_{\blacksquare_1}) x.$
\par}}
\end{lemma}
\begin{proof}
We only prove the first equality, the proof of the second being similar.
\smallskip

{{\centering
    \begin{tabular}{rclr}
     $a(R_{\Box_1};_IR_{\Box_2})x$    & $ = $ & $R_{\Box_2}^{(1)}[I^{(0)}[R_{\Box_1}^{(1)}[\{1/a\}]]](x)$ & Lemma \ref{lem:fuz:equiv:I:coposition}.2 \\
         & $ = $ & $R_{\Diamondblack_2}^{(0)}[I^{(0)}[R_{\Diamondblack_1}^{(0)}[\{1/a\}]]](x)$ & Lemma \ref{lemma:elementary property adjoints mv} \\ 
         & $ = $ & $x(R_{\Diamondblack_2};_I R_{\Diamondblack_1})a$. & Lemma  \ref{lem:fuz:equiv:I:coposition}.2
    \end{tabular}
\par}}
\end{proof}
\subsection{Properties of $I$-compositions} \label{homogeneous composition section} 
In the present section, we show that $I$-composition of $I$-compatible relations of type $A \times X$ (resp. $X \times A$) has identity $I$ (resp.\ $J$), and is associative. The following proposition is a direct consequence of Lemma \ref{lemma: many valued icompatibility meaning}.
\begin{proposition} \label{homogeneous identity}
For any formal $\mathbf{A}$-context $\mathbb{P}= (A,X,I)$, any $I$-compatible relations $B$ and $D$, 
\smallskip

{{\centering
$1. \quad (B\, ;_II) =B= (I\, ;_IB) \quad\quad\quad 2. \quad (D\, ;_IJ) =D= (J\, ;_ID)$
\par}}
\end{proposition}
\begin{proof}
We prove only item 1 as the proof of item 2 is similar.
For any $a \in A$ and $x \in X$, since $B$ is $I$-compatible, by Definition \ref{def:relational composition mv} (2) and Lemma \ref{lemma: many valued icompatibility meaning} (1),
\smallskip

{{\centering
\begin{tabular}{rclcl}
$a (B ;_I I) x$ & $=$ & $\bigwedge_{y \in X}( \bigwedge_{b \in A}( b I x \rightarrow b I y)\rightarrow a B y)$ 
& $=$ & $ a B x$;  \\
$a (I ;_I B) x$ & $=$ & $\bigwedge_{y \in X}( \bigwedge_{b \in A}( b B x \rightarrow b I y)\rightarrow a I y)$ & $=$ & $ a B x$.\\
\end{tabular}
\par}}
\end{proof}

\begin{lemma} \label{lemma: homogeneous property singleton eq} 
For any  $\mathbf{A}$-polarity $(A,X,I)$, any $I$-compatible relations $B_1, B_2, D_1, D_2$, all $\alpha \in \mathbf{A}$, $a \in A$,  $x \in X$,
{{\centering
\begin{tabular}{rrcl}
	1. & $D_1^{(0)}[I^{(1)}[D_2^{(0)} [\{\alpha\slash x\}]]] =  (D_1\, ;_I D_2)^{(0)} [\{\alpha\slash x\}]$ & { and } & $D_2^{(1)}[I^{(0)}[D_1^{(1)} [\{\alpha\slash a\}]]] =  (D_1\, ;_I D_2)^{(1)} [\{\alpha\slash a\}]$; \\
	2. & $B_1^{(0)}[I^{(0)}[B_2^{(0)} [\{\alpha\slash a\}]]] = (B_1\, ;_IB_2)^{(0)} [\{\alpha\slash a\}]$ & { and } &   $B_2^{(1)}[I^{(1)}[B_1^{(1)} [\{\alpha\slash x\}]]] = (B_1\, ;_IB_2)^{(1)} [\{\alpha\slash x\}];$
\end{tabular}
\par}}
\end{lemma}
\begin{proof}
We show only the first identity of item 1, the remaining ones being shown similarly. 
For every $x\in X$,
\smallskip

{{\centering
    \begin{tabular}{r c l r}
    $(D_1\,;_I D_2)^{(0)}[\{\alpha\slash a\}](x)$ &$=$& $\bigwedge_{b\in A}(\{\alpha\slash a\}(b)\rightarrow x(D_1\,;_I D_2)b) $\\
    &$=$& $\alpha\to x(D_1\,;_I D_2)a$\\
    
    &$=$& $(\alpha\to D_1^{(0)}[I^{(0)}[D_2^{(0)}[\{1\slash a\}]]])(x)$ & Lemma \ref{lem:fuz:equiv:I:coposition}\\
    & $=$ & $D_1^{(0)}[\alpha\land I^{(0)}[D_2^{(0)}[\{1\slash a\}]]](x)$ & Lemma \ref{lem:keytouniversemysteries}\\
    & $=$ & $D_1^{(0)}[I^{(0)}[I^{(1)}[\alpha\land I^{(0)}[D_2^{(0)}[\{1\slash a\}]]]]](x)$ &  $I$-compatibility of $D_1$\\
    & $=$ & $D_1^{(0)}[I^{(0)}[\alpha\to I^{(1)}[ I^{(0)}[D_2^{(0)}[\{1\slash a\}]]]]](x)$ &  Lemma \ref{lem:keytouniversemysteries}\\
    & $=$ & $D_1^{(0)}[I^{(0)}[\alpha\to D_2^{(0)}[\{1\slash a\}]]](x)$ & $I$-compatibility of $D_2$\\
    & $=$ & $D_1^{(0)}[I^{(0)}[D_2^{(0)}[\alpha\land\{1\slash a\}]]](x)$ &  Lemma \ref{lem:keytouniversemysteries}\\
    & $=$ & $D_1^{(0)}[I^{(0)}[D_2^{(0)}[\{\alpha\slash a\}]]](x).$ &  
    \end{tabular}
\par}}
\end{proof}

\begin{proposition}  \label{prop:homogeneous property general} 
For any  $\mathbf{A}$-polarity $\mathbb{P}= (A,X,I)$, any relations $B_1,B_2,D_1,D_2$, all $f:A\to\mathbf{A}$ and $u:X \to \mathbf{A}$,

{{\centering
\begin{tabular}{rrcl}
     1. & $D_1^{(0)}[I^{(0)}[D_2^{(0)} [f]]]= (D_1\, ;_ID_2)^{(0)} [f]$ & and & $D_2^{(1)}[I^{(1)}[D_1^{(1)} [u]]]= (D_1\, ;_ID_2)^{(1)} [u];$  \\
     2.& $B_1^{(0)}[I^{(1)}[B_2^{(0)} [u]]]= (B_1\, ;_I B_2)^{(0)} [u]$ & and & $B_2^{(1)}[I^{(0)}[B_1^{(1)} [f]]]= (B_1\, ;_I B_2)^{(1)} [f]$.
\end{tabular}
\par}}
\end{proposition}
\begin{proof}
We only prove the first equation of 1, the proof of the remaining identities being similar.
\smallskip

{{\centering
    \begin{tabular}{r c l r}
       $D_1^{(0)}[I^{(0)}[D_2^{(0)} [f]]]$ & $=$ &  $D_1^{(0)}[I^{(0)}[D_2^{(0)} [\bigvee_{a \in A} \{ f(a) \slash a \}]]]$ & $f = \bigvee_{a \in A} \{ f(a) \slash a \}$\\
        & $=$ &  $D_1^{(0)}[I^{(0)}[\bigwedge_{a \in A} D_2^{(0)} [\{ f(a) \slash a \}]]]$ & Lemma \ref{lemma: basic mv} (5)\\
        & $=$ &  $D_1^{(0)}[I^{(0)}[\bigwedge_{a \in A} I^{(1)}[I^{(0)}[D_2^{(0)} [\{ f(a) \slash a \}]]]]]$ & $D_2$ is $I$-compatible \\
        & $=$ &  $D_1^{(0)}[I^{(0)}[I^{(1)}[\bigvee_{a \in A} I^{(0)}[D_2^{(0)} [\{ f(a) \slash a \}]]]]]$ & Lemma \ref{lemma: basic mv} (5)\\
        & $=$ &  $D_1^{(0)}[\bigvee_{a \in A} I^{(0)}[D_2^{(0)} [\{ f(a) \slash a \}]]]$ & Lemma \ref{lemma:equivalents of I-compatible mv} (1)({\em iii}) as $D_1$ is $I$-compatible\\
        & $=$ &  $\bigwedge_{a \in A}D_1^{(0)}[ I^{(0)}[D_2^{(0)} [\{ f(a) \slash a \}]]]$ & Lemma \ref{lemma: basic mv} (5)\\
        & $=$ &  $\bigwedge_{a \in A}(D_1;_I D_2)^{(0)} [\{ f(a) \slash a \}]$ & Lemma \ref{lemma: homogeneous property singleton eq} (1)\\
        & $=$ &  $(D_1;_I D_2)^{(0)} [ \bigvee_{a \in A} \{ f(a) \slash a \}]$ &  Lemma \ref{lemma: basic mv} (5)\\
        & $=$ &  $(D_1;_I D_2)^{(0)}[f]$ & $f = \bigvee_{a \in A} \{ f(a) \slash a \}$
    \end{tabular}
\par}}
\end{proof}
\begin{proposition} \label{prop:homogeneous associativity holds}
For any $\mathbf{A}$-polarity $(A, X, I)$, and all $I$-compatible relations $B_1,B_2,B_3,D_1,D_2,D_3$,
\[
1. \quad 
D_1\, ;_I (D_2\, ;_I D_3)=  (D_1\, ;_I D_2)\, ;_ID_3
\quad\quad\quad
2. \quad
B_1\, ;_I(B_2\, ;_I B_3)= (B_1\, ;_I B_2)\, ;_I B_3
\]
\end{proposition}
\begin{proof}
We only prove 1, proof of 2 being similar. For any $f: A \to \mathbf{A} $, 
\smallskip

{{\centering
    \begin{tabular}{r c l l}
        $(D_1\, ;_I (D_2\, ;_I D_3))^{(0)}[f]$ &=& $D_1^{(0)}[I^{(0)}[(D_2\, ;_I D_3)^{(0)}[f]]] $ & Proposition \ref{prop:homogeneous property general} (1)\\
        &=& $D_1^{(0)}[I^{(0)}[D_2^{(0)}[I^{(0)}[D_3^{(0)}[f]]]]] $ & Proposition \ref{prop:homogeneous property general} (1)\\
        &=& $(D_1 \, ;_I D_2)^{(0)}[I^{(0)}[D_3^{(0)}[f]]] $ & Proposition \ref{prop:homogeneous property general} (1)\\
        &=& $((D_1\, ;_I D_2)\, ;_I D_3)^{(0)}[f] $ & Proposition \ref{prop:homogeneous property general} (1).
    \end{tabular}
\par}}
\end{proof}

\subsection{Properties of non $I$-mediated composition} \label{heterogeneous composition section} 
\begin{proposition}
\label{prop:relational composition fails with alpha}
	For any  $\mathbf{A}$-polarity $\mathbb{P}= (A,X,I)$, any relations $B,D,L,M$, any $\alpha \in \mathbf{A}$, $a \in A$, and $x \in X$,
	\[
	\begin{array}{rlcrl}
	1.& B^{(0)}[D^{(0)} [\{\alpha\slash b\}]] \leq (B\, ;D)^{(0)} [\{\alpha\slash b\}] && 4.& D^{(0)}[M^{(0)} [\{\alpha\slash a\}]]\leq (D\, ;M)^{(0)} [\{\alpha\slash a\}] \\
	2.& D^{(0)}[B^{(0)} [\{\alpha\slash y\}]]\leq (D\, ;B)^{(0)} [\{\alpha\slash y\}] && 5.& L^{(0)}[D^{(0)} [\{\alpha\slash a\}]]\leq (L\, ;D)^{(0)} [\{\alpha\slash a\}] \\ 
	3.& M^{(0)}[B^{(0)} [\{\alpha\slash x\}]]\leq (M\, ;B)^{(0)} [\{\alpha\slash x\}] && 6.& B^{(0)}[L^{(0)} [\{\alpha\slash x\}]]\leq (B\, ;L)^{(0)} [\{\alpha\slash x\}].
	\end{array}
	\]
\end{proposition}
\begin{proof}
1.
Let $\alpha \in \mathbb{A}$, $a, b \in A$. 
\smallskip

{{\centering
\begin{tabular}{r c l l r}
		$(B\, ;D)^{(0)} [\{\alpha\slash b\}] (a) $ & =& $ \bigwedge_{c\in A}(\{\alpha\slash b\}(c)\to a (B\,;D) c)$ & Definition of $(\cdot)^{(0)}$\\
		& =& $  \alpha \rightarrow a (B\,;D) b$ & Definition of $\{\alpha \slash b\}$\\
		& =& $\alpha \rightarrow \bigwedge_{x\in X}(xDb\rightarrow aBx)$ & Definition \ref{def:relational composition mv}(3) \\
		& =& $\bigwedge_{x\in X}\left (\alpha \rightarrow (xDb\rightarrow aBx)\right)$ & distributivity of $\rightarrow$\\
		& =& $\bigwedge_{x\in X}\left ((\alpha \wedge xDb)\rightarrow aBx\right)$ & currying \\[2mm]
$B^{(0)}[D^{(0)} [\{\alpha\slash b\}]](a)$ & = & $\bigwedge_{x\in X}\left( D^{(0)} [\{\alpha\slash b\}](x)\rightarrow a B x\right)$ & Definition of $(\cdot)^{(0)}$\\
& = & $\bigwedge_{x\in X}\left( (\alpha\rightarrow  xDb )\rightarrow a B x\right).$ & Definition of $(\cdot)^{(0)}$\\
		\end{tabular}
\par}} 
\smallskip

\noindent As $\alpha \wedge xDb\leq \alpha\rightarrow  xDb$ holds in every Heyting algebra, the statement follows from the antitonicity of $\rightarrow$. The remaining statements are proven similarly.
\end{proof}
\begin{remark}
\label{counterexamples to packing via relational composition}
The converse inequalities in Proposition \ref{prop:relational composition fails with alpha} do not hold in general. 	For the converse of item 1, let us show that the following inequality does not hold in general:
\smallskip

{{\centering $(R\, ;T)^{(0)} [\{\alpha\slash b\}] (a)\leq R^{(0)}[T^{(0)} [\{\alpha\slash b\}]](a),$ \par}}
\smallskip

\noindent i.e.~that the following inequality does not hold:
\[\bigwedge_{x\in X}\left ((\alpha \wedge xTb)\rightarrow aRx)\right)\leq \bigwedge_{x\in X}\left( (\alpha\rightarrow  xTb )\rightarrow a R x\right).\]
Indeed, let $(A, X, I, R, T)$ s.t.~$A = \{a, b\}$, $X = \{x, y\}$ and $I$ maps $(a, y) $ and $ (b, x)$ to $1$ and every other pair to $0$ (hence, every subset is Galois closed, and hence every relation is $I$-compatible). Assume that   $xTb = y Tb = 1$ and $\alpha =  aRx = aRy = 0$. On this polarity-based frame, the inequality above fails, since it becomes 
\[1 = \alpha \rightarrow aRx\leq  a R x = 0.\]
\end{remark}

\begin{remark}
\label{counterexamples to associativity}
As expected, associativity laws for non $I$-mediated compositions
do not hold in general. For instance, 
	let us find some polarity $(A, X, I)$ and some $I$-compatible $R, U:A\times X \to \mathbf{A}$ and $T:X\times A \to \mathbf{A}$   such that the following inequality fails:
		\[(R\, ;T)\, ;U \leq R\, ;(T\, ; U), \] 
		i.e.~the following inequality fails:		
\[\bigwedge_{y\in X}\left (\bigwedge_{b\in A}\left( (bUx\wedge yTb)\rightarrow aRy \right) \right)\leq \bigwedge_{y\in X}\left (\bigwedge_{b\in A}\left (bUx\rightarrow yTb \right)\rightarrow aRy \right).\]
		Let $(A, X, I)$ such that $A = \{a, b\}$, $X = \{x, y\}$ and $I$ maps $(a, y) $ and $ (b, x)$ to $1$ and every other pair to $0$ (hence, every subset is Galois closed, and hence every relation is $I$-compatible). Assume that $R$ sends $(a, x)$ and $(a, y)$ to $0$, and $U$ sends $(a, x)$ and $(b, x)$ to $0$. Under these assumptions,  the inequality above fails, since it becomes 
\[1 = 0 \rightarrow 0\leq  1\to 0 = 0.\]
	\end{remark}

\subsection{Correspondents on many-valued polarities}
\label{ssec:correspondents polarities mv}
\begin{definition}
\label{def:relalgebra polarity mv}
For any LE-language $\mathcal{L}_{\mathrm{LE}} = \mathcal{L}_{\mathrm{LE}}(\mathcal{F}, \mathcal{G})$, any
$\mathbf{A}$-polarity-based $\mathcal{L}_{\mathrm{LE}}$-frame   $\mathbb{F} = (\mathbb{P}, \mathcal{R}_{\mathcal{F}}, \mathcal{R}_{\mathcal{G}})$ based on  $\mathbb{P} = (A, X,  I)$ induces the heterogeneous relation algebra 
\[\mathbb{F}^\ast: = (\mathbf{A}^{A\times X}, \mathbf{A}^{X\times A}, \mathbf{A}^{A\times A}, \mathbf{A}^{X\times X}, \mathcal{R}_{\mathcal{F}^\ast}, \mathcal{R}_{\mathcal{G}^\ast},  ; _I^{\Diamond}, ;_I^{\Box}, ;^{\Box\Diamond},  ;^{\Diamond\Box}, ;^{{\rhd}\Box}, ;^{\Diamond{\rhd}}, ;^{{\lhd}\Diamond}, ;^{\Box{\lhd}}), \]
where, abusing notation, we understand $ \mathcal{R}_{\mathcal{F}^\ast}: = \{R_{\Diamond} \mid \Diamond\in \mathcal{F}\}\cup \{R_{\Diamondblack}\mid  \Box\in \mathcal{G}\}\subseteq \mathbf{A}^{X\times A}$ and $\mathcal{R}_{\mathcal{G}^\ast}: = \{R_{\Box} \mid \Box\in \mathcal{G}\}\cup\{ R_{\blacksquare}\mid  \Diamond\in \mathcal{F}\}\subseteq \mathbf{A}^{A\times X}$, 
where $R_{\Diamondblack}$ and $R_{\blacksquare}$ are defined as indicated in the discussion above Lemma \ref{lemma:elementary property adjoints mv},
and the operations are all binary and defined as in Definition \ref{def:relational composition mv}. Superscripts in the operations indicate the types of the two inputs, which completely determines the output type. These superscripts will be dropped whenever this does not cause ambiguities.
\end{definition}

Through Definition \ref{def:relalgebra polarity mv}, we can interpret $\mathsf{PRel}_\mathcal{L}$-inequalities also on $\mathbf{A}$-polarity-based frames in the obvious way.
In the present section we prove the following
\begin{theorem}
\label{thm:verbatim crisp mv}
For any LE-signature $\mathcal{L}$ and any inductive modal reduction principle $s(p)\leq t(p)$ of $\mathcal{L}$, the $\mathsf{PRel}_\mathcal{L}$-inequality encoding its first-order correspondent on (crisp) polarity-based frames 
also encodes its first-order correspondent on $\mathbf{A}$-polarity-based frames.
\end{theorem}

\begin{lemma}
\label{lemma:molecular polarity mv}
For any $\mathcal{L}$-frame $\mathbb{F}$ based on the $\mathbf{A}$-polarity $\mathbb{P} = (A,X,I)$, any $\varphi(!z)$, $\psi(!z)$, $\chi(!z)$, $\zeta(!z)$ as in Definition \ref{def:pol-based rels associated with terms}, any $h: A \to \mathbf{A}$ and $u: X \to \mathbf{A}$, any $a \in A$, $x \in X$, and $\alpha \in \mathbf{A}$,
\smallskip

{{\centering
\begin{tabular}{rl c rl}
    1. & $\descr{\varphi(!z)}(z \coloneqq h) =  R^{(0)}_{\varphi}[h];$ & \quad \quad & 
    3. & $\descr{\chi[\nomj/!z]}(\nomj \coloneqq \{\alpha \slash a\}^{\uparrow\downarrow}) = R^{(0)}_{\chi}[\{\alpha \slash a\}^{\uparrow\downarrow}]=  R^{(0)}_{\chi}[\{\alpha \slash a\}]$ \\
    2. & $\val{\psi(!z)}(z \coloneqq u) = R^{(0)}_\psi[u]$ & &
    4. & $\val{\zeta[\cnomm/!z]}(\cnomm \coloneqq \{\alpha \slash x\}^{\downarrow\uparrow}) = R^{(0)}_{\zeta}[\{\alpha \slash x\}^{\downarrow\uparrow}] = R^{(0)}_{\zeta}[\{\alpha \slash x\}].$
\end{tabular}
\par}}
\end{lemma}
\begin{proof}
Items 1 and 2 can be shown straightforwardly by induction on $\varphi$ and $\psi$, using Proposition \ref{prop:homogeneous property general}. Items 3 and 4 can be shown  by induction on the number of alternations of concatenations of box and diamond connectives, using  items 1 and 2, and Definition \ref{def:relational composition mv}.
\end{proof}

By Proposition \ref{prop:albaoutput}, MRPs of shape (a) are equivalent to the following pure inequalities:
\begin{equation}
\label{eq: alba output a polarity mv}
\forall \nomj\left(\mathsf{LA}(\psi)[\phi[\nomj/!y]/!u] \leq \chi[\mathsf{LA}(\alpha)[\nomj/!u]/p]\right).
\end{equation}
When interpreting the condition above on $\mathbf{A}$-polarities, it becomes
\begin{equation}
\label{eq: alba output a translated polarity mv}
\forall \alpha\forall a\forall x \left(\descr{\chi[\mathsf{LA}(\alpha)[\nomj/!u]/p] }(\nomj: =\{\alpha\slash a\}^{\uparrow\downarrow})(x)\leq \descr{\mathsf{LA}(\psi)[\phi[\nomj/!y]/!u]}(\nomj: =\{\alpha\slash a\}^{\uparrow\downarrow})(x) \right).
\end{equation}
By items 1 and 3 of Lemma \ref{lemma:molecular polarity mv},
\smallskip

{{\centering
\begin{tabular}{cc}
\begin{tabular}{rcl}
     &&$\descr{\mathsf{LA}(\psi)[\phi[\nomj/!y]/!u]}(\nomj: = \{\alpha\slash a\}^{\uparrow\downarrow})(x)$ \\& $=$ &  $R_{\mathsf{LA}(\psi)[\phi/!u]}^{(0)}[\{\alpha\slash a\}^{\uparrow\downarrow}](x)$ \\
     & $=$ &  $(R_{\mathsf{LA}(\psi)} \,;_I R_{\phi})^{(0)}[\{\alpha\slash a\}](x)$ \\
     & $=$ & $\alpha \rightarrow x (R_{\mathsf{LA}(\psi)} \,;_I R_{\phi}) a $
\end{tabular}
&
\begin{tabular}{rcl}
     &&$\descr{\chi[\mathsf{LA}(\alpha)[\nomj/!u]/p] }(\nomj:= \{\alpha\slash a\}^{\uparrow\downarrow})(x)$ \\ & $=$ & $R_{\chi[\mathsf{LA}(\alpha)/p]}^{(0)}[\{\alpha\slash a\}^{\uparrow\downarrow}](x)$ \\
     & $=$ & $R_{\chi[\mathsf{LA}(\alpha)/p]}^{(0)}[\{\alpha\slash a\}](x)$ \\
     & $=$ & $\alpha \rightarrow x (R_{\chi[\mathsf{LA}(\alpha)/p]}) a$ 
\end{tabular}
\end{tabular}
\par}}
\smallskip

\noindent Hence, \eqref{eq: alba output a translated polarity mv} can be rewritten as follows:
\begin{equation}
\label{eq: alba output a polarity after lemma mv}
\forall \alpha \forall a \forall x \left(\alpha \rightarrow x (R_{\chi[\mathsf{LA}(\alpha)/p]}) a \leq \alpha \rightarrow x (R_{\mathsf{LA}(\psi)} \,;_I R_{\phi}) a \right),
\end{equation}
which, by the monotonicity of $\rightarrow$, and the fact the identity $c = 1\rightarrow c$ holds in any Heyting algebra, is equivalent to the following $\mathsf{PRel}_\mathcal{L}$-inequality:
\begin{equation}
\label{eq final a polarity mv}
R_{\chi[\mathsf{LA}(\alpha)/p]}\subseteq R_{\mathsf{LA}(\psi)} \,;_I R_{\phi}.
\end{equation}
Similarly, by Proposition \ref{prop:albaoutput} MRPs of shape (b) are equivalent to the following pure inequalities:
\begin{equation}
\label{eq: alba output b polarity mv}
\forall \cnomm\left(\zeta[\mathsf{RA}(\delta)[\cnomm/!u]/p] \leq \mathsf{RA}(\phi)[\psi[\cnomm/!y]/!u]\right),
\end{equation}
which, when interpreting the condition above on $\mathbf{A}$-polarity-based frames, it becomes
\begin{equation}
\label{eq: alba output b translated polarity mv}
\forall \alpha \forall a \forall x\left (\val{\zeta[\mathsf{RA}(\delta)[\cnomm/!u]/p]}(\cnomm: = (\{\alpha\slash x\}^{\downarrow\uparrow})(a) \leq \val{\mathsf{RA}(\phi)[\psi[\cnomm/!y]/!u]}(\cnomm: = \{\alpha\slash x\}^{\downarrow\uparrow})(a) \right ).
\end{equation}
By items 4 and 2 of Lemma \ref{lemma:molecular polarity mv}, 
\smallskip

{{\centering
\begin{tabular}{cc}
\begin{tabular}{rcl}
&&$\val{\mathsf{RA}(\phi)[\psi[\cnomm/!y]/!u]}(\cnomm:= \{\alpha\slash x\}^{\downarrow\uparrow})(a)$\\ & $=$ &  $R_{\mathsf{RA}(\phi)[\psi/!u]}^{(0)}[\{\alpha\slash x\}^{\downarrow\uparrow}](a)$ \\
& $=$ & $(R_{\mathsf{RA}(\phi)} \,;_I R_{\psi})^{(0)}[\{\alpha\slash x\}](a)$ \\
& $=$ & $\alpha \rightarrow a(R_{\mathsf{RA}(\phi)} \,;_I R_{\psi})x$
\end{tabular}
&
\begin{tabular}{rcl}
&&$\val{\zeta[\mathsf{RA}(\delta)[\cnomm/!u]/p] }(\cnomm:= \{\alpha\slash x\}^{\downarrow\uparrow})(a)$ \\ & $=$ & $R_{\zeta[\mathsf{RA}(\delta)/p]}^{(0)}[\{\alpha\slash x\}^{\downarrow\uparrow}](a)$ \\
& $=$ & $R_{\zeta[\mathsf{RA}(\delta)/p]}^{(0)}[\{\alpha\slash x\}](a)$ \\
& $=$ & $\alpha \rightarrow a (R_{\zeta[\mathsf{RA}(\delta)/p]}) x$
\end{tabular}
\end{tabular}
\par}}
\smallskip

\noindent Therefore, we can rewrite \eqref{eq: alba output b translated polarity mv} as follows:
\[
\forall \alpha \forall a \forall x \left( \alpha \rightarrow a (R_{\zeta[\mathsf{RA}(\delta)/p]}) x \leq  \alpha \rightarrow a(R_{\mathsf{RA}(\phi)} \,;_I R_{\psi})x\right),
\]
which, as discussed above, is equivalent to
\begin{equation}
\label{eq final b polarity mv}
R_{\zeta[\mathsf{RA}(\delta)/p]}\subseteq R_{\mathsf{RA}(\phi)} \,;_I R_{\psi}.
\end{equation}
Finally, notice that, if the modal reduction principle is also analytic, the shape of $\chi(p)\neq p$ (resp.\ $\zeta(p)\neq p$)  simplifies, in case (a), to $\chi(p) = \phi_{n_\chi} \psi_{n_\chi}(p)$ with $n_\chi = 1$ and $\psi_{n_\chi}$ empty 
and, in case (b), to $\zeta(p) = \psi_{n_\zeta} \phi_{\zeta}(p)$ with $n_\zeta = 1$ and $\phi_{n_\zeta}$ empty. 
Hence, \eqref{eq final a polarity mv}  and \eqref{eq final b polarity mv}  simplify to the following inclusions, respectively:
\begin{equation}
\label{eq final analytic polarity mv}
R_{\phi_{n_\chi}}\, ;_IR_{\mathsf{LA}(\alpha)}\subseteq R_{\mathsf{LA}(\psi)} \,;_I R_{\phi} \quad\quad R_{\psi_{n_\zeta}}\, ;_I R_{\mathsf{RA}(\delta)}\subseteq R_{\mathsf{RA}(\phi)} \,;_I R_{\psi}.
\end{equation}

\subsection{Examples} 
\label{examples from rough concepts}
Let us illustrate the main result of the previous subsection by providing direct proof of the many-valued counterpart of \cite[Proposition 4.3]{conradie2021rough} by instantiating the general strategy described in the proof of Theorem \ref{thm:verbatim crisp mv}. The items of the following proposition are verbatim the same as those of \cite[Proposition 4.3]{conradie2021rough}.
\begin{proposition}
\label{lemma:correspondences}
For any LE-language $\mathcal{L}$, and any $\mathbf{A}$-frame $\mathbb{F}$ for $\mathcal{L}$ based on the polarity $(A, X, I)$,
\smallskip

{{\centering
\begin{tabular}{rlcl c rlcl}
1. & $\mathbb{F}\models \Box\phi\vdash \Diamond\phi\quad $ & iff & $\quad R_{\Box}\, ;_IR_{\blacksquare} \leq I$. & \quad\quad\quad & 7. & $\mathbb{F}\models \Diamond\Box\phi\vdash \phi\quad $ & iff & $\quad  R_{\Diamondblack} \leq R_{\Diamond}$ \\
2. & $\mathbb{F}\models \Box\phi\vdash \phi\quad $ & iff & $\quad R_\Box\leq I$ & & 8. & $\mathbb{F}\models \phi\vdash \Box \phi\quad $ & iff & $\quad I \leq R_\Box$. \\
3. & $\mathbb{F}\models \phi\vdash \Diamond\phi\quad $ & iff & $\quad R_\blacksquare\leq I$ & & 9. & $\mathbb{F}\models \Diamond\phi\vdash \phi\quad $ & iff & $\quad I \leq R_\blacksquare$. \\
4. & $\mathbb{F}\models \Box\phi\vdash \Box\Box\phi\quad $ & iff & $\quad R_{\Box}\leq R_{\Box}\, ;_I R_{\Box}$ & & 10. & $\mathbb{F}\models \Box\Box\phi\vdash \Box\phi\quad $ & iff & $\quad R_{\Box}\, ;_I R_{\Box}\leq R_{\Box}$. \\
5. & $\mathbb{F}\models \Diamond\Diamond\phi\vdash \Diamond\phi\quad $ &iff& $\quad R_{\Diamond}\leq R_{\Diamond}\, ;_I R_{\Diamond}$ & & 11. & $\mathbb{F}\models \Diamond\phi\vdash \Diamond \Diamond\phi\quad $ & iff & $\quad R_{\Diamond}\, ;_I R_{\Diamond}\leq R_{\Diamond}$. \\
6. & $\mathbb{F}\models \phi\vdash \Box\Diamond\phi\quad $ & iff & $\quad R_{\Diamond} \leq R_{\Diamondblack}$ & & 12. &  $\mathbb{F}\models \Diamond\phi\vdash \Box\phi\quad $ & iff & $\quad I\leq R_{\blacksquare}\, ;_IR_{\Box}$.\\
\end{tabular}
\par}}
\end{proposition}

\begin{proof}
\noindent 1. The MRP $\Box p \vdash \Diamond p$ is analytic inductive,  hence  of  shapes (a) and  (b). We will consider it qua type (b), i.e.~$\phi (y): =   y$, hence $\mathsf{RA}(\phi)(v): =   v$, and $\zeta (p): = \Box p$,  and $\psi(x): =  x$, and $\delta(p) :=  \Diamond p$, hence $\mathsf{RA}(\delta)(u): = \blacksquare u$.
\smallskip

{{\centering
    \begin{tabular}{rcll}
          $\forall p(\Box p \leq \Diamond p)$ 
    &  iff   & $\forall \cnomm(\Box\blacksquare \cnomm \leq \cnomm)$ & Proposition \ref{prop:albaoutput} \\
     & iff   & $\forall \cnomm\forall a(\val{\Box\blacksquare \cnomm}(a) \leq \val{\cnomm}(a))$ &  \\
    &  i.e.   & $\forall \alpha\forall a\forall x\left (R_{\Box\blacksquare}^{(0)}[\{\alpha/x\}](a)\leq I^{(0)}[\{\alpha/x\}](a)\right)$\\
    &  iff & $\forall \alpha\forall a\forall x\left (\alpha \rightarrow a R_{\Box\blacksquare}x\leq \alpha \rightarrow a Ix\right)$ & Lemma \ref{lemma:molecular polarity mv}\\
    &  iff & $\forall a\forall x\left (a R_{\Box\blacksquare}x\leq  aIx\right)$\\
    &  iff & $ R_{\Box\blacksquare}\leq   I$\\
    &  iff & $ R_{\Box}\, ;_I\, R_{\blacksquare}\leq   I$.\\
    \end{tabular}
\par}}
\noindent 5. The MRP $\Diamond  \Diamond  p \vdash \Diamond p$ is analytic inductive, hence  of both shape (a) and  (b). We will treat it qua shape (a), i.e.~$\phi (y): =   \Diamond \Diamond y$, and $\alpha (p): =  p$, hence $\mathsf{LA}(\alpha)(u): =  u$,  and $\psi(x): =  x$,  hence $\mathsf{LA}(\psi)(v): =   v$, 
and $\chi(p) :=  \Diamond p$.

{{\centering
    \begin{tabular}{rcll}
          $\forall p(\Diamond\Diamond p \leq \Diamond p)$ 
     & iff   & $\forall \nomj(\Diamond\Diamond \nomj \leq \Diamond\nomj)$ & Proposition \ref{prop:albaoutput} \\
     & iff   & $\forall \nomj\forall x(\descr{\Diamond \nomj}(x) \leq \descr{\Diamond\Diamond \nomj}(x))$ &  \\
     & i.e.   & $\forall \alpha\forall a\forall x\left (R_{\Diamond}^{(0)}[\{\alpha/a\}](x)\leq R_{\Diamond\Diamond}^{(0)}[\{\alpha/a\}](x)\right)$\\
      &iff & $\forall \alpha\forall a\forall x\left (\alpha \rightarrow x R_{\Diamond}a\leq \alpha \rightarrow x R_{\Diamond\Diamond}a\right)$ & Lemma \ref{lemma:molecular polarity mv}\\
     & iff & $\forall a\forall x\left (xR_{\Diamond}a\leq x R_{\Diamond\Diamond}a\right)$\\
     & iff & $ R_{\Diamond}\leq  R_{\Diamond\Diamond}$\\
     & iff & $ R_{\Diamond}\leq  R_{\Diamond}\, ;_I\,R_{\Diamond}$.\\
    \end{tabular}
\par}}
\smallskip

\noindent The proofs of the remaining items are collected in Appendix \ref{sec: proof of mv correspondences}.
\end{proof}

\section{Conclusions}
\label{sec:conclusions}

\paragraph{Our contributions.}  In previous work in the unified correspondence program, we identified the classes of Inductive and Sahlqvist axioms for all logics that can be given algebraic semantics based on normal lattice expansions, viz.\ LE-logics. These classes  are characterised purely in terms of the order-theoretic properties of the algebraic interpretations of the connectives, and are impervious to any change in the  choice of particular relational semantics for the logic, as long as it is linked to the the algebraic semantics via a suitable duality.  Accordingly, the correspondence machinery is modularised: the correspondent calculated (uniformly) by ALBA is the conjunction  of pure quasi-inequalities which must then be translated into a first-order correspondent by applying the appropriate standard translation for the choice of dual relational semantics. 

The work in the present paper complements those results. In a  sense, here we approach the problem from the opposite end, by laying the groundwork for a framework in which it is possible to systematically compare the first-order correspondents of a given inductive formula across different relational semantics. Concretely, we restricted our attention to the Sahlqvist modal reduction principles (MRPs) and focused on three relational settings, namely classical Kripke frames, polarity-based frames and many-valued polarity based frames. We found that, if we wrote the first-order correspondents of Sahlqvist MRPs on Kripke frames in the right way, namely as inclusions of certain relational compositions, we could obtain their correspondents on polarity-based frames, roughly speaking, simply by reversing the direction of the inclusion and replacing the identity relation $\Delta$ with the polarity relation $I$ everywhere, also in compositions of relations. The correctness of this procedure turns on the fact that, just like the lifting from a Kripke frame to polarity-based frame preserves its complex algebra (cf.~Proposition \ref{prop:from Kripke frames to enriched polarities}), it also ``preserves'' its associated relation algebra,\footnote{In the following sense: 
for any Kripke $\mathcal{L}$-frame $\mathbb{X}$, the {\em multi-type heterogeneous representation} of its associated relation algebra $\mathbb{X}^\ast = (\mathcal{P}(W\times W), \mathcal{R}_{\mathcal{F}^{\ast}}, \mathcal{R}_{\mathcal{G}^{\ast}},  \circ, \star)$  is 
\[(\mathbb{X}^*)^m: = (\mathcal{P}(W_A\times W_X), \mathcal{P}(W_X\times W_A), \mathcal{P}(W_A\times W_A), \mathcal{P}(W_X\times W_X), \mathcal{J}_{\mathcal{R}_{\mathcal{F}^\ast}^c}, \mathcal{I}_{\mathcal{R}_{\mathcal{G}^\ast}^c}  ; _{I}^{\Diamond}, ;_I^{\Box}, ;^{\Box\Diamond},  ;^{\Diamond\Box}, ;^{{\rhd}\Box}, ;^{\Diamond{\rhd}}, ;^{{\lhd}\Diamond}, ;^{\Box{\lhd}}), \]
where the compositions of  relations are defined as in the definition \ref{def:relational composition} with $I$ set to be equal to $\Delta$. 
One can verify that, if $\mathbb{X}$ is a Kripke $\mathcal{L}$-frame, then $(\mathbb{X}^*)^m\cong (\mathbb{F}_{\mathbb{X}})^\ast$ (cf.~Definitions \ref{def:lifting kripke frames} and \ref{def:rel algebras crisp} (1)). } and so relational compositions and pseudo-compositions on Kripke frames can be systematically lifted to $I$-mediated and non $I$-mediated compositions of relations on polarity-based frames.  The relations $\Delta$ and $I$ thus play the role of parameters in the correspondence. We also observed this parametricity phenomenon when moving from {\em crisp}  polarity-based frames to {\em many-valued} polarity-based frames. Here the relevant parameter is the truth-value algebra, which changes from the Boolean algebra $\mathbf{2}$ to an arbitrary complete Heyting algebra $\mathbf{A}$ while, syntactically, the first-order correspondents of Sahlqvist MRPs remain verbatim the same. This latter result partially generalizes that of \cite{BritzMScThesis} to the polarity-based setting, and provides an analogous result lifting correspondence along the dashed arrow in the following commutative diagram: 

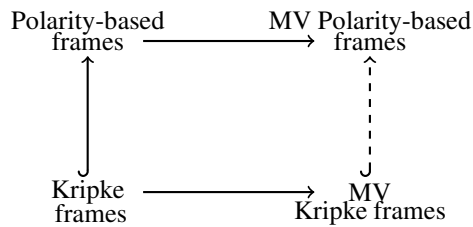
\begin{center}
\begin{tikzpicture}[scale=0.67]
\draw (5.3, 0) node {{MV}};
\draw (5.3, -0.41) node {{Kripke frames}};
\draw (-.3, 0) node {{Kripke}};
\draw (-.3, -0.41) node {{ frames}};
\draw (5.3, 3.36) node {{MV Polarity-based}};
\draw (5.3, 3) node {{frames}};
\draw (-.3, 3.36) node {{Polarity-based}};
\draw (-.3, 3) node {{frames}};
\draw[->, thick] (0.8,0) -- (4.2,0);
\draw[->, thick] (0.8,3) -- (4.2,3);
  \draw [thick, right hook->] (-.3, 0.3) -- (-.3, 2.7);
   \draw [thick, right hook->, dashed] (5.3, 0.3) -- (5.3, 2.7);
\end{tikzpicture}
\end{center}

These results and methods suggest that correspondence theories for different logics and semantic contexts can not only be methodologically {\em unified} by the same algebraic and algorithmic mechanisms, but that they can also be {\em parametrically} related in terms of their outputs.


From a methodological perspective, we note that, to be able to formulate and prove these results, we had to develop the necessary mathematical framework which allowed us to represent the first-order correspondents of MRPs on Kripke frames as inclusions of compositions and pseudo compositions of the accessibility relations. Hinging on the natural embedding of the semantic setting of Kripke frames into that of polarity-based frames, we were then in a position to transfer these inclusions to the latter via the notion of lifting, where they become inclusions of the $I$-mediated and non $I$-mediated compositions  of the relations there.  

In the remaining paragraphs, we explore some of the above mentioned points in more detail, highlight some others, suggest further directions, and conclude with a limitative counterexample and a conjecture. 

\smallskip
\noindent\textbf{Towards parametric correspondence and beyond. } The results of the present paper are of a new type,  the very formulation of which is only possible in the context of the (partial) modularization of Sahlqvist-type results for classes of logics brought about by unified correspondence theory. They represent instances of what could be called {\em parametric correspondence},
and suggest the more general question of whether other theorems and results could be parametrically transferred between LE-logics in analogous ways. For example, it seems clear that, in some instances certain truth-preserving model constructions, like bisimulations, can be seen as parametric versions of each other across relational semantics and that, perhaps, this could extend to characterisations of expressivity like the Goldblatt-Thomason theorem. A parametric approach  which is technically different from the present one, but is similar to it in aims and intended outcomes, has been fruitfully pursued when studying  coalgebras of  families of functors on sets as semantic environments for modal logic \cite{kupke2011coalgebraic}, where functors play the role of parameters. This setting has made it possible to formulate and prove parametric versions of theorems such as Goldblatt-Thomason's and van Benthem's  \cite{kurz2007goldblatt, schroder2010coalgebraic}.

\smallskip
\noindent\textbf{Transferring intuitive meaning of axioms.} We calculated the Kripke frame correspondents of a range of well-known MRPs, expressed as inclusions of relational compositions, and lifted these to polarity-based frames. The outcomes were summarised in Table \ref{fig:MPS:Famous} and are $I$- and $J$-parameterized versions of the well known classical correspondents that we are therefore  justified in calling, for example, $I$-reflexivity and $J$-transitivity.

Moreover, these contributions start laying the necessary mathematical ground to support the systematic transfer of the intuitive or conceptual meanings of LE-axioms from one semantic context to another. When reading modalities epistemically, we have used their lifted correspondents to argued that the conceptual import of the $\mathbf{T}$, $\mathbf{4}$ and $\mathbf{B}$ axioms lifts to polarity-based frames with these axioms to become analogous rationality conditions on the epistemic stances of agents reasoning about databases of objects and their features. 

This phenomenon is by no means limited to epistemic readings or to polarity-based semantics. Indeed, it has also been observed for graph-based semantics in \cite{graph-based-wollic, conradie2020non},  where epistemic meanings transfer in an analogous way into a ``hyper-constructivist'' setting. In the latter setting, modalities can also be read temporally, and the meanings of axioms enforcing e.g.~the density or unboundedness of time flows can be transferred.

\smallskip
\noindent\textbf{Representing first-order correspondents.} To achieve the results of this paper, it was crucial to express the first-order correspondents of modal axioms in a form that is appropriate across the semantic contexts of interest. Indeed, because of the marked difference in their semantic clauses, the standard translations on Kripke frames and polarity-based frames render modalities in syntactically very different ways, making the commonality in the content difficult to detect. We therefore eschewed the standard translations and expressed correspondents in a way that is itself more similar to an algebraic language, and specifically, to a language of (heterogeneous) relation algebras. Firstly, compared to standard first-order logic, this provides a more compact and computationally more felicitous way to represent MRPs' Kripke and polarity-based correspondents. Secondly, it helps greatly to reveal the relevant parameter for tracking the semantic shift. 

The appropriate representation of correspondents is sensitive both the the classes of modal formulas under consideration and the semantic contexts being compared.  For example, richer relational algebraic languages are required to express conditions corresponding to non-MRP inductive formulas.
Also, in other semantic contexts, as discussed below, it may be possible and convenient to keep closer to standard first-order syntax. For example, this is the case when the contexts being compared are Kripke frames and many-valued Kripke frames \cite{BritzMScThesis}, where standard first-order syntax works well, with the truth-value algebra already standardly incorporated as a parameter in the semantics of the first-order connectives and quantifiers. 

\smallskip
\noindent\textbf{Compositions.} A critical component of this ``relation algebra language'', mentioned in the previous paragraph,  is the introduction of several notions of compositions of binary relations on Kripke frames and on polarities. Many of these compositions can be construed as algebraic operations only in the context of heterogeneous algebras, which have been very useful as semantic structures of multi-type display calculi \cite{frittella2014multi,greco2016linear}. It would be interesting to study the properties of these operations also in this context.

\smallskip
\noindent\textbf{Other semantics contexts.}
As discussed above, the results in the present paper establish a systematic way to track the shift from one semantic context to another via the substitution of one parameter for another. This was seen in the shift from Kripke frames to polarity-based frames where the shift was tracked by a change in parameter from the identity relation $\Delta$ to the relations $I$ and $J$. In the shift from crisp to many-valued polarities the salient parameter was the truth value algebra.    

These are instances of a more general phenomenon, as it is possible to establish such systematic connections also between other semantic contexts. Indeed, this was first observed in the context of many-valued modal logic, where the first-order correspondents on many-valued Kripke frames of all Sahlqvist formulas, according to the classical definition, are verbatim the same as those on crisp Kripke frames, modulo the change in parameter from $\mathbf{2}$ to any perfect Heyting algebra $\mathbf{A}$ \cite{BritzMScThesis}. 

Closely connected to the polarity-based semantics on which we focused in the present paper, is the graph-based semantics for non-distributive LE-logics \cite{graph-based-wollic, conradie2020non} which builds on Plo\v{s}\v{c}ica's lattice representation, and many-valued versions thereof \cite{graph-based-MV}. Similar to what we showed in the present paper, it is possible to find the correspondents of Sahlqvist MRPs on graph-based frames by charging the parameter $\Delta$ in their correspondents on Kripke frames, suitably written as relational inclusions, to the edge relation $E$ of graph-based frames. 

These instances would suggest that it is natural to inquire into the relationship between first-order correspondents on any two relational semantics where the same or overlapping LE-languages can be interpreted. Among many others, one obvious cluster of candidates would be modal logic interpreted on Kripke frames, Fischer-Servi frames for intuitionistic modal logic \cite{fischerservi1977modal}, on Celani and Jansana's relational structures \cite{CelJan99} for Dunn's positive modal logic \cite{Dunn:Pos:ML}, and on many-valued versions of all of these.   

In the present paper we have laid some of the conceptual and mathematical groundwork to systematically pursue these investigations further.

\smallskip
\noindent\textbf{Beyond modal reduction principles.} We have shown that a subclass of the inductive $\mathcal{L}$-formulas, namely the Sahlqvist MRPs, have first-order correspondents on polarity-based frames which can be uniformly obtained as liftings of these inequalities's well-known first-order correspondents on Kripke frames. It is natural to ask how far this result may be extended. As we will now see, there are, in fact, Sahlqvist formulas for which this fails. In particular, consider the Sahlqvist inequality  $\Diamond(p \vee q) \leq \Diamond(p \wedge q)$. It is easy to see that, on Kripke frames,  this inequality defines the condition $\forall x \forall y \neg Rxy$, i.e. $R \subseteq \varnothing$.  The $J$-lifting  on polarity-based frames of this inclusion of binary relations is $A \times X \subseteq R_{\Diamond}$. On the other hand, processing this inequality with ALBA and interpreting the result on polarity-based frames, we find that
\begin{eqnarray}
	\forall p \forall q (\Diamond(p \vee q) \leq \Diamond(p \wedge q))\nonumber
	&\text{iff} &\forall \nomi \forall \nomj (\Diamond(\nomi \vee \nomj) \leq \Diamond(\nomi \wedge \nomj))\nonumber\\
	&\text{iff} &\forall a \forall b \left( R_{\Diamond}^{(0)}[(a^{\uparrow} \cap b^{\uparrow})^{\downarrow}] \supseteq  R_{\Diamond}^{(0)}[a^{\uparrow\downarrow} \cap b^{\uparrow\downarrow}] \right).\label{eqn:counter:example}
\end{eqnarray}
Now, \eqref{eqn:counter:example} is {\em not} equivalent on polarity-based frames to $A \times X \subseteq R_{\Diamond}$. Indeed, consider a polarity-based frame $\mathbb{F} = (\{a \}, \{x\}, I, R_{\Diamond})$ with $I = R_{\Diamond} = \varnothing$. In this frame, there is only one possible assignment to $a$ and $x$ in \eqref{eqn:counter:example}, and under this assignment the right-hand side of the inclusion in \eqref{eqn:counter:example}  becomes $R_{\Diamond}^{(0)}[a^{\uparrow\downarrow} \cap a^{\uparrow\downarrow}] = R_{\Diamond}^{(0)}[\{a\}] = \varnothing$ and hence \eqref{eqn:counter:example}  holds. However, $\mathbb{F}$ clearly violates the requirement  that $A \times X \subseteq R_{\Diamond}$.

It is worth noting that $\Diamond(p \vee q) \leq \Diamond(p \wedge q)$ contains the propositional connectives $\vee$ and $\wedge$, the interpretations of which {\em lose} order-theoretic properties (notably,  distribution over each other) 
when moving from the classical setting, represented by Kripke frames, to the non-distributive setting of polarity-based frames. Meanwhile,  the only connectives in MRPs are  boxes and diamonds, and their order-theoretic properties (namely, complete meet and join preservation, respectively)  remain completely unchanged in passing from the classical to the non-distributive environments. This leads us to the following: 

\begin{conjecture}\label{conj:ind:without:meet:join} For any LE-language $\mathcal{L} = \mathcal{L}_\mathrm{LE}(\mathcal{F}, \mathcal{G})$,
    if $\phi \leq \psi$ is an inductive  $\mathcal{L}$-inequality  which contains no occurrences of $\vee$ or $\wedge$, then its first-order correspondent over polarity-based $\mathcal{L}$-frames is a lifting of its first-order correspondent over Kripke $\mathcal{L}$-frames.
\end{conjecture}

Since crisp polarity-based frames are special $\mathbf{A}$-polarity-based frames, namely with $\mathbf{A}$ equal to the Boolean algebra $\mathbf{2}$, the counterexample above shows that, also in the $\mathbf{A}$-valued case,  Theorem \ref{prop: sahlqvist lifting} cannot be generalised beyond the inductive MRPs to all Sahlqvist formulas. However, we conjecture that it can be generalized to  $\mathcal{L}$-inequalities which contains no occurrences of $\vee$ or $\wedge$, as in Conjecture \ref{conj:ind:without:meet:join}. Moreover, the $\mathbf{A}$-valued case affords us a generalized notion of validity and, with that, of correspondence. To be precise, for $\kappa \in \mathbf{A}$, a sequent $\phi \vdash \psi$ is $\kappa$-valid on a $\mathbf{A}$-polarity-based frame if $S_A(\val{\phi},\val{\psi}) \geq \kappa$, equivalently $S_X(\descr{\psi},\descr{\phi}) \geq \kappa$, under all valuations. Accordingly, $\phi \vdash \psi$ is said to  $\kappa$-correspond to a first-order conditions $\alpha$ (expressed as an inclusion of relational compositions, or in any other way) if the $\kappa$-validity of $\phi \vdash \psi$ on any $\mathbf{A}$-polarity-based frame is equivalent to the $\kappa$-validity of $\alpha$ on that frame. $\kappa$-correspondence is studied in \cite{conradie2019logic} where it is extensively used as part of a logical treatment of categorization theory applied to construct a formal analysis of certain phenomena in multi-market competition. 

\smallskip
\noindent\textbf{Proof-theoretic developments.} The systematic study of the connections among different semantic settings such as Kripke frames, polarity-based frames, and graph-based frames \cite{conradie2020non} can also contribute to the advancement of the proof theory of LE-logics; indeed, further developments in this direction can be obtained building on the labelled calculus in the language of ALBA introduced in \cite{DDG2022} as a reformulation of the labelled calculus G3K for classical normal modal logic. This calculus can be straightforwardly adapted so to introduce e.g.~labelled calculi for the logics of Rough Concept Analysis \cite{greco2019logics}.

\bibliography{ref}
\bibliographystyle{plain}

\appendix
\section{Inductive and Sahlqvist (analytic) LE-inequalities}\label{Inductive:Fmls:Section}
In this section we  recall the notions of inductive and Sahlqvist LE-inequalities introduced in \cite{CoPa:non-dist} and their  `analytic' restrictions introduced in \cite{greco2018unified} for distributive LEs, and then generalized to arbitrary LEs  in \cite{greco2018algebraic}.  Each inequality in any of these classes 
is canonical  and elementary (cf.~\cite[Theorems 8.8 and 8.9]{CoPa:non-dist}).  

		

		\begin{definition}[\textbf{Signed Generation Tree}]
			\label{def: signed gen tree}
			The \emph{positive} (resp.\ \emph{negative}) {\em generation tree} of any $\mathcal{L}_\mathrm{LE}$-term $s$ is defined by labelling the root node of the generation tree of $s$ with the sign $+$ (resp.\ $-$), and then propagating the labelling on each remaining node as follows:
			\kern-\parskip
			\begin{itemize}
			\setlength{\itemsep}{0.2pt}
              \setlength{\parskip}{0pt}
              \setlength{\parsep}{0pt}
              \item For any node labelled with $ \lor$ or $\land$, assign the same sign to its children nodes.
			\item For any node labelled with $h\in \mathcal{F}\cup \mathcal{G}$ of arity $n_h\geq 1$, and for any $1\leq i\leq n_h$, assign the same (resp.\ the opposite) sign to its $i$th child node if $\varepsilon_h(i) = 1$ (resp.\ if $\varepsilon_h(i) = \partial$).
            \end{itemize}
            \kern-\parskip
			Nodes in signed generation trees are \emph{positive} (resp.\ \emph{negative}) if are signed $+$ (resp.\ $-$).
		\end{definition}
		
		Signed generation trees are mostly used in the context of term inequalities $s\leq t$. In this context we  typically consider the positive generation tree $+s$ for the left-hand side and the negative one $-t$ for the right-hand side. A term-inequality $s\leq t$ is \emph{uniform} in a given variable $p$ if all occurrences of $p$ in both $+s$ and $-t$ have the same sign, and  $s\leq t$ is $\epsilon$-\emph{uniform} in a (sub)array $\overline{p}$ of its variables if $s\leq t$ is uniform in $p$, occurring with the sign indicated by $\epsilon$, for every $p$ in $\overline{p}$\footnote{\label{footnote:uniformterms}Recall that if a term inequality $s(\overline{p},\overline{q})\leq t(\overline{p},\overline{q})$ is $\epsilon$-uniform in $\overline{p}$, 
		then the validity of $s\leq t$ is equivalent to the validity of $s(\overline{\top^{\epsilon(i)}},\overline{q})\leq t(\overline{\top^{\epsilon(i)}},\overline{q})$, where $\top^{\epsilon(i)}=\top$ if $\epsilon(i)=1$ and $\top^{\epsilon(i)}=\bot$ if $\epsilon(i)=\partial$. }.
		
		For any term $s(p_1,\ldots p_n)$, any order-type $\epsilon$ over $n$, and any $1 \leq i \leq n$, an \emph{$\epsilon$-critical node} in a signed generation tree of $s$ is a leaf node $+p_i$ if $\epsilon (i) = 1$ and $-p_i$ if $\epsilon (i) = \partial$. An $\epsilon$-{\em critical branch} in the tree is a branch from an $\epsilon$-critical node. Variable occurrences corresponding to $\epsilon$-critical nodes are to be solved for. 
		
		For all terms $s(p_1,\ldots p_n)$ and  order-types $\epsilon$, we say that $+s$ (resp.\ $-s$) {\em agrees with} $\epsilon$, and write $\epsilon(+s)$ (resp.\ $\epsilon(-s)$), if every leaf in   $+s$ (resp.\ $-s$) is $\epsilon$-critical.
		We  also write $+s'\prec \ast s$ (resp.\ $-s'\prec \ast s$) to indicate that the subterm $s'$ inherits the positive (resp.\ negative) sign from  $\ast s$. Finally, we will write $\epsilon(\gamma) \prec \ast s$ (resp.\ $\epsilon^\partial(\gamma) \prec \ast s$) to indicate that the signed subtree $\gamma$, with the sign inherited from $\ast s$, agrees with $\epsilon$ (resp.\ with $\epsilon^\partial$).
		We  write $\phi(!x)$ (resp.~$\phi(!\overline{x})$) to indicate that the variable $x$ (resp.~each variable $x$ in $\overline{x}$) occurs exactly once in $\phi$. Accordingly, we  write $\phi(\gamma / !x)$  (resp.~$\phi(\overline{\gamma}/!\overline{x})$) to indicate the formula obtained from $\phi$ by substituting $\gamma$ (resp.~each term $\gamma$ in $\overline{\gamma}$) for the unique occurrence of (its corresponding variable) $x$ in $\phi$.

		\begin{definition}
			\label{def:good:branch}
			Nodes in signed generation trees are called \emph{$\Delta$-adjoints}, \emph{SLR}, \emph{SRR}, and \emph{SRA}, according to the specification given in Table \ref{Join:and:Meet:Friendly:Table}.
			A branch in a signed generation tree $\ast s$, with $\ast \in \{+, - \}$, is  a \emph{good branch} if it is the concatenation of two paths $P_1$ and $P_2$, one of which may possibly be of length $0$, such that $P_1$ is a path from the leaf consisting (apart from variable nodes) only of PIA-nodes, and $P_2$ consists (apart from variable nodes) only of Skeleton-nodes. 
A branch is \emph{excellent} if it is good and in $P_1$ there are only SRA-nodes. A good branch is \emph{Skeleton} if the length of $P_1$ is $0$ (hence Skeleton branches are excellent), and  is {\em SLR}, or {\em definite}, if  $P_2$ only contains SLR nodes.
			\begin{table}
				\begin{center}
                \bgroup
                \def\arraystretch{1.2}
					\begin{tabular}{| c | c |}
						\hline
						Skeleton &PIA\\
						\hline
						$\Delta$-adjoints & Syntactically Right Adjoint (SRA) \\
						\begin{tabular}{ c c c c c c}
							$+$ &$\vee$ &\\
							$-$ &$\wedge$ \\
							\hline
						\end{tabular}
						&
						\begin{tabular}{c c c c }
							$+$ &$\wedge$ &$g$ & with $n_g = 1$ \\
							$-$ &$\vee$ &$f$ & with $n_f = 1$ \\
							\hline
						\end{tabular}
						\\
						Syntactically Left Residual (SLR) &Syntactically Right Residual (SRR)\\
						\begin{tabular}{c c c c }
							$+$ &  &$f$ & with $n_f \geq 1$\\
							$-$ &  &$g$ & with $n_g \geq 1$ \\
						\end{tabular}
						&\begin{tabular}{c c c c}
							$+$ & &$g$ & with $n_g \geq 2$\\
							$-$ &  &$f$ & with $n_f \geq 2$\\
						\end{tabular}
						\\
						\hline
					\end{tabular}
                \egroup
				\end{center}
				\caption{Skeleton and PIA nodes for $\mathrm{LE}$.}\label{Join:and:Meet:Friendly:Table}
				\vspace{-1em}
			\end{table}
		\end{definition}
We refer to \cite[Remark 3.3]{CoPa:non-dist} 
and \cite[Section 3]{CoGhPa14} for a discussion on notation and terminology.	

\begin{definition}[Inductive inequalities]\label{Inducive:Ineq:Def}
For any order-type $\epsilon$ and any irreflexive and transitive relation   $<_{\Omega}$ on $p_1,\ldots p_n$, the signed generation tree $*s$ $(* \in \{-, + \})$ of a term $s(p_1,\ldots p_n)$ is \emph{$(\Omega, \epsilon)$-inductive} if
\kern-\parskip
			\begin{enumerate}
			\setlength{\itemsep}{0.2pt}
              \setlength{\parskip}{0pt}
              \setlength{\parsep}{0pt}
				\item for all $1 \leq i \leq n$, every $\epsilon$-critical branch with leaf $p_i$ is good (cf.\ Definition \ref{def:good:branch});
				\item every $m$-ary SRR-node occurring in the critical branch is of the form $ \circledast(\gamma_1,\dots,\gamma_{j-1},\beta,\gamma_{j+1}\ldots,\gamma_m)$, where for any $h\in\{1,\ldots,m\}\setminus j$: 
\kern-\parskip
\begin{enumerate}
			\setlength{\itemsep}{0.2pt}
              \setlength{\parskip}{0pt}
              \setlength{\parsep}{0pt}
\item $\epsilon^\partial(\gamma_h) \prec \ast s$ (cf.\ discussion before Definition \ref{def:good:branch}), and
%
\item $p_k <_{\Omega} p_i$ for every $p_k$ occurring in $\gamma_h$ and for every $1\leq k\leq n$.
\end{enumerate}
\kern-\parskip
	\end{enumerate}
	\kern-\parskip
			We  refer to $<_{\Omega}$ as the \emph{dependency order} on the variables. An inequality $s \leq t$ is \emph{$(\Omega, \epsilon)$-inductive} if  $+s$ and $-t$ are $(\Omega, \epsilon)$-inductive. An inequality $s \leq t$ is \emph{inductive} if it is $(\Omega, \epsilon)$-inductive for some $<_\Omega$ and $\epsilon$.
		\end{definition}
		
		We refer to formulas $\phi$ such that only PIA nodes occur in $+\phi$ (resp.\ $-\phi$) as {\em positive} (resp.\ {\em negative}) {\em PIA-formulas}, and to formulas $\xi$ such that only Skeleton nodes occur in $+\xi$ (resp.\ $-\xi$) as {\em positive} (resp.\ {\em negative}) {\em Skeleton-formulas}\label{page: positive negative PIA}. PIA formulas $\ast \phi$ in which no nodes $+\wedge$ and $-\vee$ occur are referred to as {\em definite}. Skeleton formulas $\ast \xi$ in which no nodes $-\wedge$ and $+\vee$ occur are referred to as {\em definite}.

\begin{definition}\label{Sahlqvist:Ineq:Def}
For an order-type $\epsilon$, the signed generation tree $\ast s$, $\ast \in \{-, + \}$, of a term $s(p_1,\ldots p_n)$ is \emph{$\epsilon$-Sahlqvist} if every $\epsilon$-critical branch is excellent. An inequality $s \leq t$ is \emph{$\epsilon$-Sahlqvist} if the trees $+s$ and $-t$ are both $\epsilon$-Sahlqvist.  An inequality $s \leq t$ is \emph{Sahlqvist} if it is $\epsilon$-Sahlqvist for some $\epsilon$.
\end{definition}

\begin{definition}[Analytic inductive and  analytic Sahlqvist inequalities]
	\label{def:type5}
	For every order-type $\epsilon$ and every irreflexive and transitive relation $<_\Omega$ on the variables $p_1,\ldots p_n$,
			the signed generation tree $\ast s$ ($\ast\in \{+, -\}$) of a term $s(p_1,\ldots p_n)$ is \emph{analytic $(\Omega, \epsilon)$-inductive} (resp.~\emph{analytic $\epsilon$-Sahlqvist}) if
			\smallskip
			
			{{\centering 
			1. \quad $\ast s$ is $(\Omega, \epsilon)$-inductive (resp.~$\epsilon$-Sahlqvist); 
			\quad\quad\quad
			2. \quad every branch of $\ast s$ is good (cf.\ Definition \ref{def:good:branch}).
	        \par}}
	\smallskip       
	 
	\noindent An inequality $s \leq t$ is \emph{analytic $(\Omega, \epsilon)$-inductive} (resp.~\emph{analytic $\epsilon$-Sahlqvist})  if $+s$ and $-t$ are both analytic  $(\Omega, \epsilon)$-inductive (resp.~analytic $\epsilon$-Sahlqvist). An inequality $s \leq t$ is \emph{analytic inductive} (resp.~\emph{analytic Sahlqvist}) if is analytic $(\Omega, \epsilon)$-inductive (resp.~analytic $\epsilon$-Sahlqvist)  for some $\Omega$ and $\epsilon$ (resp.~for some $\epsilon$).
\end{definition}	

\begin{notation}\label{notation: analytic inductive}
Following \cite{chen2021syntactic},  we will sometimes represent $(\Omega, \epsilon)$-analytic inductive inequalities as follows: \[(\varphi\leq \psi)[\overline{\alpha}/!\overline{x}, \overline{\beta}/!\overline{y},\overline{\gamma}/!\overline{z}, \overline{\delta}/!\overline{w}],\] where $(\varphi\leq \psi)[!\overline{x}, !\overline{y},!\overline{z}, !\overline{w}]$ is  the skeleton of the given inequality,  $\overline{\alpha}$ (resp.~$\overline{\beta}$) denotes the positive (resp.~negative) maximal PIA-subformulas, i.e.~each $\alpha$ in $\overline{\alpha}$ and $\beta$ in $\overline{\beta}$ contains at least one $\varepsilon$-critical occurrence of some propositional variable, and moreover:\\
\kern-\parskip
\begin{tabular}{rl c rl}
1. & for all $\alpha\in \overline{\alpha}$, either 
$+\alpha\prec +\varphi$ or $+\alpha\prec -\psi$; &&
2. & for all $\beta\in \overline{\beta}$, either 
  $-\beta\prec +\varphi$ or $-\beta\prec -\psi$,
  \end{tabular}
and $\overline{\gamma}$ (resp.~$\overline{\delta}$) denotes the positive (resp.~negative) maximal $\varepsilon^{\partial}$-subformulas,  i.e.:\\
\kern-\parskip
\begin{tabular}{rl c rl}
  1. & for each $\gamma\in \overline{\gamma}$, either   $+\gamma\prec +\varphi$ or $+\gamma\prec -\psi$; &&
  2. & for each $\delta\in \overline{\delta}$, either  $-\delta\prec +\varphi$ or $-\delta\prec -\psi$.
    \end{tabular}
    \kern-\parskip
    For the sake of a more compact notation, in what follows we sometimes write $(\varphi\leq \psi)[\overline{\alpha}, \overline{\beta},\overline{\gamma}, \overline{\delta}]$ in place of \[ (\varphi\leq \psi)[\overline{\alpha}/!\overline{x}, \overline{\beta}/!\overline{y},\overline{\gamma}/!\overline{z}, \overline{\delta}/!\overline{w}]. \]
\end{notation}

\begin{definition} 
		\label{def: RA and LA}
			%
			For every definite positive PIA $\mathcal{L}_{\mathrm{LE}}$-formula $\psi = \psi(!x, \oz)$, and any definite negative PIA $\mathcal{L}_{\mathrm{LE}}$-formula $\phi = \phi(!x, \oz)$ such that $x$ occurs in them exactly once, the $\mathcal{L}^\ast_\mathrm{LE}$-formulas $\mathsf{LA}(\psi)(u, \oz)$ and $\mathsf{RA}(\phi)(u, \oz)$ (for $u \in Var - (x \cup \oz)$) are defined by simultaneous recursion as follows:
			
			{{\centering
				\begin{tabular}{r c l}
					
					$\mathsf{LA}(x)$ &= &$u$;\\
					$\mathsf{LA}(g(\overline{\psi_{-j}(\oz)},\psi_j(x,\oz), \overline{\phi(\oz)}))$ &= &$\mathsf{LA}(\psi_j)(g^{\flat}_{j}(\overline{\psi_{-j}(\oz)},u, \overline{\phi(\oz)} ), \oz)$;\\
					$\mathsf{LA}(g(\overline{\psi(\oz)}, \overline{\phi_{-j}(\oz)},\phi_j(x,\oz)))$ &= &$\mathsf{RA}(\phi_j)(g^{\flat}_{j}(\overline{\psi(\oz)}, \overline{\phi_{-j}(\oz)},u), \oz)$;\\[1.5mm]
					$\mathsf{RA}(x)$ &= &$u$;\\
					$\mathsf{RA}(f(\overline{\phi_{-j}(\oz)},\phi_j(x,\oz), \overline{\psi(\oz)}))$ &= &$\mathsf{RA}(\phi_j)(f^{\sharp}_{j}(\overline{\phi_{-j}(\oz)},u, \overline{\psi(\oz)} ), \oz)$;\\
					$\mathsf{RA}(f(\overline{\phi(\oz)}, \overline{\psi_{-j}(\oz)},\psi_j(x,\oz)))$ &= &$\mathsf{LA}(\psi_j)(f^{\sharp}_{j}(\overline{\phi(\oz)}, \overline{\psi_{-j}(\oz)},u), \oz)$.\\
				\end{tabular}
			\par}}
			\noindent Above, $\overline{\phi_{-j}}$ denotes the vector obtained by removing the $j$th coordinate of $\overline{\phi}$.
			 Thus, if $\phi(!x): = \Diamond_{i_1}\cdots \Diamond_{i_n}x$ and $\psi(!x): = \Box_{i_1}\cdots \Box_{i_n}x$, then $\mathsf{RA}(\phi)(!u) = \blacksquare_{i_n}\cdots \blacksquare_{i_1} u$ and $\mathsf{LA}(\psi)(!u) = \Diamondblack_{i_n}\cdots \Diamondblack_{i_1} u$.
\end{definition}
\begin{lemma}
\label{lemma:polarities of la-ra}
If $\phi = \phi(!x, \oz)$ (resp.~~$\psi = \psi(!x, \oz)$) are definite positive (resp.~negative) PIA $\mathcal{L}_{\mathrm{LE}}$-formulas  such that $x$ occurs in them exactly once, 
\kern-\parskip
\begin{enumerate}
\setlength{\itemsep}{0.2pt}
\setlength{\parskip}{0pt}
\setlength{\parsep}{0pt}
\item if $+x\prec +\phi$, then  $\mathsf{LA}(\phi)(u, \oz)$ is monotone in $u$ and, for each $z$ in $\oz$, it has the opposite polarity to the polarity of $\phi$ in $z$;
\item if $-x\prec +\phi$, then  $\mathsf{LA}(\phi)(u, \oz)$ is antitone in $u$ and, for each $z$ in $\oz$, it has the same polarity as $\phi$ in $z$;
\item if $+x\prec +\psi$, then  $\mathsf{RA}(\psi)(u, \oz)$ is monotone in $u$ and, for each $z$ in $\oz$, it has the opposite polarity to the polarity of $\psi$ in $z$;
\item if $-x\prec +\psi$, then  $\mathsf{RA}(\psi)(u, \oz)$ is antitone in $u$ and, for each $z$ in $\oz$, it has the same polarity as $\psi$ in $z$.
\end{enumerate}
\kern-\parskip
\end{lemma}

\section{Proof of Proposition \ref{lemma:correspondences} }
\label{sec: proof of mv correspondences}
		
		\begin{tabular}{lrcll}
		2.&	$\forall p (\Box p \leq p)$
			&iff &$\forall \bf{m} (\Box \bf{m} \leq \bf{m})$ &(Prop.~\ref{prop:albaoutput})\\
			&&iff &$\forall \alpha \forall x (R^{(0)}_{\Box}[\{\alpha/x\}^{\downarrow\uparrow}]  \leq \{\alpha/x\}^{\downarrow})$ & ($\cnomm := \{\alpha/x\}$) \\
		&	&iff &$\forall \alpha \forall x (R^{(0)}_{\Box}[\{\alpha/x\}]  \leq \{\alpha/x\}^{\downarrow})$ &($I$-compatibility of $R_{\Box}$)\\
		&&	iff &$\forall \alpha \forall x \forall a(\alpha \to a R_{\Box} x  \leq \alpha \to a I x)$ &\\
		&&	iff &$R_{\Box} \leq I$. &
		\end{tabular}
	
	\smallskip
	\begin{tabular}{lrcll}
		3.&	$\forall p ( p \leq \Diamond p)$
			&iff &$\forall \nomj ( \nomj \leq \Diamond \nomj)$ &(Prop.~\ref{prop:albaoutput})\\
			&&iff &$\forall \alpha \forall a (R^{(0)}_{\Diamond}[\{\alpha/a\}^{\uparrow\downarrow}]  \leq \{\alpha/a\}^{\uparrow})$ & ($\nomj := \{\alpha/a\}$) \\
		&	&iff &$\forall \alpha \forall a (R^{(0)}_{\Diamond}[\{\alpha/a\}]  \leq \{\alpha/a\}^{\uparrow})$ &($I$-compatibility of $R_{\Diamond}$)\\
		&&	iff &$\forall \alpha \forall a \forall x(\alpha \to x R_{\Diamond} a  \leq \alpha \to a I x)$ &\\
		&&	iff &$\forall \alpha \forall a \forall x(\alpha \to a R_{\blacksquare} x  \leq \alpha \to a I x)$ &\\
		&&	iff &$R_{\blacksquare} \leq I$. &
		\end{tabular}

\smallskip 
		\begin{tabular}{lrcll}
		4.	&$\forall p (\Box p \leq \Box\Box p)$&
			iff &$\forall \bf{m} (\Box \bf{m} \leq \Box \Box \bf{m})$ &(Prop.~\ref{prop:albaoutput})\\
			&&iff &$\forall \alpha \forall x (R^{(0)}_{\Box}[\{\alpha/x\}^{\downarrow\uparrow}]  \leq R_\Box^{(0)}[(R_\Box^{(0)}[\{\alpha/x\}^{\downarrow\uparrow}])^\uparrow])$ & ($\cnomm := \{\alpha/x\}$) \\
			&&iff &$\forall \alpha \forall x (R^{(0)}_{\Box}[\{\alpha/x\}]  \leq R_\Box^{(0)}[(R_\Box^{(0)}[\{\alpha/x\}])^\uparrow])$ &($I$-compatibility of $R_{\Box}$)\\
			&&iff &$\forall \alpha \forall x \forall a(\alpha \to a R_{\Box} x  \leq \alpha \to a R_\Box;R_\Box x)$ & 
			\\
			&&iff &$R_{\Box} \leq R_\Box;R_\Box$. & 
		\end{tabular}

\smallskip 
		\begin{tabular}{lrcll}
		6.	&$\forall p (p \leq \Box\Diamond p)$&
			iff &$\forall \bf{j} (\Diamondblack \bf{j} \leq \Diamond\bf{j})$ &(Prop.~\ref{prop:albaoutput})\\
			&&iff &$\forall \alpha \forall a (R^{(0)}_\Diamond[\{\alpha/a\}^{\uparrow\downarrow}] \leq R^{(0)}_\Diamondblack[\{\alpha/a\}^{\uparrow\downarrow}])$ & ($\nomj := \{\alpha/a\}$) \\
		&&	iff &$\forall \alpha \forall a (R^{(0)}_\Diamond[\{\alpha/a\}] \leq R^{(0)}_\Diamondblack[\{\alpha/a\}])$ &($I$-compatibility of $R_{\Diamond}$ and $R_\Diamondblack$)\\
			&&iff &$\forall \alpha \forall a \forall x(\alpha \to x  R_\Diamond a  \leq \alpha \to x R_\Diamondblack a)$ & 
			\\
			&&iff &$R_\Diamond \leq R_\Diamondblack$. & 
		\end{tabular}
	
\smallskip
	
		\begin{tabular}{lrcll}
		7.	&$\forall p (\Diamond\Box p \leq p)$&
			iff &$\forall \bf{j} (\Diamond\bf{j} \leq \Diamondblack\bf{j})$ &(Prop.~\ref{prop:albaoutput})\\
			&&iff &$\forall \alpha \forall a (R^{(0)}_\Diamondblack[\{\alpha/a\}^{\uparrow\downarrow}] \leq R^{(0)}_\Diamond[\{\alpha/a\}^{\uparrow\downarrow}])$ & ($\nomj := \{\alpha/a\}$) \\
			&&iff &$\forall \alpha \forall a (R^{(0)}_\Diamondblack[\{\alpha/a\}] \leq R^{(0)}_\Diamond[\{\alpha/a\}])$ &($I$-compatibility of $R_{\Diamond}$ and $R_\Diamondblack$)\\
			&&iff &$\forall \alpha \forall a \forall x(\alpha \to x  R_\Diamondblack a  \leq \alpha \to x R_\Diamond a)$ & 
			\\
			&&iff &$R_\Diamondblack \leq R_\Diamond$. & 
		\end{tabular}
		
\smallskip
		\begin{tabular}{lrcll}
		8.	&$\forall p (p \leq \Box p)$
			&iff &$\forall \bf{m} (\bf{m} \leq \Box\bf{m})$ &(Prop.~\ref{prop:albaoutput})\\
			&&iff &$\forall \alpha \forall x (\{\alpha/x\}^\downarrow \leq R^{(0)}_\Box[\{\alpha/x\}^{\downarrow\uparrow}])$ & ($\cnomm := \{\alpha/x\}$) \\
		&&	iff &$\forall \alpha \forall x (\{\alpha/x\}^\downarrow \leq R^{(0)}_\Box[\{\alpha/x\}])$ &($I$-compatibility of $R_{\Box}$)\\
			&&iff &$\forall \alpha \forall x \forall a(\alpha \to a  I x  \leq \alpha \to a R_\Box x)$ & 
			\\
			&&iff &$I \leq R_\Box$. &
		\end{tabular}
	
\smallskip 
		\begin{tabular}{lrcll}
		9.	&$\forall p (\Diamond p \leq p)$&
			iff &$\forall \bf{m} (\bf{m} \leq \blacksquare\bf{m})$ &(Prop.~\ref{prop:albaoutput})\\
			&&iff &$\forall \alpha \forall x (\{\alpha/x\}^\downarrow \leq R^{(0)}_\blacksquare[\{\alpha/x\}^{\downarrow\uparrow}])$ & ($\cnomm := \{\alpha/x\}$) \\
			&&iff &$\forall \alpha \forall x (\{\alpha/x\}^\downarrow \leq R^{(0)}_\blacksquare[\{\alpha/x\}])$ &($I$-compatibility of $R_{\blacksquare}$)\\
			&&iff &$\forall \alpha \forall x \forall a(\alpha \to a  I x  \leq \alpha \to a R_\blacksquare x)$ & 
			\\
			&&iff &$I \leq R_\blacksquare$. & 
		\end{tabular}
	
\smallskip		
		\begin{tabular}{lrcll}
		10.	&$\forall p (\Box\Box p \leq \Box p)$
		&	iff &$\forall \bf{m} (\Box\Box\bf{m} \leq \Box\bf{m})$ &(Prop.~\ref{prop:albaoutput})\\
		&&	iff &$\forall \alpha \forall x (R^{(0)}_\Box[(R^{(0)}_\Box[\{\alpha/x\}^{\downarrow\uparrow}])^\uparrow] \leq R^{(0)}_\Box[\{\alpha/x\}^{\downarrow\uparrow}])$ & ($\cnomm := \{\alpha/x\}$) \\
			&&iff &$\forall \alpha \forall x (R^{(0)}_\Box[(R^{(0)}_\Box[\{\alpha/x\}])^\uparrow] \leq R^{(0)}_\Box[\{\alpha/x\}])$ &($I$-compatibility of $R_{\Box}$)\\
			&&iff &$\forall \alpha \forall x \forall a(\alpha \to a  (R_\Box;R_\Box) x  \leq \alpha \to a R_\Box x)$. & 
			\\
			&&iff &$R_\Box;R_\Box \leq R_\Box$. & 
		\end{tabular}
\smallskip
	
		\begin{tabular}{lrcll}
		11. &	$\forall p (\Diamond p \leq \Diamond\Diamond p)$
			&iff &$\forall \bf{j} (\Diamond\bf{j} \leq \Diamond\Diamond\bf{j})$ &(Prop.~\ref{prop:albaoutput})\\
			&&iff &$\forall \alpha \forall a (R^{(0)}_\Diamond[(R^{(0)}_\Diamond[\{\alpha/a\}^{\uparrow\downarrow}])^\downarrow] \leq R^{(0)}_\Diamond[\{\alpha/a\}^{\uparrow\downarrow}])$ & ($\nomj := \{\alpha/a\}$) \\
			&&iff &$\forall \alpha \forall a ([R^{(0)}_\Diamond[(R^{(0)}_\Diamond[\{\alpha/a\}])^\downarrow] \leq R^{(0)}_\Diamond[\{\alpha/a\}]])$ &($I$-compatibility of $R_{\Diamond}$)\\
			&&iff &$\forall \alpha \forall a \forall x(\alpha \to x  (R_\Diamond;R_\Diamond) a  \leq \alpha \to x R_\Diamond a)$ & 
			\\
			&&iff &$R_\Diamond;R_\Diamond \leq R_\Diamond$. & 
		\end{tabular}
	
\smallskip

		\begin{tabular}{lrcll}
12. &			$\forall p (\Diamond p \leq \Box p)$
			&iff &$\forall \bf{m} (\bf{m} \leq \blacksquare\Box\bf{m})$ &(Prop.~\ref{prop:albaoutput})\\
			&&iff &$\forall \alpha \forall x (\{\alpha/x\}^\downarrow \leq R^{(0)}_\blacksquare[(R^{(0)}_\Box[\{\alpha/x\}^{\downarrow\uparrow}])^\uparrow])$ & ($\cnomm := \{\alpha/x\}$) \\
			&&iff &$\forall \alpha \forall x (\{\alpha/x\}^\downarrow \leq R^{(0)}_\blacksquare[(R^{(0)}_\Box[\{\alpha/x\}])^\uparrow])$ &($I$-compatibility of $R_{\Box}$)\\
			&&iff &$\forall \alpha \forall x \forall a(\alpha \to a I x  \leq \alpha \to a (R_\blacksquare;R_\Box) x)$ &
			\\
			&&iff &$I \leq R_\blacksquare;R_\Box.$ &
		\end{tabular}

\end{document}